\documentclass[preprint]{revtex4}
\usepackage{epsfig,graphics,subfigure,amssymb,amsmath,subeqnarray,color,xcolor,amssymb}
\usepackage[colorlinks=true, citecolor=red, linkcolor=blue,urlcolor=blue ]{hyperref}
\usepackage[mathscr]{euscript}
\usepackage{amsthm}
\usepackage{booktabs}
\usepackage{graphicx}
\usepackage{mathtools}
\usepackage{siunitx}
\usepackage{upgreek}
\usepackage{xcolor}
\usepackage{cancel}
\usepackage{setspace}
\usepackage{bbm}
\usepackage{float}
\usepackage{enumitem}
\usepackage{hyperref}

\usepackage[T1]{fontenc}
\usepackage[utf8]{inputenc}

\DeclareSymbolFont{rsfs}{U}{rsfs}{m}{n}
\DeclareSymbolFontAlphabet{\mathscrsfs}{rsfs}

\begin{document}
\title{Hydrodynamic Instabilities of Active Jets}

%
%
%
%
%

\author{Marco Vona$^1$, Isabelle Eisenmann$^2$ , Nicolas Desprat$^2$, Rapha\"el Jeanneret$^{2,*}$, Takuji Ishikawa$^3$, Eric Lauga$^{1,*}$ } 
\affiliation{
	$^1$Department of Applied Mathematics and Theoretical Physics, University of Cambridge, Wilberforce Road, Cambridge CB3 0WA, UK.\\ 
	$^2$Laboratoire de Physique de l'\'Ecole normale sup\'erieure, ENS, Universit\'e PSL, CNRS, Sorbonne Universit\'e, Universit\'e Paris Cit\'e, F-75005 Paris, France.\\
	$^3$Department of Biomedical Engineering, Tohoku University, 6-6-01, Aoba, Aramaki, Aoba-ku, Sendai 980-8579, Japan.}	
\email[Correspondence: ]{raphael.jeanneret@phys.ens.fr, e.lauga@damtp.cam.ac.uk}

\begin{abstract}

Using a combination of theory, experiments, and numerical simulations, we investigate the stability of coherent structures in a suspension of strongly aligned active swimmers. We show that a dilute jet of {pullers} undergoes a  {{pearling}} instability, while a jet of {pushers} exhibits a helical (or, in two dimensions, zigzag) instability. We further characterise the nonlinear evolution of these instabilities, deriving exact and approximate solutions for the spreading and mutual attraction of puller clusters, as well as the wavelength coarsening of the  helical instability. Our theoretical predictions closely match  the experimentally observed wavelengths, timescales, and flow fields in suspensions of photophobic algae, as well as results from direct numerical simulations. These findings reveal the intrinsic instability mechanisms of aligned active suspensions and demonstrate that coherent structures can be destabilised by the flows they generate.
\end{abstract}
\maketitle

\setlength\abovedisplayskip{15pt}
\setlength\belowdisplayskip{15pt}
\def\u{{\bf u}}\def\n{{\bf n}}\def\f{{\bf f}}
\def\t{{\bf t}}\def\d{{\rm d}}\def\Pe{{\rm Pe}}
\def\r{{\bf r}}\def\x{{\bf x}}
\def\e{{\bf e}}\def\1{{\bf 1}}\def\0{{\bf 0}}
\def\p{{\partial}}
\def\C{{\boldsymbol C}}\def\D{{\boldsymbol D}}
\def\x{{\boldsymbol x}}
\def\v{\vspace {1cm}}
\def\textheight{25cm}

\newcommand{\bb}[1]{\boldsymbol{#1}}
\newtheorem{definition}{Definition}
\newtheorem{theorem}{Theorem}
\newtheorem{lemma}{Lemma}
\newtheorem{proposition}{Proposition}
\newtheorem{example}{Example}
\definecolor{mycolour}{RGB}{0,128,0}
\section{Introduction}
Suspensions of  {active} particles  {exhibit} rich and diverse behaviours~\cite{koch2011collective, ishikawa2008coherent},  {such as accumulation near boundaries \cite{elgeti2015physics}, flocking \cite{vicsek2012collective}, enhanced aggregation and cluster formation \cite{kraikivski2006enhanced}, and phase-transitioning between solid-like and gas-like states \cite{theurkauff2012dynamic}. Active swimmers have even been proposed as a {vessel} for the direct transport of cargo on the microscale, with applications to drug delivery and nanoscale assembly \cite{wang2013nanomachines}.} Due to long-range hydrodynamic interactions, {actively-driven flows} often present much larger  {length scales} than those characterising the individual active particles~\cite{saintillan2008instabilitiesA, saintillan2008instabilitiesB,saintillan2013active}, leading to collective motion and well-studied phenomena such {as} bacterial turbulence~\cite{dunkel2013fluid}, enhanced tracer diffusivity~\cite{stenhammar2017role},  {and long-range density modulations in the presence of inclusions \cite{arnoulx2024anomalous}}.    

Much is known about collective dynamics induced by biased motility in the presence of an external stimulus, such as light~\cite{williams2011photo}, magnetic fields~\cite{thery2020self}, hydrodynamic shear~\cite{durham2009disruption}, or gravity~\cite{pedley1988growth, pedley1990new, bees2020advances, fung2020bifurcation}. Much attention has in particular been given to the formation of  {macroscopic}, beam-like concentrated structures with large particle density in active suspensions~\cite{pedley1988growth, bees2020advances}.  {A classical mechanism leading to the formation of such structures is sedimentation~\cite{pedley1988growth, kessler1985hydrodynamic, denissenko2007velocity, fung2023analogy, font2019collective}, which, in combination with the particles' geometry, can result in self-enriching downwelling regions of high shear~\cite{koch1989instability}.}  {Numerical experiments have further revealed that the sole  tendency of particles  to align in response  {to}  {external} stimuli, such as light~\cite{williams2011photo}, magnetic fields~\cite{thery2020self}, oxygen concentration~\cite{garcia2013light, jibuti2014self}, or the direction of gravity~\cite{bees2020advances, metcalfe2001falling, kessler1984gyrotactic}, may be sufficient for the formation of coherent slender structures, even in the absence of sedimentation. {In the paper,} we will  refer to these structures as {\it active jets}.
 }

 {To our knowledge, a controlled experimental realisation of such active jets is still lacking. The main difficulty is that cells must be consistently well-aligned to prevent the jet from flying apart. This is only possible if cells are made to reorient on a timescale which is much faster than that of the flow itself. For experiments involving gravitaxis or gyrotaxis, typical reorientation timescales for the model organism \textit{C.~reinhardtii} are known to be on the order of seconds~\cite{pedley1990new}, which is comparable to the observed flow timescales. In this situation, the jet would therefore quickly become incoherent}.

As a result of these experimental challenges, studies of active jets {have} been mostly computational in nature~\cite{jibuti2014self, ishikawa2022instability}. An intriguing feature of such jets is that they are fundamentally unstable, exhibiting instabilities driven by cell activity.  {Notably,} numerical studies~\cite{ishikawa2022instability} of concentrated gyrotactic plumes of strongly active sedimenting squirmers~\cite{pak2014generalized} 
demonstrated {{actively-driven} shape instabilities that depended} on whether the swimmers were ``pushers''~\cite{drescher2011fluid} or ``pullers''~\cite{drescher2010direct}. Specifically, pusher jets were prone to a  {helical}  {instability}~\cite{ishikawa2022instability}, while puller jets presented a  {pearling} instability~\cite{rayleigh1879stability}. A similar numerical treatment recovered a pearling instability for phototactic pullers under confinement~\cite{jibuti2014self}.  {The origin of these instabilities can be {rationalized} theoretically by means of simplified one-dimensional setups~\cite{lauga2021zigzag, lauga2017clustering}, with the underlying physical mechanism  related to the active stresses exerted by the swimmers.} 
 
\begin{center}
 \begin{figure}[t]
\includegraphics[width=0.9\textwidth]{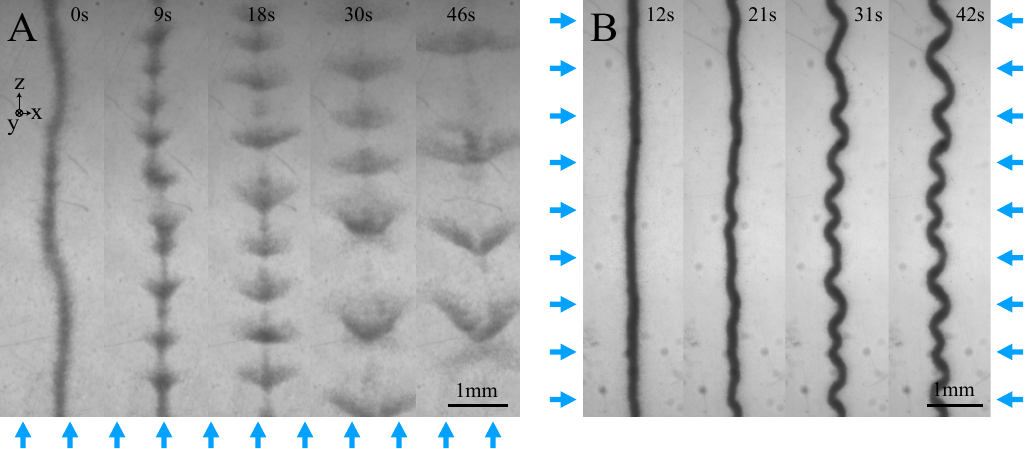}
\caption{{Empirical observations motivating the theory developed in the paper}. (A) Experiments with {jets} of negatively phototactic \textit{C.~reinhardtii} show that, upon setting the {jet} into motion by means of an axial light source (light direction $+\mathbf e_z$), the shape of the jet becomes unstable, breaking up into  {clusters}. {Over time, these} spread out perpendicularly to the  axis of the jet, {before undergoing a ``V''-shaped instability (top view)}. (B) {Conversely}, if the algae are rotated perpendicularly to the jet axis by means of side illumination, the  {jet}   buckles  {(top view)}.  Adapted from Ref.~\cite{shortpaper}.}
\label{Fig: Experiments_Numerics_Active_Jet}
\end{figure}
\end{center}

 {In the joint article}~\cite{shortpaper}, 
 {we propose a novel experimental way to  create stable active jets by steering phototactic cells via strong illumination.} Negatively
phototactic \textit{C.~reinhardtii} were loaded into a chamber, which was placed between two
parallel arrays of LED lights. After  {a few minutes}, the light-fleeing cells formed a dense band
at the centre of the chamber. If the side LEDs were switched off and a collimated LED light
parallel to the band was switched on, the jet was set into motion and quickly broke up
into  {clusters} (time-lapse images shown in Fig.~\ref{Fig: Experiments_Numerics_Active_Jet}A), consistently with predictions for the first instability (pearling) from  simulations~\cite{ishikawa2022instability}. In order to reproduce the {zigzag} instability for pusher cells~\cite{ishikawa2022instability},  {we} next exploited the fact that  {increasing the intensity of the lateral LEDs strengthened the alignment of algae perpendicularly to the band's axis, producing microscopic dipolar flows similar to those of pushers aligned with the jet axis.} Remarkably, this setup was then able to reproduce the pusher  {zigzag} instability 
{(the experimental results are summarised in Fig.~\ref{Fig: Experiments_Numerics_Active_Jet}B)}.

In the current article, we provide a fully three-dimensional dilute continuum model explaining the numerically~\cite{ishikawa2022instability, ishikawa2008coherent} and experimentally~\cite{shortpaper} observed instabilities of an active  {jet of aligned cells}. Extending previous one-dimensional studies~\cite{lauga2021zigzag, lauga2017clustering} to  finite-sized  {jets} of active particles, we compute the linearised time evolution of a perturbation of the jet geometry.  {We recover} the aforementioned {{pearling}} and  {helical/zigzag} instabilities for both cylindrical jets and two-dimensional ``sheets'' of swimmers, and determine the corresponding wavelengths and growth rates.  {We then extend our theoretical framework to investigate the long-term jet dynamics. We numerically and analytically determine the time evolution of the clusters resulting from the breakup of a puller jet, deriving both exact solutions in special cases as well as long-term similarity solutions. We furthermore obtain approximate evolution equations for the {nonlinear} dynamics of a buckled pusher jet, determining the driving flows by means of asymptotic methods}. We further compare our findings with experiments~\cite{shortpaper}, demonstrating quantitative agreement between the wavelength, growth rate and flow fields.  {To corroborate the theory, we finally} perform agent-based  {numerical simulations} implementing the Stokesian dynamics method, and validate our prediction for the {wavelength} and growth rate of each instability type, as well as the long-term dynamics.

The  paper is structured as follows{:} in Section~\ref{Continuum Description of the Jet}, we introduce the different jet geometries considered and derive the continuum model for the suspension, consisting of effective equations for the flow and the swimmer volume fraction. We then {specialise} these equations {to} the limiting case of a jet with a sharp interface and a constant swimmer volume fraction {(Section~\ref{Constant-Concentration Solution})}. We hence carry out a linear stability analysis in the three-dimensional case {(Section~\ref{Solving for the Flow})}, and in the two-dimensional setup {(Section~\ref{Sheet of Swimmers})}. We next consider the long-term evolution of the puller jet {(Section~\ref{Stretching of Puller Clusters})}, and the evolution of a buckled pusher jet {(Section~\ref{Long-Term Dynamics of the 3D Waving Instability})}.  {We finally compare our analytical results to  numerical simulations and experiments {(Sections~\ref{Comparison with Numerical Simulations} and \ref{Comparison with Experiments}, respectively)}.}

\section{Continuum modelling of active jets}\label{Continuum Description of the Jet}
Our first goal is to establish the continuum model we will work within, and to specialise  {it} to  {the} case of active  {jet}s in order to derive effective evolution equations for their shape.  {Our analysis  relies on the assumptions of zero diffusion for the swimmers (both in position and orientation), and of constant swimming direction (which will be physically justified by considering the various timescales involved). The effect of the swimmers on the flow is coarse-grained into an effective active stress~\cite{saintillan2008instabilitiesA, saintillan2008instabilitiesB, fung2022local} capturing the average bulk {stresses} {exerted by} the particles on the fluid~\cite{batchelor1970stress, pedley1990new, saintillan2018rheology}. As we will see, under these assumptions, the time evolution of a swimmer-laden region is driven  {by} the Stokes equations, paired with suitable continuity and force-balance boundary conditions.}

\subsection{Jet geometry}
Throughout this paper, we consider the case of a dilute suspension of neutrally buoyant (and hence force-free), identical spherical swimmers in an unbounded fluid.  {The swimmers align in response to an external stimulus}, such as light, gravity,  {magnetic fields}, or chemical  {gradients}. We denote the orientation of each swimmer  by a unit vector $\mathbf p$, and 
take the preferred direction for alignment to be $\mathbf e_z$. We let the swimmer radius be $a_s$, and we denote by $U_s$ the (constant) swimming speed of an isolated swimmer. Finally, we {use} $\phi(\mathbf x,t)$ {to indicate} the swimmer volume fraction, with $\phi\ll 1$ in the dilute limit. We assume that all swimmers can initially be found inside a specified region of space, which corresponds to the base state of the  {jet}. 

\begin{center}
 \begin{figure}[t]
\includegraphics[scale=0.3]{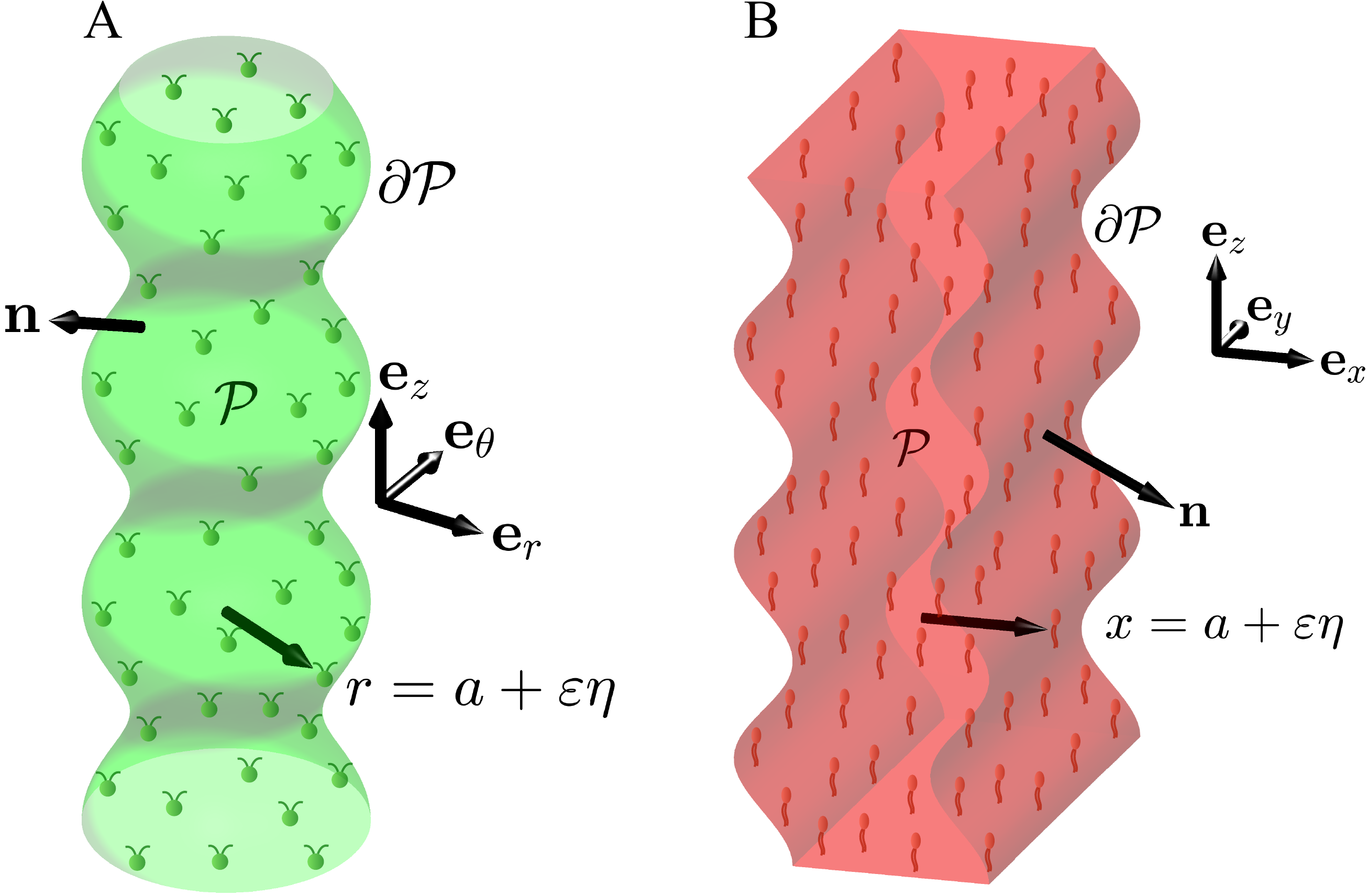}
\caption{Schematic illustration of the two  {jet} geometries considered in our modelling. (A) Cylindrical active  {jet} of radius $a$. (B) Two-dimensional  active {jet} of half-width $a$. The fluid outside the  {jet} is a simple Newtonian solvent with no active particles. In each case, we analyse the growth of a boundary perturbation of magnitude  $\varepsilon \eta\ll a$.  {Our analysis concerns both pushers and pullers for each geometry (the above sketch only illustrates particular examples).}}
\label{Fig: Jet Sketch}
\end{figure}
\end{center}

Two different geometries will be considered (Fig.~\ref{Fig: Jet Sketch}), motivated by experiments:
\begin{enumerate}
    \item Cylindrical  {jet}s: all swimmers are initially positioned in the domain given by $0\leq r\leq a$, $-\infty<z<\infty$ in cylindrical coordinates. This is the case illustrated in Fig.~\ref{Fig: Jet Sketch}A.
    \item Two-dimensional  {jet}s: all swimmers are initially positioned in {a} sheet of finite thickness, given by $-a\leq x\leq a$, $-\infty<y,z<\infty$ in Cartesian coordinates.  {The sheet is taken to be homogeneous in the $y$ direction, so that the dynamics is effectively restricted to the $xz$ plane.} This case is sketched in Fig.~\ref{Fig: Jet Sketch}B.

\end{enumerate}

\subsection{Modelling assumptions}\label{Assumptions}
 {{As  mentioned above,} in order for our model to accurately describe the experimental jet, which maintains a sharp boundary at all times, two further assumptions are needed:}
\begin{enumerate}
    \item Swimmers are clamped so that their swimming direction is constantly along the axis of the unperturbed  {jet}. In other words, $\mathbf p\equiv \mathbf e_z$ at all times.
    \item Spatial diffusion is negligible, so that swimmers do not leave the  {jet} as a result of stochastic effects.
\end{enumerate}

{If swimmers were not aligned (assumption \#1), variations in swimming directions due to the flow or rotational diffusion would cause the jet to fly apart over time. Similarly, translational diffusion (assumption \#2) would smear out the  boundary of the jet, which is at odds with experiments}.  {Physically,} assuming that $\mathbf p\equiv \mathbf e_z$ at all times is appropriate whenever the reorientation timescale $B$ is much shorter {than the flow timescale $T$ (corresponding to the onset of the instability)}. Indeed, in the dilute limit, the {evolution of the} swimming direction  can be described by the Fax\'en equation~\cite{saintillan2018rheology, koch2011collective, bees2020advances}
\begin{align}
\dot{\mathbf p}&= \frac{1}{2B}(\mathbf I-\mathbf p\mathbf p)\cdot\mathbf e_z+\frac{1}{2}\boldsymbol{\omega}(\mathbf x,t)\times\mathbf p,\label{Orientational Flux}
\end{align}
{where} $\boldsymbol{\omega}=\nabla\times \mathbf u$ is the {flow's} vorticity. The clamped approximation {($\mathbf p\equiv  \mathbf e_z$)} is valid whenever  {$\lVert\boldsymbol{\omega}\rVert \ll B^{-1}$ or, in other words, when $B\ll L/U$. The right-hand side coincides with the advective timescale, which, as we will later see, coincides with the flow timescale $T$. {Assumptions \#1 and \#2 are sufficient to formulate a theoretical framework for the study of active jets. The applicability of these assumptions to our specific experimental setup will be discussed in Section~\ref{Comparison with Experiments}.}}

\subsection{Equations of motion}

Having specified the geometry, we now turn our attention to the derivation of the flow equations. 

\subsubsection{{The active stresslet}}
On the scale of an individual swimmer, the activity and  {stimulus}-driven taxis are associated {with} a stresslet $S_{ij}$ and, in the {cases (e.g.)}~of magnetotaxis and gravitaxis {(but not for phototaxis or chemotaxis)}, a torque $L_i$ on the fluid.  {Indeed, while for magnetotaxis and gravitaxis swimmers are subject to a physical, orientation-dependent torque (provided by the magnetic field or gravity, respectively), in the case of phototaxis or chemotaxis, reorientations are achieved purely by swimming, which is inherently torque-free.}

It is a classical result that the stresslet and torque  {of a given swimmer} are  explicitly given by~\cite{saintillan2018rheology}
\begin{align}
S_{ij}&= {\int_{\p B}\left[\frac{1}{2}(\sigma_{ik}x_j+\sigma_{jk}x_i)b_k-\mu(v_ib_j+v_jb_i)-\frac{1}{3}(x_k\sigma_{kl}b_l)\delta_{ij}\right]\mathrm dA,} \label{S_ij}\\ 
L_i&= {\varepsilon_{ijk}\int_{\p B}\sigma_{jl}x_kb_l\mathrm dA,} \label{L_ij}
\end{align}
where $\p B$ is the closed surface of the swimmer, $\mu$ is the dynamic  viscosity of the fluid, $v_i$ is  {the variation in the background flow due to the swimmer}, $\sigma_{ij}$ is  {the variation in the background hydrodynamic stress due to the swimmer},  {$b_i$ is the local unit normal to the swimmer's body}, and $\varepsilon_{ijk}$ is the antisymmetric Levi-Civita tensor.

The leading-order  values of $S_{ij}$ and $L_i$ are the same as for an isolated swimmer, up to corrections (caused by particle-particle interactions) proportional to the swimmer volume fraction, which can be neglected in the dilute limit. In particular, if the swimming direction is clamped, the torque is purely a response to particle-particle interaction, so $L_i=\mathcal O(\phi)$. We can therefore assume $L_i=0$ at leading order in $\phi$, regardless of whether organisms are torque-free or not. In this sense, {while our experiments focus on the special case of phototaxis,} our dilute modelling approach  provides a unifying description of a multitude of taxes under different physical mechanisms.

A further simplification may be inferred by noting that, while $S_{ij}$ is \textit{a priori} a traceless symmetric tensor with $5$ independent components, in the case of steady axisymmetric swimmers the number of independent components reduces to just one. This allows us to write~\cite{saintillan2018rheology}
\begin{equation}
S_{ij}=S\left(p_ip_j-\frac{1}{3}\delta_{ij}\right)    \label{Axisymmetric stresslet expression},
\end{equation}
where $S$ is a scalar parameter intrinsic to the swimmer: ``puller'' swimmers have $S>0$, while ``pushers'' have $S<0$~\cite{saintillan2008instabilitiesA, saintillan2008instabilitiesB}.

\subsubsection{{Collective flow equations}}
The superposition of all {the dipolal} flows created by the swimmers results in a macroscopic, continuum flow with velocity $\mathbf u(\mathbf x,t)$. A classical calculation by Batchelor~\cite{batchelor1970stress} shows that one may define an effective stress {for this flow}, {given by a} divergence-free stress tensor {obtained by suitable averaging over} the fluid-particle system within the {jet}. {When the swimming direction is clamped, the result is}
\begin{equation}
\Sigma^-_{ij}=-q^-\delta_{ij}+\mu\left(\frac{\p u^-_i}{\p x_j}+\frac{\p u^-_j}{\p x_i}\right)+\frac{S\phi}{V} p_ip_j\label{Sigma^-_ij expression},\end{equation}
where  {$q$ denotes the pressure and} a $``-"$  {superscript} denotes the interior of the jet. Outside the  {jet}, the particle concentration is $0$ and the stress tensor reduces to the Newtonian value
\begin{equation}
\Sigma^+_{ij}=-q^+\delta_{ij}+\mu\left(\frac{\p u^+_i}{\p x_j}+\frac{\p u^+_j}{\p x_i}\right),
\end{equation}
where a $``+"$  {superscript} denotes the  {exterior} of the jet.

\v

We note that, in order to be able to identify Eq.~\eqref{Axisymmetric stresslet expression} with the particle stress, we implicitly assumed the suspension to be locally statistically homogeneous~\cite{batchelor1970stress}. In other words, the properties of the suspension should vary on a length scale much larger than the swimmer size. While this assumption is apparently violated at the sharp boundary of the jet, {in Appendix~\ref{Matched Asymptotic Solution} we explicitly show} that one may think of the discontinuity as the leading-order approximation {of} a smooth concentration field, for which the continuum assumption is valid.

Effective equations for the interior and exterior of the  {jet} may now be derived by requiring stresses to balance on every material element (Eqs.~\ref{Bulk Jet Stress Balance}, \ref{Bulk Exterior Stress Balance}, \ref{Table Boundary Stress Balance}). {In addition, we impose} incompressibility conditions on the velocity fields (Eq.~\ref{Bulk Jet Incompressibility}, \ref{Bulk Exterior Incompressibility}), as well as continuity of velocity at the  {jet} boundary (Eq.~\ref{Table Boundary Continuity of Velocity}) in order to avoid infinite stresses{:}

\vspace{-1cm}
\begin{minipage}[t]{0.3\textwidth}
\begin{subequations}
\begin{align}
&\text{ {Jet}}\nonumber\\\hline
&\p_j\Sigma^-_{ij}=0\label{Bulk Jet Stress Balance}\\
&\p_i u_i^-=0 \label{Bulk Jet Incompressibility}
\end{align}
\end{subequations}    
\end{minipage}
\begin{minipage}[t][4cm][t]{0.3\textwidth}
\begin{subequations}
\begin{align}
&\text{Exterior}\nonumber\\\hline
&\p_j\Sigma^+_{ij}=0 \label{Bulk Exterior Stress Balance}\\
&\p_i u_i^+=0 \label{Bulk Exterior Incompressibility}
\end{align}
\end{subequations}    
\end{minipage}
\begin{minipage}[t][4cm][t]{0.3\textwidth}
\begin{subequations}
\begin{align}
&\text{Boundary}\nonumber\\\hline
&\Sigma^-_{ij}n_j=\Sigma^+_{ij}n_j\label{Table Boundary Stress Balance}\\
&u_i^-=u_i^+\label{Table Boundary Continuity of Velocity}
\end{align}
\end{subequations}    
\end{minipage}
where $\mathbf n$ denotes the local unit normal to the  {jet} boundary (Fig.~\ref{Fig: Jet Sketch}). In Appendix~\ref{Matched Asymptotic Solution}, we show that Eqs.~\eqref{Bulk Jet Stress Balance} through \eqref{Table Boundary Continuity of Velocity} may be derived via the method of matched asymptotics as the leading-order flow equations for a {smooth but rapidly varying} concentration profile.  {Alternatively, {Eqs}.~\eqref{Bulk Jet Stress Balance} through \eqref{Table Boundary Continuity of Velocity} may be postulated empirically by writing the bulk flow as the superposition of {the swimmers'} dipolar flows~\cite{saintillan2013active}, without appealing to statistical homogeneity or stress-balance arguments; such an alternative derivation is presented in Appendix~\ref{Continuum equations alternative derivation}.}

\subsubsection{ {Swimmer conservation}}
Finally, in order to close the system, we need an evolution equation for the swimmer concentration $\phi$. Individual swimmer motion is a combination of self-propulsion along the constant swimming direction, and advection by the flow. Therefore, if $\mathbf x(t)$ denotes the position of a swimmer, then
\begin{equation}
\dot{\mathbf x}(t)=\mathbf u(\mathbf x,t)+U_s \mathbf e_z \label{Positional Flux} .
\end{equation}
The corresponding evolution of the volume fraction is found by imposing swimmer conservation, {i.e.}~$\mathrm d(\phi\mathcal W)/\mathrm dt=0$, where $\mathcal W$ is a small volume. Because the right-hand side of Eq.~\eqref{Positional Flux} is divergence-free, $\mathrm d\mathcal W/\mathrm dt=0$ and the volume fraction is constant along swimming trajectories, i.e.
\begin{equation}
 {\frac{\mathrm d}{\mathrm dt}\phi[\mathbf x(t),t]=0.} \label{Swimmer Conservation}    
\end{equation}
 {{Note that} the result in Eq.~\eqref{Swimmer Conservation}  {is equivalent to} the standard Smoluchowski equation~\cite{saintillan2008instabilitiesA, saintillan2008instabilitiesB} for a smooth field $\phi$, given by $\phi_t+\nabla\cdot[(\mathbf u+U_s\mathbf e_z)\phi]=0$,  {which is} the same as $\mathrm d\phi/\mathrm dt=0$.


\subsection{Constant-concentration solution}
\label{Constant-Concentration Solution}
We now restrict our attention to a special class of solutions of Eqs.~\eqref{Bulk Jet Stress Balance}-\eqref{Table Boundary Continuity of Velocity}, \eqref{Swimmer Conservation}. We denote the region of space occupied by the  {jet} at time $t$ by $\mathcal P(t)$. {As  a model for our   experiments~\cite{shortpaper}}, we restrict our attention to the case in which, initially, the swimmer concentration  {is} uniform inside the  {jet}. Mathematically, at time $t=0$, the swimmer volume fraction is therefore taken to be
\begin{align}
 \phi(\mathbf x,t=0)=
 \begin{cases}
 \phi_0 & \mathbf x\in \mathcal P(0),\\
 0 & \mathbf x\not\in \mathcal P(0),
 \end{cases}
\end{align}
for some constant $\phi_0$. The swimmer conservation 
Eq.~\eqref{Swimmer Conservation} then admits the solution
\begin{align}
 \phi(\mathbf x,t)=
 \begin{cases}
 \phi_0 & \mathbf x\in \mathcal P(t),\\
 0 & \mathbf x\not\in \mathcal P(t),
 \end{cases}
\end{align}
where $\mathcal P(t)$ is the advected image of $\mathcal P(0)$ under Eq.~\eqref{Positional Flux}.  The fact that the concentration is constant inside the  {jet} offers considerable simplifications to our model. Indeed, the bulk stress-balance equation {\eqref{Bulk Jet Stress Balance}} in the {jet} {(with stress tensor in Eq.~\eqref{Sigma^-_ij expression})} {reduces} to the Stokes equation $\p_i q^-=\mu\p_j\p_j u^-_i$, and the active component of the stress only appears in boundary condition~\eqref{Table Boundary Stress Balance}. {Referring to} Fig.~\ref{Fig: Jet Sketch} for the notation, the simplified equations that we work with are therefore 
\vspace{-1.5cm}
\begin{center}
\begin{minipage}[t]{0.3\textwidth}
\begin{subequations}
\begin{align}
&\mathbf x\in \mathcal P(t)\nonumber\\\hline
&\p_i q^--\mu \p_{j}\p_ju^-_i=0 \label{Plume Bulk Stokes}\\
&\p_i u_i^-=0 \label{Plume Bulk Incompressibility Simplified}
\end{align}
\end{subequations}    
\end{minipage}
\begin{minipage}[t]{0.3\textwidth}
\begin{subequations}
\begin{align}
&\mathbf x\not\in \mathcal P(t)\nonumber\\\hline
&\p_i q^+-\mu \p_{j}\p_ju^+_i=0 \label{Exterior Bulk Stokes}\\
&\p_i u_i^+=0 \label{Exterior Bulk Incompressibility Simplified}
\end{align}
\end{subequations}    
\end{minipage}
\begin{minipage}[t]{0.3\textwidth}
\begin{subequations}
\begin{align}
&\mathbf x\in \p\mathcal P(t)\nonumber\\\hline
&\Sigma^-_{ij}n_j=\Sigma^+_{ij}n_j \label{Stress Continuity Simplified}\\
&u_i^-=u_i^+ \label{Velocity continuity Simplified}\\
& \dot{x}_i=u_i+U_s p_i \label{Boundary Advection Simplified}
\end{align}
\end{subequations}    
\end{minipage}
\end{center}
 {Physically, the active stress only appears at the boundary (Eq.~\ref{Stress Continuity Simplified}) because the bulk dipoles cancel each other in a uniform concentration field. Consequently, only the boundary forces remain unbalanced, leading to {fluid flow}. This intuition is made precise in Appendix~\ref{Continuum equations alternative derivation}.}

\section{Linear stability  of a cylindrical active jet}\label{Solving for the Flow}
We now specialise the above framework to a particular  {jet} geometry, specifically to the case where $\mathcal P(0)$ is the cylinder $0\leq r\leq a$, $-\infty<z<\infty$ (Fig.~\ref{Fig: Jet Sketch}A). {The applicability of this idealised geometry to our specific experimental setup is discussed in Section~\ref{Comparison with Experiments}}. As a reminder, the swimmer volume fraction inside the {jet} is constant ($\phi\equiv \phi_0$), and the swimming direction is vertically clamped ($\mathbf p\equiv \mathbf e_z$).  {Our} goal is to study the evolution of the shape of the {jet}, as given by Eqs.~\eqref{Plume Bulk Stokes}--\eqref{Boundary Advection Simplified}, when the {location of the jet boundary} is perturbed. In particular, we are interested in whether a given boundary perturbation grows or decays with time. We define a boundary perturbation as ``stable'' if it decays to zero over time, and ``unstable'' if it grows. The term ``stability'' will therefore always refer to the shape of the {jet}.

\subsection{Base state}\label{Sec: Base State}
The base state corresponds to {an everywhere-vanishing velocity field}, i.e.
\begin{align}
\mathbf u_0^{\pm}&=\mathbf 0, & q_0^{\pm}&=0, & \mathbf{\Sigma}^+_0&=\mathbf 0, & \mathbf{\Sigma}^-_0&=\frac{S\phi_0}{V}\mathbf e_z\mathbf e_z, \label{3d Case Base State}  
\end{align}
which is easily verified to be a solution of 
Eqs.~\eqref{Plume Bulk Stokes}--\eqref{Boundary Advection Simplified}. In particular, the stress boundary condition {\eqref{Stress Continuity Simplified}} is satisfied, as the orientation of the force dipoles exerted by the swimmers is purely axial, with no component perpendicular to the jet boundary. The kinematic boundary condition {\eqref{Boundary Advection Simplified}} is likewise satisfied, as points at the boundary are advected according to $\dot{\mathbf x}=U_s\mathbf e_z$, which leaves the shape of the {jet} unchanged. Notably, if $\mathbf p$ was not fixed and, instead, diffused rotationally, boundary swimmers would migrate outwards, and \eqref{3d Case Base State} would not be a steady solution of Eqs.~\eqref{Plume Bulk Stokes}--\eqref{Boundary Advection Simplified}.

\subsection{$\mathcal O(\varepsilon)$ Flow: linear perturbation}\label{Sec: Linear Perturbation}
We now introduce a small boundary perturbation {of order $\varepsilon$,} which leaves the active stresses unbalanced at the boundary, resulting in a net ambient flow and motion of the jet. More precisely, we let the {time-dependent} deformed jet boundary be given by $r=R(z,\theta,t)$, with cylindrical coordinates $-\infty\leq z\leq \infty$ and $0\leq \theta <2\pi$ (Fig.~\ref{Fig: Jet Sketch}A). In the limit of a small  {distortion} ($|\varepsilon|\ll 1$), we may classically exploit linearity to assume that the perturbation consists of a single Fourier mode, i.e.
\begin{align}
\displaystyle R=a+\varepsilon  {\eta(z,\theta,t)}=a(1+\varepsilon e^{st+\mathrm ikz+\mathrm in\theta})\label{Perturbed Radius Fourier modes}.
\end{align}
 {We may assume $k\geq 0$, $n\in \mathbb Z_{\geq 0}$, as different sign combinations are equivalent to simply reflecting the setup.}
We now aim to solve equations  Eqs.~\eqref{Plume Bulk Stokes}--\eqref{Boundary Advection Simplified} to determine the growth rate {$\Re(s)$} as a function of the axial and angular wavenumbers, $k$ and $n$. As is standard, we may then deduce which perturbations grow and which decay by noting that stable Fourier modes have ${\Re(s)}<0$, while unstable modes have ${\Re(s)}>0$.

In order to solve for $s$, we expand the velocities, pressures, and stresses in the small parameter $\varepsilon$ as
\begin{subequations}
\begin{align}
\mathbf u^{\pm}&=\mathbf u_0^{\pm}+\varepsilon\mathbf u_1^{\pm}+\mathcal O(\varepsilon^2),\\
q^{\pm}&=q_0^{\pm}+\varepsilon q_1^{\pm}+\mathcal O(\varepsilon^2),\\
\mathbf \Sigma^{\pm}&=\mathbf \Sigma^{\pm}_0+\varepsilon\mathbf \Sigma^{\pm}_1+\mathcal O(\varepsilon^2),
\end{align}
\end{subequations}
where the base state corresponds to Eq.~\eqref{3d Case Base State}. 
To study the evolution of the boundary disturbance $\displaystyle\varepsilon\eta$, we must expand the bulk flow equations and boundary conditions \eqref{Plume Bulk Stokes}--\eqref{Boundary Advection Simplified} up to and including $\mathcal O(\varepsilon)$. Following notation shown in  Fig.~\ref{Fig: Jet Sketch}A, {at order $\mathcal O(\varepsilon)$}

\vspace{-1.5cm}
\begin{center}
\begin{minipage}[t]{0.26\textwidth}
\begin{subequations}
\begin{align}
&0\leq r\leq a\nonumber\\
\hline \nonumber& \\[-3.5ex]
&\mu\nabla^2\mathbf u^-_1=\nabla q^-_1\label{Order Epsilon Stokes Equation} \\
& \nabla\cdot\mathbf u_1^-=0 \label{Order Epsilon Incompressibility Inner}
\end{align}
\end{subequations}    
\end{minipage}
\hspace{0.2cm}
\begin{minipage}[t]{0.26\textwidth}
\begin{subequations}
\begin{align}
&a< r<\infty\nonumber\\
\hline \nonumber& \\[-3.5ex]
&\mu\nabla^2\mathbf u^+_1=\nabla q^+_1 \label{Order Epsilon Stokes Equation Outer}\\
&\nabla\cdot\mathbf u_1^+=0 \label{Order Epsilon Incompressibility Outer}\\
&\lim_{r\to\infty}\mathbf u^+_1=\mathbf 0 \label{Order Epsilon Decay at Infinity}
\end{align}
\end{subequations}    
\end{minipage}
\hspace{0.2cm}
\begin{minipage}[t]{0.41\textwidth}
\begin{subequations}
\begin{align}
&r=a\nonumber\\
\hline \nonumber& \\[-4ex]
&(\mathbf \Sigma_1^+-\mathbf \Sigma_1^-)\cdot\mathbf e_r=-\frac{S\phi_0}{V}\eta_z\mathbf e_z\label{Order Epsilon Boundary Stress}\\
&\mathbf u_1^+=\mathbf u_1^-\label{Order Epsilon Continuity of Velocity}\\
& \frac{\p \eta}{\p t}+U_s\frac{\p\eta}{\p z}=\mathbf u^{\pm}_1\cdot \mathbf e_r \label{Order Epsilon Boundary Evolution}
\end{align}
\end{subequations}    
\end{minipage}
\end{center}
 Physically, Eq.~\eqref{Order Epsilon Boundary Stress} shows that the flow is driven by the boundary stresslets, while Eq.~(\ref{Order Epsilon Boundary Evolution}) reflects the fact that, for a clamped swimming direction, the swimming speed only causes the {jet} to rigidly translate  {along $\mathbf e_z$}. 
 
 We may integrate Eqs.~\eqref{Order Epsilon Stokes Equation}--\eqref{Order Epsilon Boundary Evolution} {exactly, thus determining} the velocity and pressure fields $\mathbf u_1^{\pm}$, $q_1^{\pm}$ up to \textcolor{blue}{a} multiplicative constant; Eq.~\eqref{Order Epsilon Boundary Evolution} then provides a condition for the existence of non-zero solutions to Eqs.~\eqref{Order Epsilon Stokes Equation} through \eqref{Order Epsilon Continuity of Velocity}, which uniquely determines the growth rate. {The details of the calculation are provided in Appendix~\ref{Appendix A}, and here we only present} the final result for the growth rate ${\Re(s)}$, which is 
\begin{align}
  {\Re(s)}=\frac{S\phi_0}{2\mu V}\frac{I_n(\xi) \left[K_n(\xi) \left(2n^2 +\xi^2 \right)-n\xi K_{n+1} (\xi)\right]-I_{n+1} (\xi)\left[\xi^2 K_{n+1} (\xi)-n\xi K_n (\xi)\right]}{\xi I_n (\xi)K_{n+1} (\xi)+\xi K_n (\xi)I_{n+1} (\xi)} \label{Most General Growth Rate} ,
\end{align}
 where $\xi=ak$ is the dimensionless axial wavenumber, while the $I_n$ and $K_n$ are modified Bessel functions \cite{arfken2011mathematical, watson1922treatise}. {Our analysis also reveals that $\Im(s)=-\mathrm ikU_s$. Therefore, as suggested by Eq.~\eqref{Order Epsilon Boundary Evolution}, the swimming speed only leads to net upward motion of the jet without affecting the extend of the boundary distortion, {and therefore the stability characteristics}.}

\subsection{Analysis of the growth rate}\label{Analysis of the Growth Rate}

 \begin{figure}[t]
 \hspace{-2.5cm}
\includegraphics[scale=0.3]{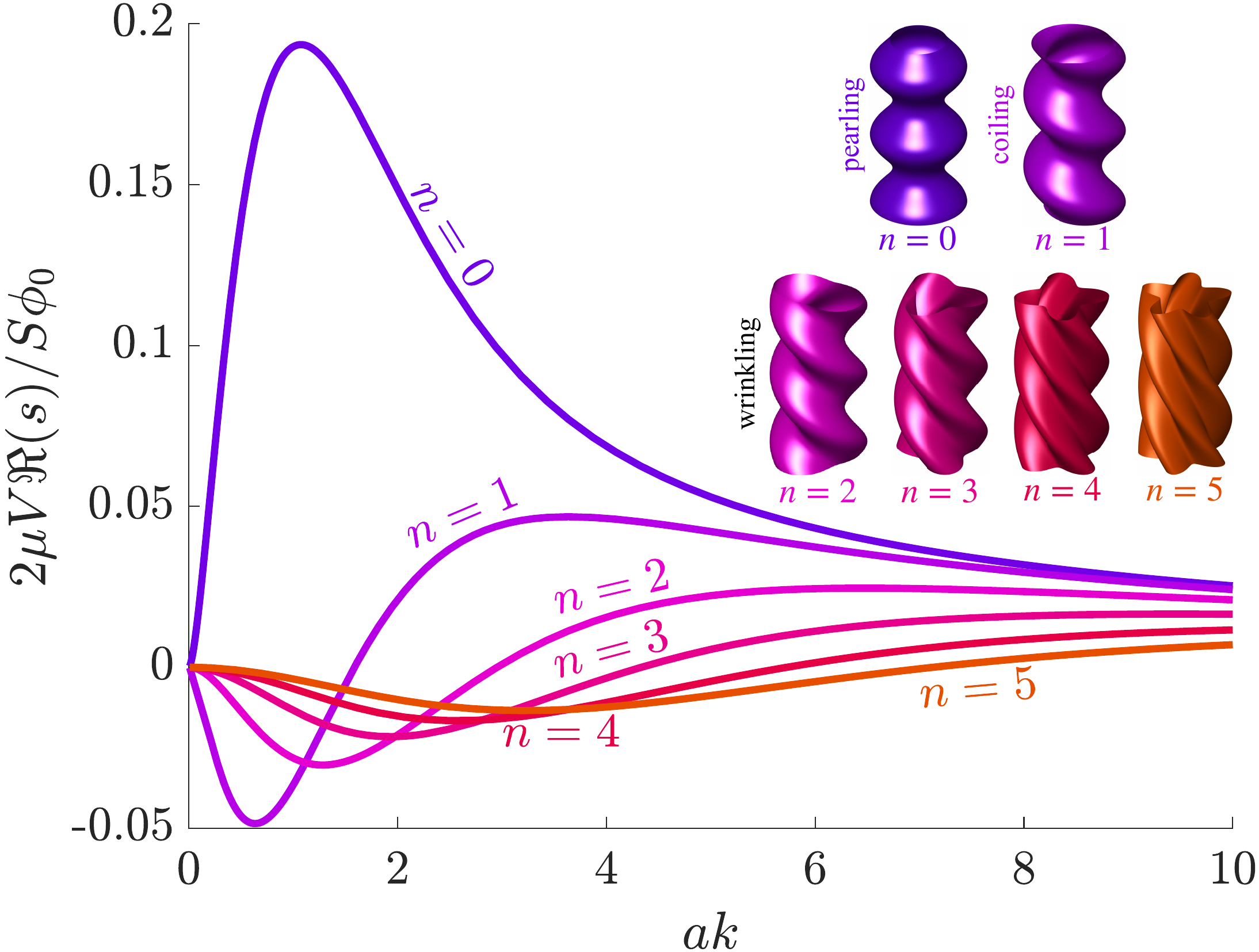}
\caption{Plot of dimensionless growth rate $2\mu V\Re(s)/S\phi_0$ as a function of the dimensionless wavenumber for $0\leq n\leq 5$. Inset shows the corresponding {shapes} of the jet boundary.  {The $n=0$ mode corresponds to axisymmetric pearling, $n=1$ corresponds to helical buckling (with circular cross-section), while $n\geq 2$ are higher-order wrinkling modes with centreline parallel to $\mathbf e_z$. {Note that, if $S<0$ (pushers), the graph of $\Re(s)$ as a function of $ak$ has the opposite sign to the curves plotted above, {so the stable/unstable regions are flipped}.}} 
}
\label{Fig: Dispersion_Relation_6_Modes}
\end{figure}

As expected,  Eq.~\eqref{Most General Growth Rate} shows that the timescale of the instability is set by the stresslet strength $S$. The dependence of the growth rate ${\Re(s)}$ on the wavenumbers $\xi$ and $n$ is shown in Fig.~\ref{Fig: Dispersion_Relation_6_Modes}. {We may thus} fully determine the modes $(\xi,n)$ that cause the perturbation to grow or decay for pushers ($S<0$) and pullers ($S>0$). We next provide a summary of the results, with further {details of the asymptotic calculations given in} Appendix~\ref{Growth Rate Asymptotics}. The two main stability results are:
\begin{enumerate}
    \item Jets of pullers (i.e.~with $S>0$) are unstable to all axisymmetric modes $(\xi,0)$, and to non-axisymmetric modes $(\xi, n)$ with $n\geq 1$ for small axial wavelengths ($\xi\gtrsim 2^{1/2}n$).
    \item Jets of pushers ($S<0$) are only unstable to non-axisymmetric modes $(\xi, n)$ with $n\geq 1$ and large axial wavelengths ($\xi\lesssim 2^{1/2}n$).
\end{enumerate}
In other words, puller jets are unstable whenever the axial wavelength is much smaller than the angular wavelength, while the situation is reversed for pushers. This may be rationalised by noting that the puller dipole flow tends  {to} relax angular distortions, since pullers eject fluids sideways. On the other hand, pushers drive the opposite flow to pullers, thereby enhancing wrinkly modes. Therefore, if most of the distortion is in the $\theta$ direction (i.e.~$\xi\ll n$), then pullers have a stabilising effect, and pushers have a destabilising effect. The opposite occurs if most of the distortion is axial (i.e.~$\xi\gg n$). In particular, since angular {variations are much slower than axial ones (and hence negligible)} for $\xi\to\infty$, the resulting jet is unstable for pullers and stable for pushers, with 
\begin{equation}
{\Re(s)}\sim \frac{S\phi_0}{8\mu V\xi}, 
\qquad \xi\to\infty    
\end{equation}
regardless of $n$.

\subsection{Fastest-growing modes} \label{Fastest-growing modes}

A second,  more experimentally relevant question, concerns the nature of the instability observed in  {a laboratory setting}. In the context of linearised theory, any perturbation in the shape of a  {real-world} {jet} may be decomposed into Fourier modes of the form \eqref{Perturbed Radius Fourier modes}, each growing or shrinking over time with growth rate given by Eq.~\eqref{Most General Growth Rate}. The observed mode in a given jet will therefore be the pair $(\xi,n)$ which maximises ${\Re(s)}$. This allows us to predict the experimentally observed instability types for pushers and pullers. Like before, we refer to  Appendix~\ref{Growth Rate Asymptotics} for the mathematical details. The main results are:
\begin{enumerate}
    \item The fastest-growing mode for a jet of pullers ($S>0$) is the axisymmetric one ($n=0$) with wavenumber $\xi=\xi^*\sim 1.0750$. This corresponds to  {a {{pearling}}} instability. The selected instability wavelength evaluates to $\lambda^*\sim 5.8447\times a$, while the growth rate is $\Re[s^*]\sim 0.0968\times S\phi_0/\mu V$.
    \item The fastest-growing mode for a jet of pushers ($S<0$) is the {helical} one ($n=1$) with wavenumber $\xi=\xi^*\sim 0.6350$. The selected instability wavelength evaluates to $\lambda^*\sim 9.8943\times a$, while the growth rate is $\Re[s^*]\sim -0.0243 \times S\phi_0/\mu V$. \end{enumerate}

Interestingly, the wavelength is in each case only set by the initial {jet} radius $a$, and is independent of $\phi_0$ (the only dimensionless quantity in the problem).  {This is because, in the dilute limit, the flow scales linearly with $\phi_0$, leaving wavelengths unaffected}. The growth rate for the puller jet is also faster (by a factor of about $4$) than the growth rate of {a} jet of pusher with the same value of $|S|$. Finally, for both swimmer types, higher-ordered wrinkling modes grow on progressively slower timescales. In Appendix~\ref{Growth Rate Asymptotics}, we show in particular that {the} $n$th mode grows like {${\Re(s)}\sim 0.02\times S\phi_0/(\mu V n)$ for pullers and ${\Re(s)}\sim 0.03\times |S|\phi_0/(\mu V n)$ for pushers.}\\

\subsection{Oscillating stresslets}
{In the real biological system, the stresslet strength $S$ for \textit{C.~reinhardtii} is not time-independent. Rather, $S(t)$ oscillates around its mean value during the algae's beating pattern, switching from $S>0$ during the power stroke to $S<0$ during the recovery stroke \cite{klindt2015flagellar}. Furthermore, due to the large beat frequency ($50$ Hz), the microscopic flows driven by the swimmers are unsteady beyond about $20$ swimmer radii of each alga \cite{klindt2015flagellar}.}

{While accurately modelling such effects appears very challenging, we comment on the simplest case of spatially-independent stresslet strength $S(t)$ and swimming speed $U_s(t)$, i.e.~synchronous beating of all swimmers. {Assuming the   viscosity of the fluid to be sufficiently large} as to overdamp any inertial effects, the stability of the active jet from Section~\ref{Sec: Base State} and Section~\ref{Sec: Linear Perturbation} {can be analysed} by replacing the term $e^{st}$ in Eq.~\eqref{Perturbed Radius Fourier modes} with a generic function $e^{s(t)}$ (with $s(0)=0$). Linear stability analysis then readily provides
\begin{align}
  \Re(s)=\frac{\phi_0}{2\mu V}\frac{I_n(\xi) \left[K_n(\xi) \left(2n^2 +\xi^2 \right)-n\xi K_{n+1} (\xi)\right]-I_{n+1} (\xi)\left[\xi^2 K_{n+1} (\xi)-n\xi K_n (\xi)\right]}{\xi I_n (\xi)K_{n+1} (\xi)+\xi K_n (\xi)I_{n+1} (\xi)}\int_0^t S(t')\mathrm dt'. \label{Growth oscillating stresslets}
\end{align}
For a rapidly oscillating stresslet with period-averaged value $\langle S(t)\rangle$, Eq.~\eqref{Growth oscillating stresslets} is approximately equivalent to Eq.~\eqref{Most General Growth Rate} with $S$ replaced by $\langle S(t)\rangle$. In other words, our analysis remains approximately valid (at least in simple cases) for time-dependent swimming gaits, with an ``effective''  stresslet strength $\langle S(t)\rangle$. We expect a similar conclusion to hold when inertia is present, since rapidly oscillating swimmers should interact primarily via their period-averaged flows.}

\section{Linear stability  of a 2D active sheet}\label{Sheet of Swimmers}
{As a natural extension of the previously described cylindrical jet, in this section we consider the stability of an infinite sheet jet given by 
$-a\leq x\leq a$, $-\infty<y,z<\infty$ in Cartesian coordinates (setup shown in Fig.~\ref{Fig: Jet Sketch}B). We assume that the sheet stays homogeneous in the $y$ direction, so that the dynamics is effectively confined to the $xz$ plane. The relevance of this geometry to our experiments will be addressed in Section~\ref{Comparison with Experiments}.}

As in Section~\ref{Solving for the Flow}, we aim to understand if a given perturbation to the boundary of the sheet tends to grow or shrink with time, depending on whether the swimmers are pushers or pullers. Mathematically, we perturb the sheet boundary so that, for $t>0$, the swimmers are located in $X^-(z,t)\leq x\leq X^+(z,t)$. We consider two forms of the perturbation (Fig.~\ref{Fig: Sinuous Varicose Perturbation 2D}):
\begin{enumerate}
    \item Sinuous (i.e.~in phase): $X^+=a\left(1+\varepsilon e^{st+\mathrm ikz}\right)$, $X^-=a\left(-1+\varepsilon e^{st+\mathrm ikz}\right)$;
    \item Varicose ({i.e.}~antiphase): $X^+=a\left(1+\varepsilon e^{st+\mathrm ikz}\right)$, $X^-=a\left(-1-\varepsilon e^{st+\mathrm ikz}\right)$.
\end{enumerate}

 \begin{figure}[t]
\includegraphics[width=0.6\textwidth]{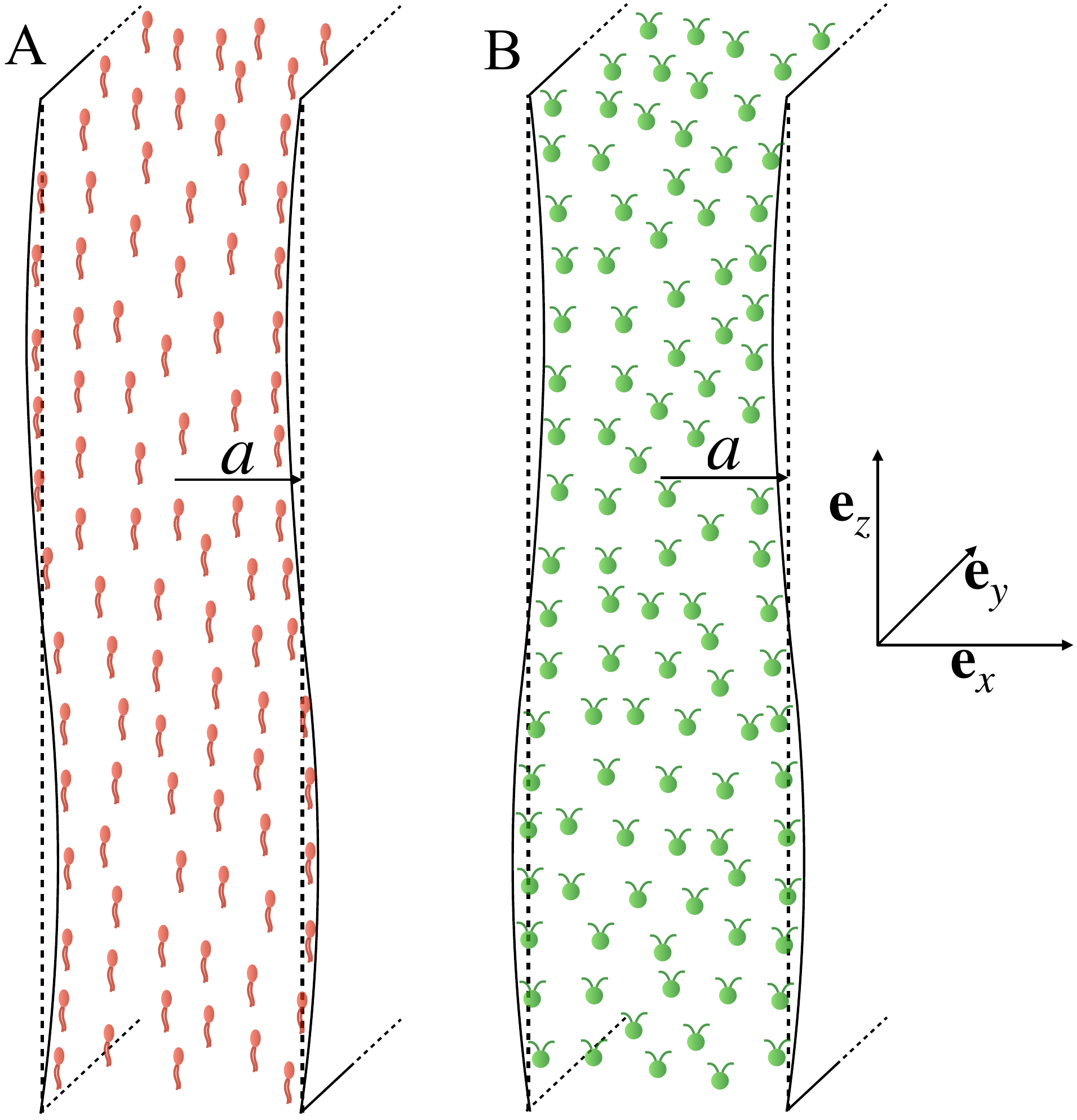}
\caption{Schematic representation of the sheet geometry and the two perturbations considered: (A) sinuous perturbation (i.e.~in phase). (B) Varicose perturbation (i.e.~antiphase).}
\label{Fig: Sinuous Varicose Perturbation 2D}
\end{figure}

In each case, we seek to determine the growth rate ${\Re(s)}$, which {dictates} the stability of the perturbation depending on whether ${\Re(s)}>0$ (unstable) or ${\Re(s)}<0$ (stable). As before, we provide the details of the calculation in Appendix~\ref{Appendix C: Growth Rate of the Quasi two-dimensional Sheet} and only {state} the main results:
\begin{enumerate}
    \item In the case of a sinuous perturbation, the growth rate is given by
    \begin{align}
      \Re(s)&=-\frac{S\phi_0}{2\mu V}\xi e^{-2\xi}, \label{2d Growth rate sinuous}
    \end{align}
      {where the dimensionless wavenumber is again $\xi\coloneqq ak$}. Therefore, a sinuous perturbation is always unstable for pushers, consistently with the three-dimensional case {(as a sinuous perturbation is the two-dimensional equivalent of a helical mode in three dimensions)}. The most unstable wavenumber, which corresponds to the one measured in experiments, is $\xi=\xi^*=1/2$. The selected instability wavelength is therefore $\lambda^*=4\pi a\sim 12.5664\times a$, while the growth rate is predicted to be $\Re[s^*]\sim -0.0920\times S\phi_0/\mu V$.

   \item  In the case of a varicose perturbation, the growth rate is given by
    \begin{align}
      \Re(s)&=\frac{S\phi_0}{2\mu V}\xi e^{-2\xi}.\label{2D Growth Rate Varicous} 
    \end{align}
    Therefore, a varicose perturbation is always unstable for pullers, consistently with the three-dimensional case {(as a varicose perturbation is the two-dimensional equivalent of three-dimensional pearling)}. As in the previous case, the most unstable wavenumber, is $\xi=\xi^*=1/2$. The selected instability wavelength is therefore $\lambda^*=4\pi a\sim 12.5664\times a$, while the growth rate is given by $\Re[s^*]\sim 0.0920\times S\phi_0/\mu V$.
\end{enumerate}

{For both instabilities, $\Im(s)=-\mathrm ikU_s$, corresponding to net upwards translation of the jet, as was the case in Section~\ref{Solving for the Flow}.}

{We conclude with the following important observation: suppose that the particles had zero swimming speed and were oriented along the $x$ axis instead of the $z$ axis, so that $\mathbf p=\pm\text{sgn}(x)\mathbf e_x$ and $\mathbf S=S(\mathbf e_x\mathbf e_x-\mathbf I/3)$. In this case, the two instabilities would still be observed, but they would be reversed, {i.e.~a sheet of rotated pushers would now be prone to pearling and a sheet of rotated pullers would now buckle}. This follows straightforwardly from the calculation in Appendix~\ref{Appendix C: Growth Rate of the Quasi two-dimensional Sheet}, and it is primarily a consequence of the fact that a sheet of rotated stresslets exerts a boundary stress}
\begin{align}
{S\mathbf e_x(\mathbf e_x\cdot \mathbf n)=S\mathbf n-S\mathbf e_z(\mathbf e_z\cdot\mathbf n)}.
\end{align}
{Since the $S\mathbf n$ term merely corresponds to a constant pressure, the flow driven by the rotated stresslet is the same as the flow driven by axial stresslet of strength $-S$. This reflect the fact that, in two dimensions, the flow field of a pusher and a puller stresslet of the same absolute magnitude are identical up to a $90^{\circ}$ rotation. Therefore, a varicose instability is unstable for a sheet of rotated pushers (identical to axial pullers), and vice versa for a sinuous perturbation.} 
\section{Long-term nonlinear evolution of puller clusters}\label{Stretching of Puller Clusters}
After a puller jet breaks up into  {clusters}, these are observed to flatten and stretch laterally under the effect of the internal active stresses {(Fig.~\ref{Fig: Experiments_Numerics_Active_Jet}A)}. This regime sets in once the initial varicose perturbation is well-developed. 

We now demonstrate that exact solutions for this long-term evolution may still be found, both for two- and three-dimensional jets (referred to as 2D and 3D below). In our theory, we assume clusters to be {well-}separated and hydrodynamic interactions between them to be negligible. In particular, we ignore the {fact} that, rather than pinching off completely, small cylindrical threads of swimmers {temporarily} linger between clusters (Fig.~\ref{Fig: Experiments_Numerics_Active_Jet}A). {Indeed, such threads are expected to eventually be unstable to pearling like the initial jet, and not meaningfully alter the hydrodynamics due to their small size.}

We now provide a summary of our results, while the details can be found in the specified Appendix sections. In what follows, we set the swimming speed to zero, as a finite value only rigidly translates the clusters upwards without changing their shape (since all swimmers are aligned).

\subsection{Similarity solutions}\label{Sec: Similarity solutions}
We first show in Appendices~\ref{Thin Limit of a Stretching 3D Cluster} and \ref{Thin Limit of a Stretching 2D Cluster} that the clusters eventually stretch in a self-similar fashion (neglecting cluster-cluster interactions), attaining an oblate ellipsoidal shape in 3D and an elliptical shape in 2D. In the self-similar limit, independently  of the initial shape, we find that the semi-major axes $R_{\text{3d}}$, $R_{\text{2d}}$ of the 3D and 2D clusters eventually grow in time like \begin{subeqnarray}\label{eq:similarity-solutions}
R_{\text{3d}}(t)&\sim&\displaystyle  \left(\frac{9}{32}\frac{S\phi_0 V_0}{\mu V}\right)^{1/3}t^{1/3}\slabel{3d similarity solution},\\
R_{\text{2d}}(t)&\sim&\displaystyle  \left(\frac{S\phi_0 A_0}{\pi \mu V}\right)^{1/2} t^{1/2},\slabel{2d similarity solution}
\end{subeqnarray}
where $V_0$ is the initial cluster volume in 3D, and $A_0$ is the initial cluster area in 2D. {In both the $3$D and the $2$D geometry, clusters stretch perpendicularly to the $z$ axis.} The scalings in Eq.~\eqref{eq:similarity-solutions}  are expected to hold as long as $R_{\text{3d}}$, $R_{\text{2d}}$ are much smaller than the typical cluster-cluster separation, in each case giving a cut-off time {of order} $\phi_0^{-1}$. The semi-minor axes are found by imposing that the volume of the oblate ellipsoid (3D) and ellipse (2D) are $V_0$ and $A_0$ at all times, since clusters stretch under incompressible flow (Eq.~\ref{Positional Flux}). The semi-minor axes therefore shrink like $t^{-2/3}$, $t^{-1/2}$ in the respective cases. The power laws in Eq.~\eqref{eq:similarity-solutions} may be rationalised by noting that a dipolar flow decays like $1/r^2$ in 3D, meaning that particle-particle separation grows like $\dot r\sim 1/r^2$, or $r\sim t^{1/3}$. Similarly, because the dipole flow decays like $1/r$
 in 2D, we expect $\dot r\sim 1/r$, or $r\sim t^{1/2}$. Incidentally, the result in Eq.~\eqref{2d similarity solution}  provides an asymptotic solution for the spreading of 2D pusher clusters~\cite{OppenheimerVShape}, which stretch into ellipses along an axis perpendicular to that of puller clusters. 

\subsection{Exact solution for spherical and cylindrical clusters}\label{Sec: Exact solution for spherical and cylindrical clusters}
Under the simplifying assumption that the initial cluster shape  {is a sphere (3D), or a circle (2D)}, we further show in Appendices \ref{Analytical Solution for the Spreading of a Spherical Cluster} and \ref{Analytical Solution for the Spreading of a Cylindrical Cluster} that the  {cluster} shape is \textit{exactly} an oblate ellipsoid (3D) or an ellipse (2D) at all times. In the 3D setup, the time evolution of the semi-major axis, non-dimensionalised by the initial value $R_0$, is governed by 
\begin{subeqnarray}\label{eq:29}
\frac{\mathrm dR}{\mathrm dt}&=&\frac{S\phi_0}{4\mu V}\frac{R^7}{(R^6-1)^2}\left[-3+\frac{R^6+2}{(R^6-1)^{1/2}}\cos^{-1}\left(\frac{1}{R^3}\right)\right],\slabel{R ode 3d main}\\
R(0)&=&1,\slabel{ICS 3d main}
\end{subeqnarray}
while in 2D it is given  by
\begin{subeqnarray}\label{eq:28}
\frac{\mathrm d R}{\mathrm dt}&=& \frac{S\phi_0}{2\mu V}\frac{R^3}{(1+R^2)^2}\slabel{R ode 2d main},\\
R(0)&=&1{.}\slabel{ICS 2d main}
\end{subeqnarray}
 We can confirm the analytical findings through direct numerical simulations based on the integral solution of the Stokes equations~\cite{happel1983low, kim2013microhydrodynamics} (details of the numerical schemes are provided in Appendices~\ref{Evolution of a 3D Cluster}, \ref{Evolution of a Quasi-2D Cluster}), as well as agent-based simulations of a 2D sheet of pullers {(Section~\ref{Comparison with Numerical Simulations})}. A summary of our results is shown in Fig.~(\ref{Fig: Cluster Spreading}),  {where we compute the dimensionless cluster radius versus dimensionless time for the $3$D (Fig.~\ref{Fig: Cluster Spreading}A) and $2$D (Fig.~\ref{Fig: Cluster Spreading}B) systems. In each case, the solution of Eqs.~\eqref{eq:29} and \eqref{eq:28} is compared with the values from the boundary integral method, showing {excellent} agreement}.  {In particular,} we visually demonstrate the transition between the {initial-conditions-dependent} evolution and the self-similar dynamics for $t\gg \mu V/S\phi_0$. {We note that, as the shapes determined from the boundary integral method are subject to numerical errors, these solutions are prone to developing a zigzag instability analogous to the one in Section~\ref{Sheet of Swimmers} (see Section~\ref{Sec: Long-term evolutions of clusters and zigzags}). From Eqs.~\eqref{2d Growth rate sinuous} and \eqref{2D Growth Rate Varicous} with $\xi=1/2$ (the most unstable wavenumber), we know that this instability grows on a timescale $4e\times \mu V/s\phi_0$; stopping numerical integration at the earlier time $8\times \mu V/s\phi_0$ (Fig.~\ref{Fig: Cluster Spreading}) therefore prevents this instability from developing.}

 \begin{figure}[t]
\centerline{\includegraphics[scale=0.45]{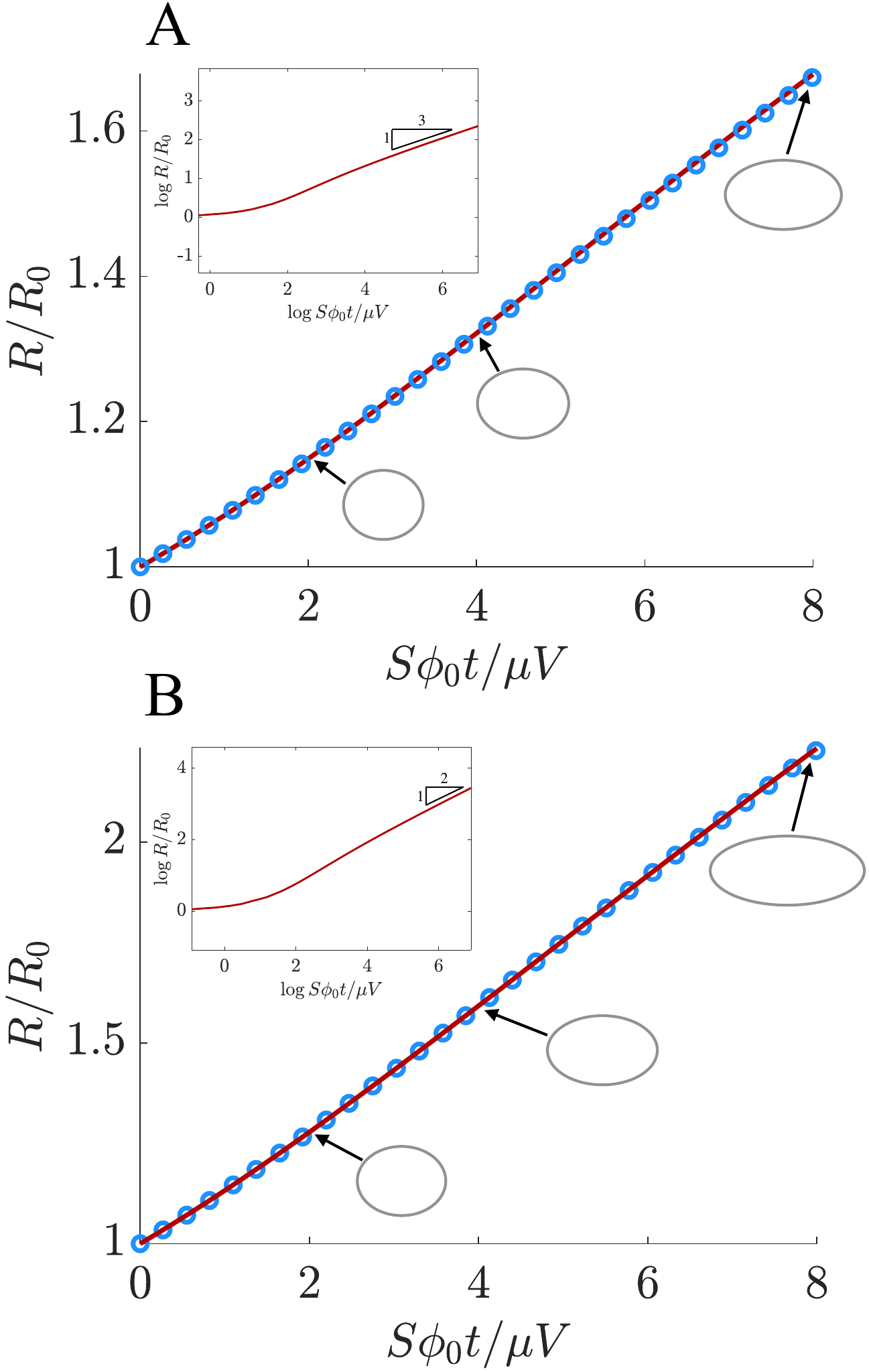}}
\caption{Evolution of puller clusters with time. (A) An initially spherical cluster stretches exactly into an oblate ellipsoid of {semi-major axis} $R$. The solid red line represents the predicted evolution obtained by solving for the flow (i.e.~Eqs.~\ref{R ode 3d main}-\ref{ICS 3d main}); the light blue circles represent {numerical data} (boundary integral method). Inset shows the long-term evolution, wherein the cluster stretches in a similarity solution with time dependence $R\sim t^{1/3}$, as predicted theoretically in Eq.~\eqref{3d similarity solution}. (B) Corresponding evolution of a  {$2$D cluster}  of pullers, which stretches into an elliptical cylinder. The red line is the numerical solution of Eqs.~\eqref{R ode 2d main} and \eqref{ICS 2d main}, while {the} light blue circles are numerical data. Inset: the cluster eventually reaches a similarity solution with $R\sim t^{1/2}$, as predicted in Eq.~\eqref{2d similarity solution}. {In both cases, clusters stretch perpendicularly to the axis of the original jet.}}
\label{Fig: Cluster Spreading}
\end{figure}

\subsection{Interactions between clusters}\label{Interactions Between Clusters}

So far, we have considered the behaviour of clusters in isolation, assuming {cluster-cluster} interactions to be negligible. We may now use the leading-order external flows, determined in Appendices~\ref{Analytical Solution for the Spreading of a Spherical Cluster}-\ref{Analytical Solution for the Spreading of a Cylindrical Cluster}, to analyse cluster-cluster interactions, in the limit where clusters only interact in the far field (i.e.~when the typical cluster size is much smaller than the typical separation length). In this case, because the  leading-order external flow is the same as the (negative) perturbation flow of an oblate ellipsoid (3D) or an ellipse (2D) in strain flow, {clusters interact as stresslets at leading order}. In Appendices~\ref{Flow Outside a 3D Cluster}--\ref{Flow Outside a 2D Cluster}, we demonstrate that the stresslet strengths are given by
\begin{subeqnarray}
\mathbf S&=&SN\left(\mathbf e_z\mathbf e_z-\frac{1}{3}\mathbf I\right), \qquad\text{(3D)}\\ 
\mathbf S&=&S\tilde{N}\left(\mathbf e_z\mathbf e_z-\frac{1}{2}\mathbf I_2\right), \qquad\text{(2D)}
\end{subeqnarray}
where $N$ is the total number of swimmers inside the 3D cluster, and $\tilde N$ is the number   per unit {extent in the $y$ direction} inside the 2D cluster. 

Remarkably, this leading-order flow does not depend on the clusters' elongation. Therefore, assuming the separation between clusters to be much larger than the typical cluster size, we expect clusters themselves to interact like a line of puller stresslets~\cite{lauga2017clustering} of strength $SN$ or $S\tilde N$, {arranged along the $z$ axis (Fig.~\ref{Fig: Experiments_Numerics_Active_Jet}A)}. In particular, the spacing between the clusters is unstable, and we expect them to merge over time. Specifically, if $\lambda$ is the typical cluster-cluster separation (on the order of the {{pearling}} wavelength), the typical external velocity field felt by each cluster has typical size $SN/\mu\lambda^2$ in 3D and $SN/\mu\lambda$ in $2D$, so that clusters are expected to merge on a timescale $\mu\lambda^3/SN$ in 3D and $\mu\lambda^2/S\tilde N$ in 2D.

\subsection{Two-dimensional cluster rotation and the ``V'' instability}\label{Sec: 2-dimensional Cluster Rotation and the ``V'' instability}

 \begin{figure}[t]
\includegraphics[width=\textwidth]{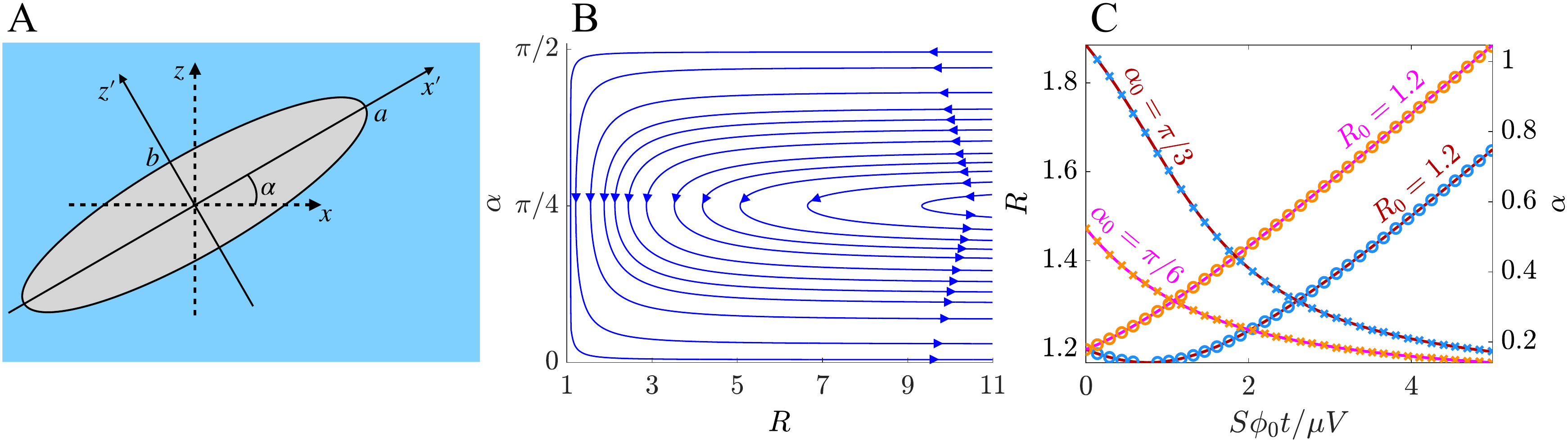}
\caption{{Dynamics of tilted puller clusters. (A) Clusters are rotated ellipses with semi-major/minor axes $a$ and $b$ (respectively)- Clusters are endowed with an $(x',z')$ coordinate system rotated through an angle $\alpha$ from the horizontal. (B) Phase diagram of evolution of dimensionless {semi-major axis} $R$ and angle $\alpha$. (C) Representative trajectories for initial conditions: $R_0=1.2$ and $\alpha_0=\pi/6$, $\alpha_0=\pi/3$. Solid lines represent theoretical predictions from Eq.~\eqref{Evolution equations tilted cluster}, while scatter plots represent numerical values obtained with the boundary-integral method in Section~\ref{Evolution of a Quasi-2D Cluster}.}}
\label{Fig: Tilted_Cluster_Sketch_Main_Text}
\end{figure}

{We have so far considered the dynamics of puller clusters extending perpendicularly to the alignment direction $\mathbf e_z$. A similar, but more general, calculation may  be carried out when the major axis of a two-dimensional puller cluster is tilted at an angle $\alpha\neq 0$ relative to the $x$ direction. As for the case $\alpha=0$ (Section~\ref{Sec: Exact solution for spherical and cylindrical clusters}), we only summarise the main results of the computation, with further details explained in Appendix \ref{Appendix: Analytical Solution for a Tilted Puller Cluster}.  }

{
In the idealised case where the cluster is initially an ellipse, we analytically show that its shape remains exactly elliptical at all times, while undergoing both stretching and  rotation. Endowing the cluster with rotated coordinates $(x,',z')$, and defining its semi-major and minor axes to be $a$ and $b$, respectively (see Fig.~\ref{Fig: Tilted_Cluster_Sketch_Main_Text}A), the flow $\mathbf u$ within the cluster takes the form
\begin{align}
\mathbf u=E(x'\mathbf e'_x-z'\mathbf e'_z)+\Omega_1\left(\frac{x'}{a^2}\mathbf e_z'-\frac{z'}{b^2}\mathbf e'_x\right)+\Omega_2(x'\mathbf e'_z-z'\mathbf e'_x).    \label{u^- ansatz tilted cluster}
\end{align}
The three terms correspond, in order, to transverse straining motion, elliptical recirculation (``treadmilling''), and rigid-body rotation. Matching boundary stresses, the corresponding flow strengths can be found in terms of the cluster's tilt angle $\alpha$ (Fig.~\ref{Fig: Tilted_Cluster_Sketch_Main_Text}A):
\begin{subequations}
\begin{align}
E&=\frac{S\phi_0}{2\mu V}\frac{ab\cos \left(2\alpha \right)}{\left(a+b\right)^2 },\\
\Omega_1&=\frac{S\phi_0}{2\mu V}\frac{a^2 b^2{\left(a^2 +b^2 \right)}\sin \left(2\alpha \right)}{{{\left(a+b\right)}}^3{\left(a-b\right)}},\\
\Omega_2&=-\frac{S\phi_0}{\mu V}\frac{a^2 b^2 \sin \left(2\alpha \right)}{{{\left(a+b\right)}}^3 {\left(a-b\right)}}.
\end{align}
\label{E Omega_1 Omega_2 tilted cluster}
\end{subequations}}

 {Denoting by $\mathcal A_0\equiv ab$ the constant cluster area, it is thus found that the dimensionless semi-major axis $R=\mathcal A_0^{-1/2}a$ and angle $\alpha$ obey   the   coupled evolution equations
\begin{subequations}
\begin{align}
\frac{\mathrm dR}{\mathrm dt}&=\frac{S\phi_0}{2\mu V} \frac{R^3\cos(2\alpha)}{(R^2+1)^2},\label{dR/dt tilted puller cluster}\\
\frac{\mathrm d\alpha}{\mathrm dt}&=-\frac{S\phi_0}{\mu V}\frac{R^4\sin(2\alpha)}{(R^2+1)^3(R^2-1)}.
\end{align}
\label{Evolution equations tilted cluster}
\end{subequations}
The initial conditions are of the form $\alpha(0)=\alpha_0$, $R(0)=R_0>1$ (as $R_0$ corresponds to the longer axis of the starting ellipse). }

{Assuming (without loss of generality)  that $\alpha_0\in [0,\pi/2]$, the phase portrait for both $R(t)$ and $\alpha(t)$ is plotted in Fig.~\ref{Fig: Tilted_Cluster_Sketch_Main_Text}B, while representative trajectories are depicted in Fig.~\ref{Fig: Tilted_Cluster_Sketch_Main_Text}C. We note the presence of constant-angle (equilibrium) solutions with $\alpha\equiv 0$ or $\alpha\equiv \pi/2$. Both of these regimes correspond to the cluster contracting along $\mathbf e_z$ and expanding along $\mathbf e_x$, with late-time behaviours $R\sim t^{1/2}$, $R\sim t^{-1/2}$, respectively. }

{If, on the other hand, $\alpha_0\in (0,\pi/2)$, then $\alpha$  decreases monotonically from its initial value, eventually approaching some limiting value $\alpha_{\infty}$ as $R\to\infty$. This limit be found by directly integrating Eq.~\eqref{Evolution equations tilted cluster}:
\begin{align}
\alpha_{\infty}=\frac{1}{2}\sin^{-1}\left(\frac{R_0^2-1}{R_0^2+1}\sin 2\alpha_0\right).
\end{align}
Since $\alpha_{\infty}\in (0,\pi/4)$ for any initial $\alpha_0$, all cluster angles are eventually found in this range. Finally, the radial deformation is expected to appear slowest when $\alpha_{\infty}\sim \pi/4$, corresponding to $\alpha_0\sim \pi/4$ and $R_0\gg 1$. In other words, while no exact steady state exists for both $R$ and $\alpha$, cluster evolve the slowest when they are long and tilted at $\pm 45^{\circ}$, consistently with the ``V'' instability observed in Ref.~\cite{OppenheimerVShape} for elongating clusters of aligned active particles. We remark that, while the simulations in Ref.~\cite{OppenheimerVShape} involved pusher stresslets oriented along the $x$ axis (see Fig.~\ref{Fig: Tilted_Cluster_Sketch_Main_Text}A), the argument in Section~\ref{Sheet of Swimmers} shows that this system evolves identically to a cluster of puller stresslets oriented along the $z$ axis (Fig.~\ref{Fig: Tilted_Cluster_Sketch_Main_Text}A).}
{We finally remark that, at late times, specifically in the limit $t\gg \mu V/S\phi_0$, the radial dynamics is expected to follow the self-similar behaviour
\begin{align}
R(t)\sim\displaystyle  \left(\frac{S\phi_0 \cos 2\alpha_{\infty}}{\mu V}\right)^{1/2} t^{1/2},\label{Tilted cluster similarity solution} 
\end{align}
corresponding to the $R\gg 1$ limit of Eq.~\eqref{dR/dt tilted puller cluster}. Eq.~\eqref{Tilted cluster similarity solution} generalises Eq.~\eqref{2d similarity solution} to the case of a tilted cluster. {A plot of the flows inside and outside a tilted $2$D elliptical cluster is provided in Fig.~\ref{Fig: Long Term Experiments Image}B, where we see good agreement with experimental measurements.}}

\section{Long-term nonlinear evolution of the 3D {helical} pusher instability}\label{Long-Term Dynamics of the 3D Waving Instability}

We finally turn our attention to the long-term evolution of the 3D pusher jet. While we have established that the initially cylindrical {jet} buckles into a helical mode, the linearised theory cannot capture the evolution of the jet after the instability has fully developed. We may, however, obtain an approximate description of a buckled jet in the experimentally relevant limit of {a slender structure, i.e.,~when the jet radius is much smaller than the buckling wavelength}. We can once again ignore the swimming speed, which only results in a rigid translation of the jet. 

\subsection{Slender-body theory}
While obtaining exact solutions outside the linear regime appears  difficult, a considerable simplification is offered by the observation that the selected wavelength $\lambda^*\sim 9.8943\times a$ is considerably larger than the  radius of the jet. It therefore seems legitimate to approximate the flows associated with large zigzags  via an asymptotic expansion in the small slenderness $a/\lambda^*\sim 0.1$. 

{Informally, while the ambient flow created by each section of the jet varies on a {length scale} $a$, the curvature of the jet only becomes apparent on a much larger {length scale} $L\sim\lambda^*$ (Fig.~\ref{Fig: Tilted Cylinder Schematics}A). The flow around the jet is therefore locally the same, with corrections of order $a/L$, as the flow around an infinite straight cylinder.}
At leading order in {$a/L$}, the slender jet {therefore} appears as a collection of non-interacting straight cylinders with different cross-sections {(Fig.~\ref{Fig: Tilted Cylinder Schematics}B)}, each driving a flow via the active boundary stress {$S\phi_0(\mathbf e_z\cdot\mathbf n)\mathbf e_z/V$}. We can determine the flow associated with each cylinder by solving Eqs.~\eqref{Plume Bulk Stokes}--\eqref{Boundary Advection Simplified}, while simultaneously identifying the local {jet} shape $\mathcal P$. In the framework of matched asymptotics, this leading-order flow {in} $\lVert\mathbf x-\mathbf x_0\rVert=\mathcal O(a)$, {where $\mathbf x_0$ is the centreline point of interest,} should then be matched onto the {$\mathcal O(a/L)$} ambient flow generated by sections of the jet with {$\lVert\mathbf x-\mathbf x_0\rVert\gg a$}. 
In order to determine the leading-order inner solution, we endow the  centreline {$\mathbf x(s)$} of the jet ($s$ being arclength) with a local orthonormal material frame $\{\mathbf t_1,\mathbf t_2,\mathbf t_3\}$, defined as (see Fig.~\ref{Fig: Tilted Cylinder Schematics}B)
\begin{subeqnarray}
\mathbf t_1(s)&=&\sin {\zeta}(s)\cos{\psi}(s)\mathbf e_x+\sin {\zeta}(s)\sin{\psi}(s)\mathbf e_y-\cos{\zeta}(s)\mathbf e_z,\\
\mathbf t_2(s)&=&-\sin{\psi}(s)\mathbf e_x+\cos{\psi}(s)\mathbf e_y,\\
\mathbf t_3(s)&=&\cos {\zeta}(s)\cos{\psi}(s)\mathbf e_x+\cos {\zeta}(s)\sin{\psi}(s)\mathbf e_y+\sin{\zeta}(s)\mathbf e_z.
\end{subeqnarray}
Here, $\mathbf t_3=\mathbf x'(s)$ is the local unit tangent, $\mathbf t_1$ is a local unit normal, and $\mathbf t_2$ is the binormal parallel the $xy$ plane. The angles ${\zeta}$, ${\psi}$ correspond to the local pitch (i.e.,~the slope relative to the $xy$ plane) and yaw (i.e.,~the twist around the $z$ axis), respectively (Fig.~\ref{Fig: Tilted Cylinder Schematics}{B, C}). Notice that $\{\mathbf t_1,\mathbf t_2,\mathbf t_3\}$  {do not} correspond to the usual Frenet frame. As the next step, we assume that the local {cross-section} $\mathcal C(s)$ {of the jet boundary}, spanned by $\mathbf t_1$ and $\mathbf t_2$, is at leading order an ellipse given by
\begin{equation}
\mathcal C(s)=\{\mathbf x(s)+a(s)\cos(\eta) \mathbf t_1(s)+b(s)\sin(\eta)\mathbf t_2(s): 0\leq\eta<2\pi\},    
\end{equation}
where $a$, $b>0$ are the (unknown) semi-axes. We will later see that, at late times, $a>b$. Intuitively, the cross-section at each point is a tilted ellipse such that the plane spanned by the surface normal $\mathbf t_3$ and the major axis $a\mathbf t_1$ is parallel to $\mathbf e_z$ (see illustration in Fig.~\ref{Fig: Tilted Cylinder Schematics}B, C). The inner solution at each point $\mathbf x(s_0)$ is then the solution of Eqs.~\eqref{Plume Bulk Stokes}--\eqref{Boundary Advection Simplified} with $\mathcal P$ being an infinite elliptical cylinder with axis $\mathbf x(s_0)+s \mathbf t_1(s_0)$ ($-\infty<s<\infty)$ and constant cross-section $\mathcal C(s_0)$. 

  \begin{figure}[t]
 \hspace{1cm}
\includegraphics[width=0.75\textwidth]{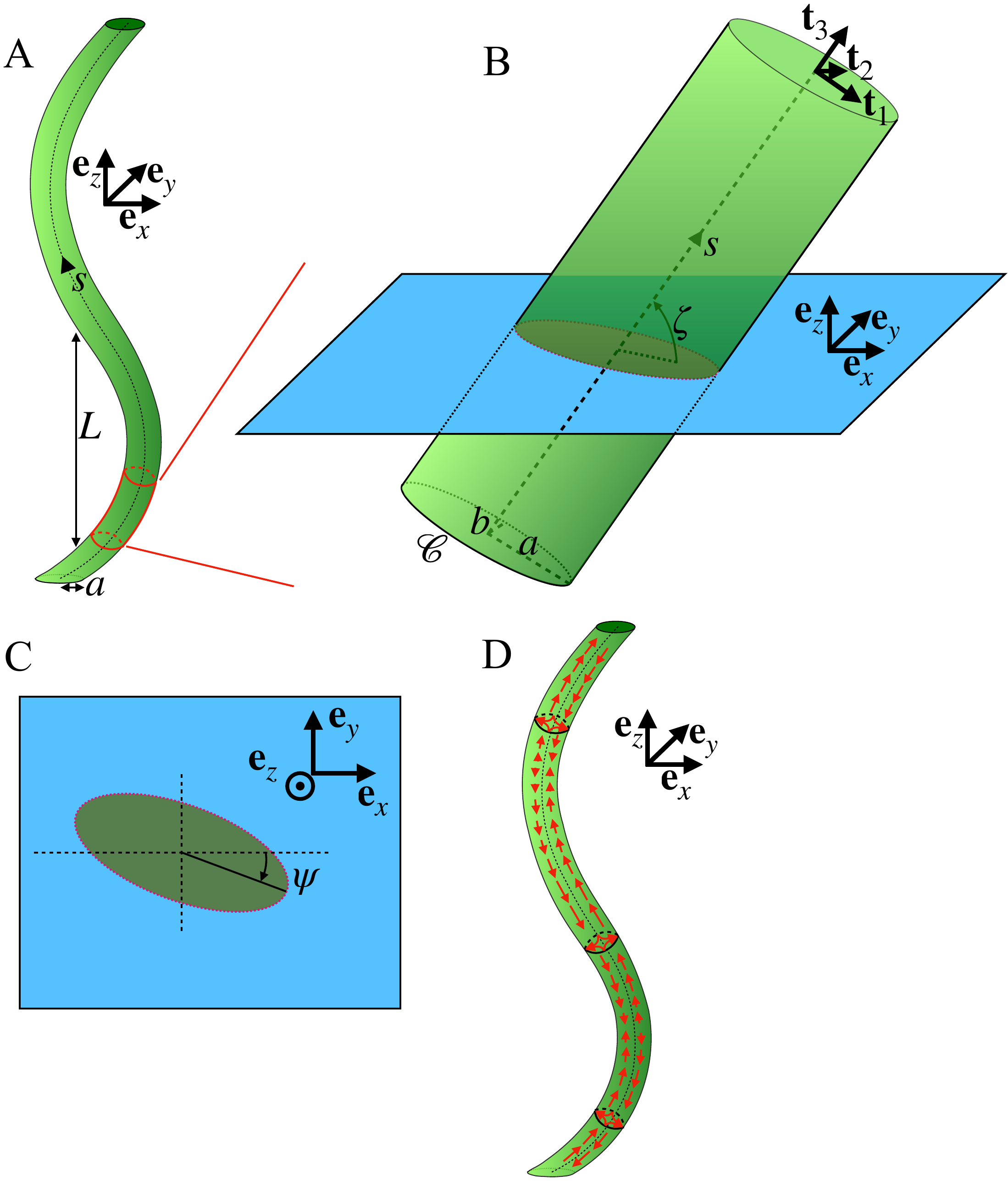}
\caption{Model of titled pusher jet. {(A) The jet is assumed to be slender, i.e.,~its local radius $a$ is much smaller than the {length scale} $L$ from its curvature.
(B) In this limit, the jet locally looks like an infinite straight cylinder, which we find to be elliptical in cross-section. The local semi-major and minor axes are labelled $a$ and $b$,  respectively. The cylinder is endowed with a local material frame $\{\mathbf t_1,\mathbf t_2,\mathbf t_3\}$, with $\mathbf t_1$, $\mathbf t_2$ pointing along the semi-axes, and $\mathbf t_3$ aligned with the axis of the cylinder. The local pitch and yaw are denoted by ${\zeta}$ and ${\psi}$, respectively (with ${\psi}$ defined as shown in C), and the cross-section boundary is $\mathcal C(s)$, with $s$ being the axial arclength. (D) Schematic representation of the flow (red arrows) inside a three-dimensional buckled jet of pushers. The axial component of the pushing force drives an axial shear flow, while the transverse force component generates a straining flow which deforms the jet. The shearing flow is strongest when ${\zeta}=\pi/4$ or ${\zeta}=3\pi/4$, and it vanishes when ${\zeta}=\pi/2$ (vertical sections). Arrow sizes are used solely to  qualitatively illustrate the flow strength.}
}
\label{Fig: Tilted Cylinder Schematics}
\end{figure}

Defining local Cartesian coordinates $\mathbf x-\mathbf x(s_0)=x_1\mathbf t_1+x_2\mathbf t_2+x_3\mathbf t_3$, the local flow inside the jet associated with this configuration is assumed to be a combination of axial shear and cross-sectional strain (Fig.~\ref{Fig: Tilted Cylinder Schematics}D){:}
\begin{subeqnarray}
\mathbf u&=&\gamma x_1\mathbf t_3+E(x_1\mathbf t_1-x_2\mathbf t_2),\\
\boldsymbol{\sigma}&=&-p_0\mathbf I+\gamma \mu(\mathbf t_1\mathbf t_3+\mathbf t_3\mathbf t_1)+2\mu E(\mathbf t_1\mathbf t_1-\mathbf t_2\mathbf t_2).
\end{subeqnarray}
Imposing continuity of velocity and stresses at the jet boundary {allows us to determine} the  straining rate $E$ and the shearing rate $\gamma$ as 
\begin{subeqnarray}
 E&=&-\frac{S\phi_0 \cos^2{\zeta}}{2\mu V}\frac{\chi}{(1+\chi)^2},\slabel{Straining Rate Zigzags}\\
\gamma&=&\frac{S\phi_0 \sin 2{\zeta}}{2\mu V}\frac{\chi}{1+\chi},\slabel{Shearing Rate Zigzags}
\end{subeqnarray}
where {$\chi=b/a$}  is the aspect ratio (see details in Appendix~\ref{Slender-Body Theory for 3D Pusher Jet}). Eq.~\eqref{Straining Rate Zigzags} reflects the fact that the projection of the pusher dipole on the $\{\mathbf t_1,\mathbf t_3\}$ plane has magnitude $S\cos^2{\zeta}$, resulting in an in-plane dynamics similar to that of stretching puller clusters (Section~\ref{Stretching of Puller Clusters}). {A jet of pushers $(S<0)$ therefore stretches along $\mathbf t_1$ (Eq.~\ref{Straining Rate Zigzags}). The $(\mathbf t_1,\mathbf t_3)$ plane, in which extension occurs, corresponds to the plane slicing the jet in half and parallel to the $z$ axis.}

\subsection{Evolution equation for a buckled jet}
Our results indicate that, at leading order in {$a/L$}, a solution exists wherein the  centreline of the jet is frozen in space, while the cross-section deforms dynamically. Importantly, this does not contradict the shape evolution computed in our previous linear stability analysis (Section~\ref{Solving for the Flow}), since leading-order slender-body theory is insensitive to deformations of order $\mathcal O(\varepsilon)\ll {\mathcal O(a/L)}$. Once the perturbation has fully developed, the small initial $\mathcal O(\varepsilon)$ perturbation is forgotten and our theory suggests the existence of a steady solution for the shape. Once the jet has reached a steady state, our analysis suggests that the structure is still highly dynamical{:} cells are stirred by the shearing flow, which carries them in loops from peak to peak, where $\gamma=0$ (Fig.~\ref{Fig: Tilted Cylinder Schematics}D). 

Our theory also predicts that each cross-section $\mathcal C(s)$ progressively deforms into a slender ellipse of constant area under the transverse straining flow. Specifically, if $R(t)$ is the local semi-major axis, non-dimensionalised by the initial value (i.e.~$R(t)=a(t)/a(0)$), then Eq.~\eqref{Straining Rate Zigzags} implies that
\begin{equation}
\frac{\mathrm dR}{\mathrm dt}={-}\frac{S\phi_0 \cos^2{\zeta}}{2\mu V} \frac{R^3}{(1+R^2)^2}\label{R evolution equation zigzags} .
\end{equation}
At large times, Eq.~\eqref{R evolution equation zigzags} indicates   that $\dot R\sim R^{-1}$, or $R\sim t^{1/2}$. Therefore, the local cross-section becomes increasingly stretched over time. 


\section{Comparison with experiments and simulations}
\label{sec:comparison}

In order to validate the theory developed in this paper, {in this section} we compare our predictions with our experiments  and with  numerical simulations. A detailed comparison of the predicted wavelengths and growth rates is carried out in Ref.~\cite{shortpaper}, {to which the reader is referred} for further details. Here, our main focus will {instead be} on validating the predicted long-term dynamics of each instability.

\subsection{Comparison with numerical simulations}\label{Comparison with Numerical Simulations}

\subsubsection{Numerical setup and method}
We carried out agent-based simulations for the 2D setup described in Section~\ref{Sheet of Swimmers}, as well as the 3D setup in Section~\ref{Solving for the Flow}. 
In the two-dimensional case (Fig.~\ref{Fig:Numerics_Results_Image}A, B, D), active particles of radius $a_s$ were seeded uniformly at random in the domain $-a<x<a$.  The particles were taken to be squirmers with a prescribed surface slip velocity
\begin{align}
\mathbf u_s=\frac{3}{2}U\beta\sin \theta\cos \theta \mathbf e_{\theta} \label{Slip Velocity} 
\end{align}
($U$ being a typical velocity) in body-fixed spherical polar coordinates centered around the swimming direction $\mathbf p$. Because the activation has front-back reflectional symmetry, the swimming speed of such squirmers is $U_s=0$, making them ``shakers''. This choice made the jet  structure more robust, and allowed the instabilities to be tracked for longer; furthermore, from our theoretical analysis, we do not expect the swimming speed to change our conclusions when the swimming direction is held fixed. A shaker endowed with slip velocity \eqref{Slip Velocity} exerts a stresslet on the fluid of scalar magnitude~\cite{lauga2016stresslets}
\begin{equation}
S=6\pi\beta\mu a_s^2 U.  \label{S function of beta}
\end{equation}
In addition, particles were assumed to reorient in response to an aligning torque, which, as we argued in Section~\ref{Continuum Description of the Jet}, does not affect the leading-order dynamics, and is thus an appropriate mechanism for simulating experiments with phototaxis~\cite{shortpaper}. In {2D} simulations, we always took $\beta>0$ (so that $S>0$, appropriate for \textit{C.~reinhardtii} algae). To simulate the {{pearling}} instability (Fig.~\ref{Fig:Numerics_Results_Image}A, D), the restoring  torque was taken to be axial, with
\begin{equation}
\mathbf G= G_{\text{bh}}\mu a_s^2 U\mathbf p\times \mathbf e_z,   \label{aligning axial torque}
\end{equation}
and dimensionless strength $G_{\text{bh}}>0$. Conversely, to reproduce the zigzag instability (Fig.~\ref{Fig:Numerics_Results_Image}B, D) observed  experimentally with pullers~\cite{shortpaper}, active particles were instead rotated towards the  axis of the jet ($x=0$) by means of a torque
\begin{equation}
\mathbf G= -\mu a_s^2 U G_{\text{bh}}\text{sgn}(x)\mathbf p\times \mathbf e_x,   
\end{equation}
where $\text{sgn}(x)$ is the sign function. {From Section~\ref{Sheet of Swimmers}, we thus infer that this sheet should have the same evolution as a sheet of axially oriented pushers of strength $-S$.}

 \begin{figure}[t]
\centerline{\includegraphics[width=0.95\textwidth]{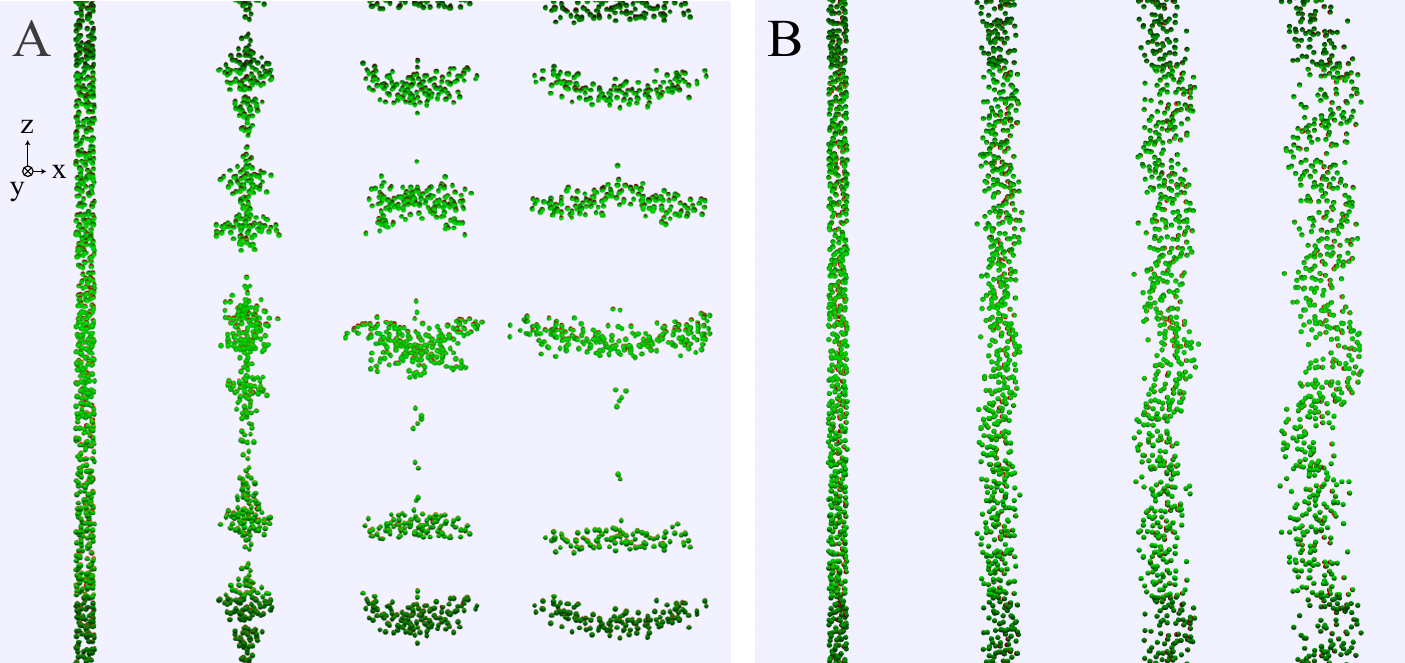}}
\caption{{Numerical realisation of the phenomena captured experimentally in Fig.~\eqref{Fig: Experiments_Numerics_Active_Jet}. (A) A sheet of squirmers, {subject} to an external torque aligning swimmers along the sheet axis, breaks up into  {clusters}, which stretch horizontally over time before developing a ``V''-shaped instability. (B) When squirmers are  subject to a torque that tries to rotate them towards the axis of the sheet, the zigzag instability emerges instead.}}
\label{Fig: Numerics_Sheet_Jet}
\end{figure}
 
The motion of the squirmers in response to the collective flow generated by the active stresses was computed with the aid of Stokesian Dynamics \cite{ishikawa2012vertical}, supplemented with periodic boundary conditions for the flow in the $y$ and $z$ directions. Such boundary conditions were implemented purely to allow for numerical resolution, and were chosen to approximate an unbounded domain. {In our simulations, we took the 2D jet to occupy the domain $96.5a_s<x<103.5a_s$, $0<y<20 a_s$, $0<z<200a_s$, while the box was instead taken to be $0<x<200 a_s$, $0<y<20 a_s$, $0<z<200 a_s$. For the chosen volume fraction, the average inter-swimmer distance at $t=0$ was about $2.7a_s$}. In order to keep the jet coherent, the reorientation timescale $B$ (estimated by balancing the applied torque with the torque on a rotating rigid sphere)
\begin{equation}
B\sim \frac{a_s}{G_{\text{bh}}U}    
\end{equation}
was always taken to be much smaller than the instability timescale $T$ (Eqs.~\ref{Most General Growth Rate}, \ref{2d Growth rate sinuous}, \ref{2D Growth Rate Varicous})
\begin{equation}
 T\sim \frac{\mu V}{S\phi_0}\sim\frac{a_s}{|\beta| U \phi_0}.  
\end{equation}
This was achieved by choosing $G_{\text{bh}}\gg |\beta| \phi_0$. In typical simulations $G_{\text{bh}}\sim 100$, $\phi_0\sim 0.1$, $|\beta|\sim 1$, so reorientation was essentially instantaneous on the timescale of the flow.

 \begin{figure}[t]
 \hspace{-1cm}
\centerline{\includegraphics[scale=0.35]{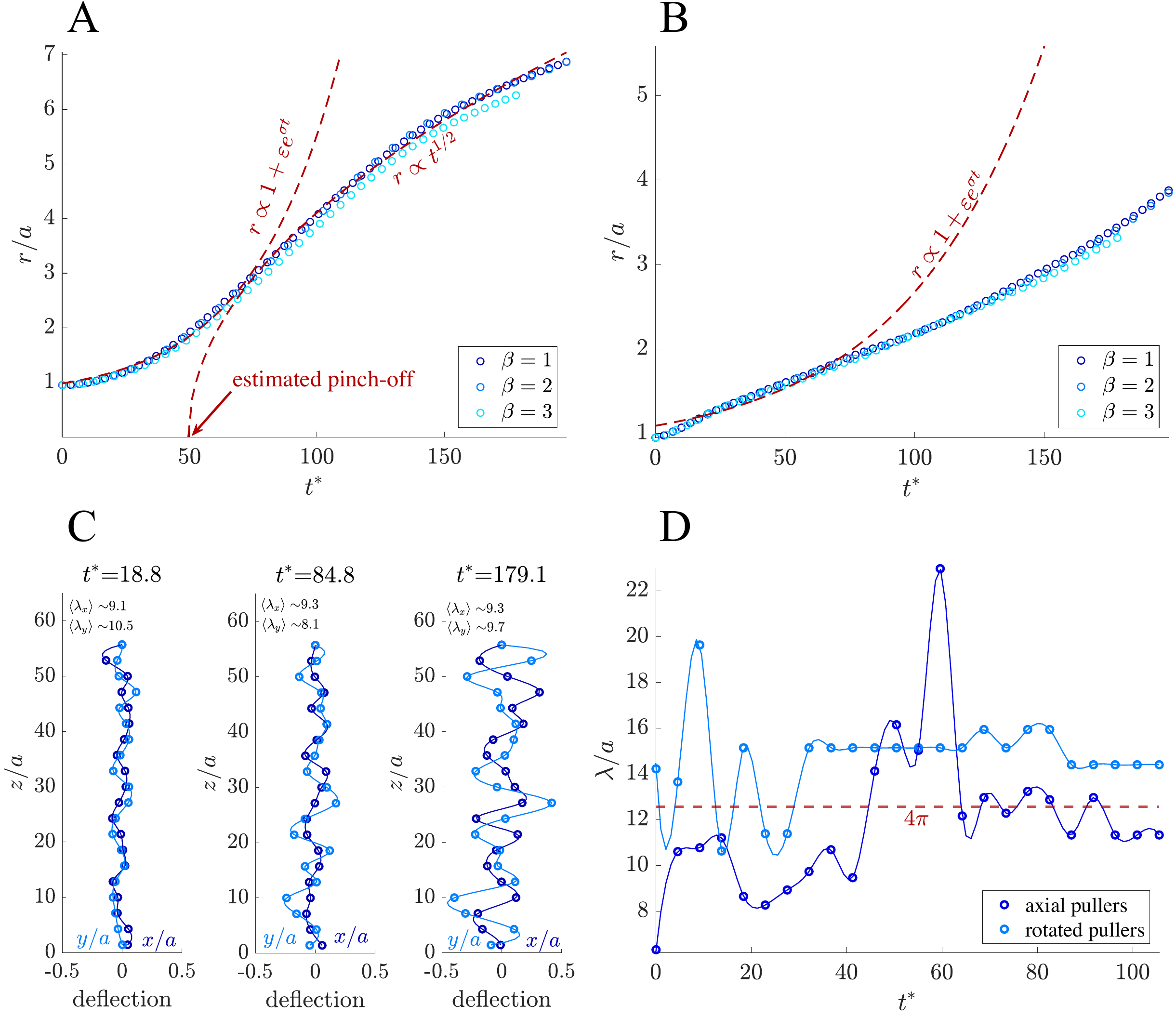}}
\caption{Summary of results of numerical simulations. (A) Time evolution of a two-dimensional sheet of axial pullers. Circles represent the axial deflection in the $x$ direction {(numerical data)} as a function of dimensionless time $t^*=6\pi \beta\phi Ut/a_s$ for various values of $\beta$, showing  collapse of the data. The dimensionless axial deflection was computed by averaging $|x|/a$ over the $15\%$ outermost swimmers. We fitted the early-time behaviour with $r/a\sim 1+\varepsilon \exp(\sigma t)$ and the late-time with {$r/a\sim p(t-q)^{1/2}$} (see Section~\ref{Stretching of Puller Clusters}). (B) Similar analysis as (A), but for a sheet of rotated pullers; curves once again collapse neatly for varying $\beta$. (C) Time evolution of a cylindrical jet of axial pushers. The jet was divided into vertical sections{,} and the average $x$ and $y$ positions of the swimmers in each section were computed (circles; solid lines serve as guides for the eye). When plotted again the section-averaged $z$, data show that the jet buckles into a helical mode. For each of $x(z)$, $y(z)$, time-dependent wavelengths $\lambda_x$, $\lambda_y$ were estimated as the average distance between local maxima of the corresponding curves. (D) Evolution of the instability wavelength for a sheet of axial and rotated pullers. The wavelength was computed from the discrete Fourier transform of the swimmers' axial displacement. Numerical data (circles), show that $\lambda/a\sim 4\pi$,  {in agreement with the theory (see Section Section~\ref{Sheet of Swimmers})}. Solid lines serve as guides for the eye.}
\label{Fig:Numerics_Results_Image}
\end{figure}

 \subsubsection{{Comparison with theory at early and late times}}
Numerical simulations of axial and rotated puller sheets exhibited the expected {{pearling}}/{zigzag} instabilities (respectively), with wavelengths and timescales compatible with the theoretical predictions in Section~\ref{Sheet of Swimmers}. The temporal evolution of the sheet is seen to scale linearly with $\beta$, as demonstrated by the collapse of the growth curves in Fig.~\ref{Fig:Numerics_Results_Image}A and B when plotted against the dimensionless time $t^*=6\pi\beta\phi Ut/a_s$. Exponential fitting of the growth curves in Fig.~\ref{Fig:Numerics_Results_Image}A, B provides the estimates {$r/a\sim 1+0.37\times \exp(0.029\times t^*)$} {($r$ being the average position $|x|$ of the 15$\%$ outermost swimmers}) for axial pullers (Fig.~\ref{Fig:Numerics_Results_Image}A), and {$r/a\sim 1+0.35\times \exp(0.019\times t^*)$} for rotated pullers (Fig.~\ref{Fig:Numerics_Results_Image}B). {The number of significant figures in the fitted values reflects the corresponding $95\%$ confidence intervals.}
{The dimensionless growth rates extracted from simulations may now be compared with Eqs.~\eqref{2d Growth rate sinuous} and \eqref{2D Growth Rate Varicous} for $\xi=1/2$ (the most unstable wavenumber). After non-dimensionalising the corresponding growth rates by $6\pi \beta\phi U/a_s$, we find that, for both instabilities, $s^*=3/(16\pi e)\sim 0.0220$, which closely resembles the numerical estimates}. The fitted values of $\varepsilon$ are instead not as significant, as the initial perturbation is inherently random.

The long-term behaviour of the puller sheet in Fig.~(\ref{Fig:Numerics_Results_Image}A) was further investigated. {As clusters stretch thin, the zigzag instability predicted 
in Section~\ref{Sheet of Swimmers} is indeed observed (Fig.~\ref{Fig: Numerics_Sheet_Jet}A), and the resulting clusters attain a shape reminiscent of the ``V'' instability (Section~\ref{Sec: 2-dimensional Cluster Rotation and the ``V'' instability}, \cite{OppenheimerVShape}).} {Before this happens, the evolution of the cluster is approximately given by Eq.~\eqref{2d similarity solution}. In particular, the  {numerical} long-term growth can be fitted by $r/a\sim 0.578\times (t^*-49)^{1/2}$ (Fig.~\ref{Fig:Numerics_Results_Image}A), suggesting that the sheet pinches off at $t^*\sim 49$, consistently with the  {theoretical dimensionless} instability timescale $1/\sigma^*\sim 45.5452$.}  
{We remark that, since the clusters are not precisely ellipses to start with, the exact solutions derived in Sections~\ref{Sec: Exact solution for spherical and cylindrical clusters} and \ref{Sec: 2-dimensional Cluster Rotation and the ``V'' instability} are only in qualitative agreement with the numerical clusters {(Fig.~\ref{Fig: Numerics_Sheet_Jet}A)}. On the other hand, as cluster continue to stretch, they transition to the self-similar behaviours from Section~\ref{Sec: Similarity solutions} and Section~\ref{Sec: 2-dimensional Cluster Rotation and the ``V'' instability}, since a thin cluster has no memory of its initial shape.} 

The long-term evolution of the pusher instability was  {further} investigated in Fig.~\ref{Fig:Numerics_Results_Image}C by considering a three-dimensional setup similar to the one studied theoretically in Sections~\ref{Solving for the Flow} and \ref{Long-Term Dynamics of the 3D Waving Instability}.
{This time,} spherical pushers were initially distributed uniformly at random inside a cylindrical domain {$r\leq 3.5a_s$}  (Fig.~\ref{Fig:Numerics_Results_Image}C), and an axial torque (Eq.~\ref{aligning axial torque}) was applied to maintain {their} axial orientation. {The box size was in this case taken to be $-200a_s<x,y<200 a_s$, $-100a_s<z<100a_s$, and the average inter-swimmer distance at $t=0$ was about $2.3a_s$.} After the jet buckled, we proceeded to track the displacement of the $x$ and $y$ coordinates of the jet centreline, calculated by breaking up the jet into sections and calculating the average swimmer position within each section (Fig.~\ref{Fig:Numerics_Results_Image}C).  {This revealed helical buckling, with $x(z)$ and $y(z)$ showing {out-of-phase} periodic variations.} 
The instability wavelengths measured in Fig.~(\ref{Fig:Numerics_Results_Image}C) suggest $8\leq \lambda/a\leq 10$, consistently with our theoretical prediction $\lambda/a\sim 9.8943$ derived in Section~\ref{Fastest-growing modes}. We further show in Fig.~\ref{Fig:Numerics_Results_Image}C that the shape of the jet does not change meaningfully over time, consistently with the slender-body analysis in Section~\ref{Long-Term Dynamics of the 3D Waving Instability}.

Finally, in Fig.~\ref{Fig:Numerics_Results_Image}D {we show} the measured wavelength for a sheet of axial and rotated pullers. Numerical results are in very good agreement with the theoretical prediction $\lambda/a\sim 4\pi$, derived in Section~\ref{Sheet of Swimmers}. 



\subsection{{Comparison with experiments}}\label{Comparison with Experiments}

We now compare the predictions of our theory with the experimental  data from Ref.~\cite{shortpaper}. As a reminder, in our experiments we employ a strain  {of} \textit{C.~reinhardtii} (CC125) in photophobic conditions to study both instability types. The puller instability was obtained when cells were aligned with the jet axis while, in order to generate the instability for pushers, algae were oriented perpendicularly to the jet axis by means of strong parallel arrays of LEDs. By doing so, the flow field around each swimmer resembled that of a pusher oriented along the  vertical axis of the jet. 

\subsubsection{Parameters}
{In our experiments, due to the illumination geometry and the presence of the tank, the geometry of the jet is halfway between the cylindrical and sheet-like cases in Fig.~\eqref{Fig: Jet Sketch}. Specifically, the jet takes the shape of  a flattened cylinder, with a width of about $100-200\ \mu$m in the $x$ direction and about $50\ \mu$m in the $y$ direction \cite{shortpaper,eisenmann_light-induced_2024}. We thus expect the dynamics to be a hybrid of the limit cases  discussed in Section~\ref{Solving for the Flow} (cylindrical geometry) and Section~\ref{Sheet of Swimmers} (sheet-like geometry)}. 
{The experimental jet satisfies the fast-reorientation limit detailed in Section~\ref{Assumptions}, given the typical experimental {values} $B\sim 1$~s, $T\sim 10$ s~\cite{shortpaper}. Likewise, the persistent sharp drop in concentration outside the jet \cite{shortpaper} indicates that the Péclet number is large enough for the zero-diffusion approximation in Section~\ref{Assumptions} to be valid.} {On the other hand, by virtue of the mechanism for jet formation, the experimental jets are highly concentrated {($\phi\sim 0.5$ for the zigzag instability, $\phi\sim 0.2$ for the pearling instability)}.  As such, Eqs.~\eqref{Most General Growth Rate}, \eqref{2d Growth rate sinuous} and \eqref{2D Growth Rate Varicous} provide upper bounds for the experimental growth rates, rather than precise estimates, {as expected from the extra resistance added by particle-particle interactions}. Indeed, further numerical simulations  with a more concentrated jet than in Section~\ref{Comparison with Numerical Simulations} ($\phi_0=0.124$ instead of $\phi_0=0.07$, not shown) confirm that continuum theory overestimates the growth rate in the non-dilute case.}

{Despite being well outside the dilute limit assumed in the theoretical framework, the experiments are nonetheless well-described by both the 3D and 2D models, which correctly predict the type of instability and the characteristic wavelengths for axial and rotated pullers}.   {This is likely because the instability wavelengths only depend weakly on the concentration.}
We refer {the reader} to the joint paper \cite{shortpaper} for {a} full quantitative  comparison {of the initial instability dynamics}{;} in what follows, we focus on the flows associated with the onset of each instability, as well as the long-term evolution.

 \begin{figure}[t]
 \hspace{-0.5cm}
\includegraphics[width=\columnwidth]{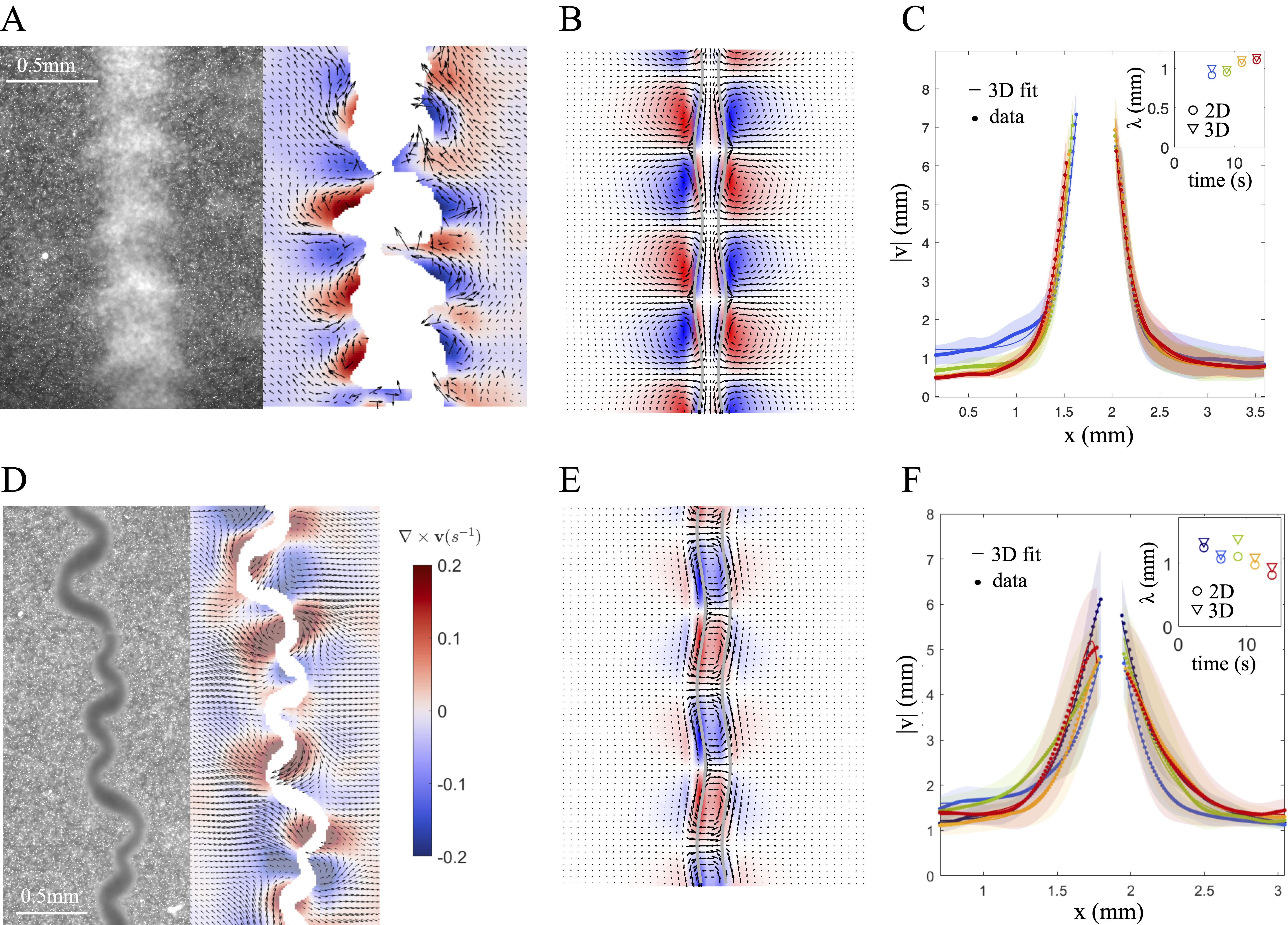}
\caption{Comparison of the flow structure around the jet. (A) \& (D) Dark-field images of the jet of pullers (A) and rotated pullers (D) (left), along with PIV in the laboratory frame, with background colour showing the out-of-plane component of the vorticity (right, {same colourbar for A and D}).  (B) \& (E) Theoretical predictions for the flow direction and vorticity for the most unstable mode of a  {3D} jet of pushers and pullers, respectively. The vorticity fields near the jets match the ones in the PIV images A and D. (C) \& (F) Flow magnitude away from the jets at different times. Solid lines indicate {fits} with the three-dimensional decay rate $a(x-x_0)^{1/2}\exp[-k(x-x_0)]+d${, with $a, x_0, k$ and $d$ as free fitting parameters}. Insets show the wavelength $2\pi/k$ obtained by fitting the 2D and 3D model to the experimental data.}
\label{Fig: Flow_Fields_Comparison}
\end{figure}

\subsubsection{Flow field at {the onset of} the instability}
{In Fig.~\ref{Fig: Flow_Fields_Comparison} we show a comparison between the theoretical flow fields derived from Eqs.~\eqref{U_1^+ Expression}, \eqref{V_1^+ Expression}, \eqref{W_1^+ Expression}, \eqref{U_1^- Expression}, \eqref{V_1^- Expression}, and \eqref{W_1^- Expression}  (derivation in Appendix~\ref{Appendix A}), and} the {experimental} measurements (Movie S1 and S2).
{We measured the fluid flow created by the cells by particle image velocimetry (PIV) analysis of suspensions laden with passive tracer particles. We used polystyrene beads of $2\ \mu$m diameter as tracers. Movies were recorded under the microscope at 8 fps, using the sources driving the instability to achieve darkfield illumination. As we needed to distinguish algae from tracers to measure the flows, the exposure time was chosen high enough to blur the fast-swimming algae, but low enough to keep the much slower beads crisp (Fig.~\ref{Fig: Flow_Fields_Comparison}A, D). This allowed us to simultaneously track the jet destabilization (for which single-cell resolution is not necessary), and the flow fields. We then used Matlab's PIVlab, after running our images through a spatial band-pass filter to ensure only the beads' movements were taken into account. }

We see from Fig.~\ref{Fig: Flow_Fields_Comparison}  that our model quantitatively agrees with the experiments, both in terms of the flow circulation (Fig.~\ref{Fig: Flow_Fields_Comparison}A, D versus Fig.~\ref{Fig: Flow_Fields_Comparison}B, E), as well as the flow decay rate (Fig.~\ref{Fig: Flow_Fields_Comparison}C, F).  Specifically, from the results in Eqs.~\eqref{U_1^+ Expression}, \eqref{V_1^+ Expression} and \eqref{W_1^+ Expression} for the external flow (Appendix~\ref{Appendix A}), we expect the flow field around a 3D jet to have a characteristic decay {rate} {$\lVert\mathbf u_1^+\rVert^{\rm (3D)} \sim rK_n(kr)\sim r^{1/2}\exp(-kr)$}, which agrees very well with the data in Fig.~\ref{Fig: Flow_Fields_Comparison}C, F. Similarly, the flow around a 2D jet is expected to decay like {$\lVert\mathbf u_1^+\rVert^{\rm (2D)}\sim xe^{-kx}$} (see Eqs.~\ref{U_1+ 2d} and \ref{V_1+ 2d} in Appendix~\ref{Appendix C: Growth Rate of the Quasi two-dimensional Sheet}) which also   matches the flow data reasonably well, with fitted wavelength $2\pi/k$ comparable to the one obtained from the 3D fit. This  reflects the fact that the  geometry of the jet is a hybrid between our 3D and 2D setups.

\subsubsection{Long-term evolutions of clusters and zigzags}\label{Sec: Long-term evolutions of clusters and zigzags}

{We next set out to validate our theoretical predictions for the long-term evolutions of the instabilities. Visual comparison of the theoretical and experimental data is provided in Fig.~\ref{Fig: Long Term Experiments Image}. {The expected elliptical cluster shape (Section~\ref{Sec: Exact solution for spherical and cylindrical clusters}) after pinch-off is   in qualitative agreement with the experiments (Fig.~\ref{Fig: Experiments_Numerics_Active_Jet}A). After $\sim$1 min, the clusters destabilised into the ``V'' shape (see Fig.~\ref{Fig: Experiments_Numerics_Active_Jet}A and \ref{Fig: Long Term Experiments Image}C) already shown in simulations (Fig.~\ref{Fig: Numerics_Sheet_Jet}A). The tilted clusters resulting from this instability are nearly elliptical in shape, in agreement with Section~\ref{Sec: 2-dimensional Cluster Rotation and the ``V'' instability}.}  
This secondary instability can also be understood purely in terms of hydrodynamic interactions: {from Section~\ref{Sheet of Swimmers}, we know that the flow within a stretching two-dimensional cluster is mathematically identical whether the swimmers are pullers oriented along the $z$ axis, or pushers oriented along the $x$ axis. Therefore, a spreading puller cluster behaves like a pusher jet oriented along the $x$ axis. As such, two-dimensional puller clusters are subject to the zigzag instability (Section~\ref{Sheet of Swimmers}), which becomes apparent as soon as the major axis of the cluster} is on the order of the wavelength $\lambda^*_{\rm 2D}=4\pi w$ ($w$ being the minor axis of the cluster). A single wavelength develops, giving the droplets their final characteristic ``moustache'' shape. These structures are stable and highly dynamic, with cells circulating {in vortices} over hundreds of micrometers (Fig.~\ref{Fig: Long Term Experiments Image}C). Such stability, as well as the presence of the treadmilling flow (Fig.~\ref{Fig: Long Term Experiments Image}A and especially Fig.~\ref{Fig: Long Term Experiments Image}C), match the steady-state predictions from Section~\ref{Sec: 2-dimensional Cluster Rotation and the ``V'' instability}} (Fig.~\ref{Fig: Long Term Experiments Image}B).}

{Remarkably, as shown in Ref.~\cite{OppenheimerVShape}, a strongly aligned active suspension of microswimmers exhibits symmetries that restrict the allowed distribution of particles. In particular, the striking macroscopic ``V''  structures observed in experiments can be understood through a Hamiltonian formalism. The Hamiltonian for such a system is notably scale-invariant and symmetric with respect to $\pm 45^{\circ}$,  leading to lines of particles at angles $\pm 45^{\circ}$ - the same result we obtained through our flow calculation.}

 {{Furthermore, measured cluster  {aspect ratios} (Fig.~\ref{Fig: Long Term Experiments Image}D) appear to {decay} at an {average} rate $\mathcal O(t^{-1})$, even though the data show high variability.}} This is in agreement with our theory (Section~\ref{Sec: Similarity solutions}), predicting the aspect ratio to decay like $t^{-1}$ in both 3D and 2D. {The experimental variability is {likely} a consequence {of noise, as well as} interactions between the clusters, since {the} clusters are not isolated as assumed by the analytical prediction.}

\begin{figure}[t]
	\centerline{\includegraphics[width=\columnwidth]{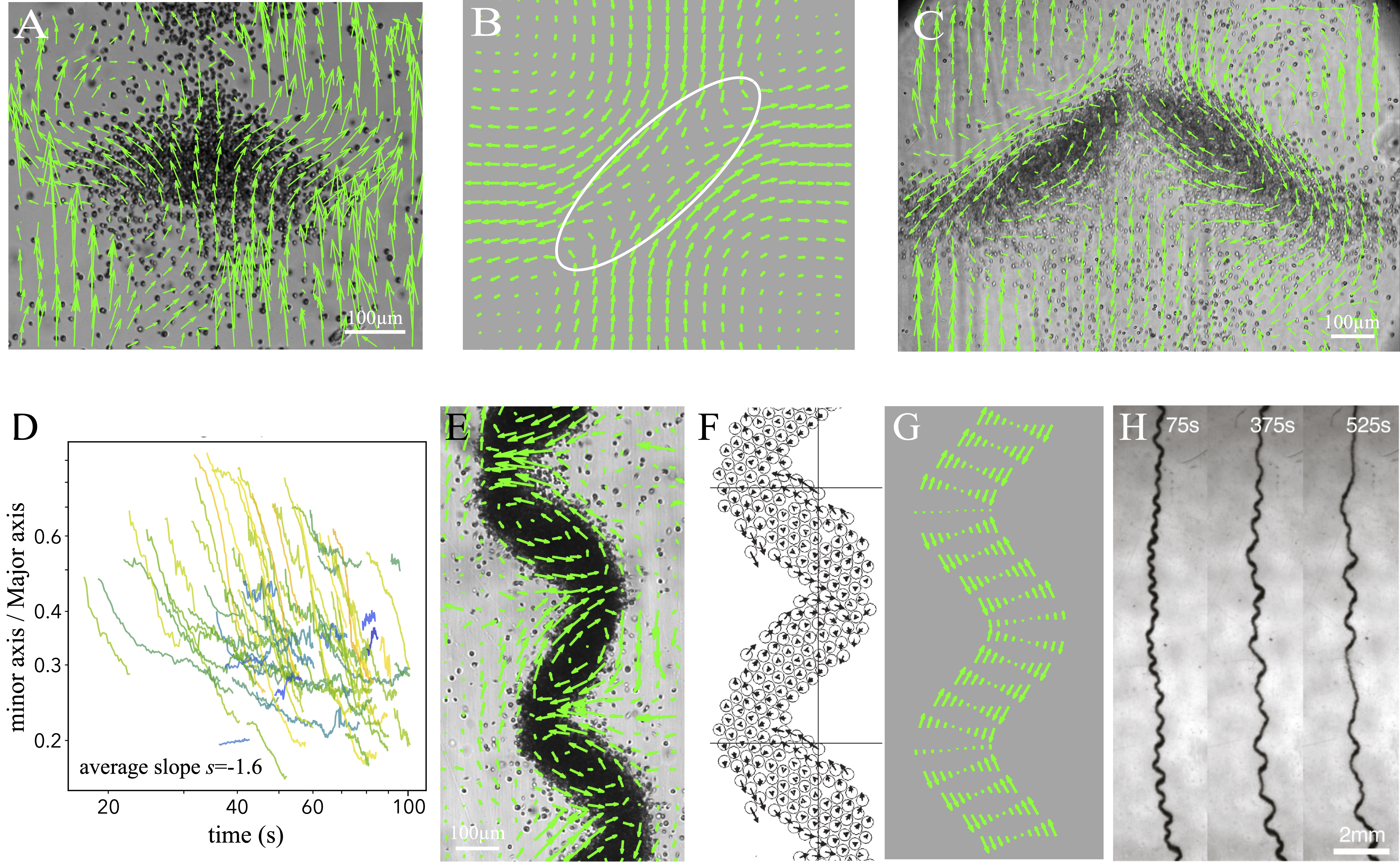}}
	\caption{Evolution of clusters and zigzags  at large times. (A) Cell displacement field inside a puller cluster after the instability has first developed, highlighting strain-like flow inside the cluster and stresslet-like flow outside (PIV on the cells in the reference frame of the cluster; {the background flow corresponds to the global velocity of the cluster. Top view, light coming from the top of the image}). (B) {Theoretical flow field for a 2D puller cluster tilted at $45^{\circ}$ (Section~\ref{Sec: 2-dimensional Cluster Rotation and the ``V'' instability})}. (C) Cell displacement field in the final \textcolor{blue}{``V ''}-shaped {cluster} (PIV on the cells in the co-moving frame, light coming from the top of the image). (D) Time evolution of experimental cluster aspect ratio. The color {encodes} the slope of the curves (yellow to blue). (E) Cell displacement field inside a jet of rotated pullers after the instability has first developed. (F): Same, numerics \cite{ishikawa2008coherent}. (G) Theoretical flow inside 3D zigzags at late times, consisting of axial shear and transverse strain. (H) Experiments showing coarsening of zigzag wavelength {on the timescale of $5$-$10$ minutes}.} 
	\label{Fig: Long Term Experiments Image}
\end{figure}


Finally, we consider the late-time evolution of the zigzags. As predicted by the theory (Section~\ref{Long-Term Dynamics of the 3D Waving Instability}), the amplitude of the zigzags is seen to rapidly stabilize after the initial exponential growth \cite{shortpaper} (Fig.~\ref{Fig: Long Term Experiments Image}E). The theoretical flow, approximately consisting of a superposition of axial shear and transverse strain (Fig.~\ref{Fig: Long Term Experiments Image}G), agrees  well with the detailed measurements in Fig.~\ref{Fig: Long Term Experiments Image}E. {Because the shape evolution is slower where the jet is nearly vertical (i.e.,~$\cos{\zeta}\sim 0$ in Eq.~\ref{R evolution equation zigzags}), we expect a less wavy jet to stay unchanged for longer. This is reflected in our experiments, where shape deformations lead to mergers of the zigzags, producing a slowly-evolving structure with large wavelength (Fig.~\ref{Fig: Long Term Experiments Image}H). The experimental jet is seen to slowly evolve over a period of minutes} (Fig.~\ref{Fig: Long Term Experiments Image}H). Such a large time scale is justified by Eq.~\eqref{R evolution equation zigzags}, wherein $\cos^2{\zeta} \ll 1$ for a nearly-vertical jet (${\zeta}\sim \pi/2$). For instance, taking $\cos{\zeta}\sim a/\lambda^*\sim 0.1$, the {helical mode} evolves on a timescale $100$ times {longer than} the instability timescale. For an instability timescale of about $10$ s, it would {thus} take on the order of $10^3$ s ($\sim 15$ minutes) to observe any significant changes (Fig.~\ref{Fig: Long Term Experiments Image}H).

\subsubsection{Microscopic origin of an effective stresslet strength depending on light intensity}

\begin{figure}
	\includegraphics[width=1\columnwidth]{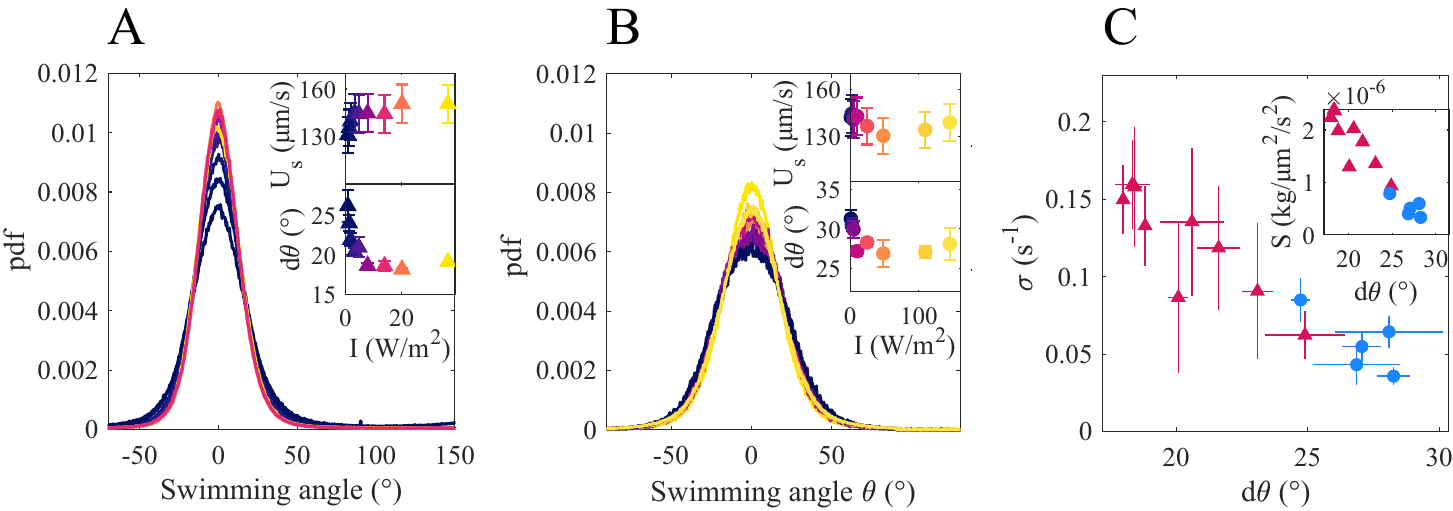} 
	\caption{{
		Instantaneous swimming angle distributions of algae at the light intensities used in the macroscopic experiments. \textbf{A}: zigzags setting with light coming from one side LED array, \textbf{B}: {pearling} setting with a single collimated LED. Insets: angular dispersion $d\theta$, extracted from a Gaussian fit, and average swimming speed $U_s$ vs light intensity. Colors correspond to light intensity.
		\textbf{C}: The growth rates $\sigma$ of the {pearling} (blue circles) and zigzags (red triangles) instabilities decrease with the angular swimming dispersion $d\theta$ of the cells. For each data point, $\sigma$ and $d\theta$ were independently measured at a given light intensity. Inset: effective stresslets $S_{\rm eff}$ extracted from the growth rates using the 3D model with {$\phi_{0}^{\rm zigzag}=0.5$} and $\phi_{0}^{\rm pearling}=0.2$.}
	}
	\label{fig:growthrate}
\end{figure}

{The driving force {behind} the instabilities (and the main parameter of the model) is the stresslet exerted by the cells on the fluid. In the experiments, however, the growth of the instabilities was controlled by the intensity of the axial (for clusters) or side (for zigzags) LEDs ~\cite{shortpaper}. In order to rationalize the experimentally-measured growth rates at different light intensities in the framework of the model, we investigated how cell swimming was related to the light intensity. To do so, we measured the trajectories of dilute suspensions of algae { at different light intensities,} under the same illumination geometries as Fig.~\ref{Fig: Experiments_Numerics_Active_Jet} in Ref.~\cite{shortpaper}. The resulting swimming angle distributions were seemingly Gaussian, which allowed us to extract their standard deviation $d\theta(I)$, capturing the angular swimming dispersion around the light's main direction at a given intensity (Fig.~\ref{fig:growthrate}A, B). {Importantly, the observably large spread of the distributions were mainly due to the cells' helical trajectories \cite{leptos2023phototaxis}}. In both settings, cells followed the light direction more consistently as the intensity was increased.
The algae swimming speed $U_s$ also slightly changed with $I$ (insets of Fig.~\ref{fig:growthrate}A, B). From the measured dependence of the instability growth rates on the angular dispersion $d\theta$ (Fig.~\ref{fig:growthrate}C), we could estimate a corresponding ``effective stresslet'' from the theoretical predictions {$\sigma^{\text{pearling}}_{\text{3D}}\sim 0.097\times\frac{|S|\phi_0}{\mu V_s}$} and  $\sigma^{\text{zigzag}}_{3D}\sim 0.024\times\frac{|S|\phi_0}{\mu V_s}$ (Section~\ref{Fastest-growing modes}), with the experimental values {$\phi_{0}^{\rm pearling}=0.2$ and $\phi_{0}^{\rm zigzag}=0.5$}. Remarkably, the estimates all fell on the same curve (Inset of Fig.~\ref{fig:growthrate}C), suggesting that the growth rate is indeed well-described by our model, with $S$ replaced by an effective stresslet $S_{\text{eff}}(d\theta)$. The magnitude of this effective stresslet decreased significantly as the angular dispersion $d\theta$ increased, which could not be justified solely in terms of the small change in swimming speed with light intensity. We instead propose that, when cells wobble around their main swimming direction, their average stresslet is smaller than when they swim along straight paths.} 

\section{Conclusion}
In this paper, we employed a coarse-grained continuum theory to investigate the shape instabilities of {jet}s and sheets of active swimmers. {Taken together, our results describe several geometric features of the invasion of clear fluid by a strongly aligned active suspension \cite{miles2019active}.} 
In Section~\ref{Continuum Description of the Jet}, we derived  {continuum equations} for the evolution of {coherent active structures}, under the  assumptions of negligible diffusion and a constant swimming direction. We then conducted a linear stability analysis of the shape of 3D {jet}s (Section~\ref{Solving for the Flow}), and 2D sheets (Section~\ref{Sheet of Swimmers}). We recovered the {{pearling}} instability for pullers and the {helical/zigzag} instability for pushers, and predicted observable features of such instabilities, such as the growth rates and wavelengths (the comparison of our predictions  with experiments is detailed in our companion paper~\cite{shortpaper}).  {We then extended our continuum theory to study the long-term evolution of {a} puller cluster both in a 3D and in a 2D setup (Section~\ref{Stretching of Puller Clusters}). We recovered exact {solutions} for the evolution of a spherical or cylindrical cluster, and derived {universal} long-term similarity solutions in the two cases. In particular, the scalings $r\sim t^{1/3}$ in 3D and $r\sim t^{1/2}$ in 2D reflected the underlying dipolar flows. We finally employed slender-body theory to derive approximate equations for the evolution of the 3D zigzags (Section~\ref{Long-Term Dynamics of the 3D Waving Instability}), showing that, while the pusher jet eventually stops buckling, its features continue evolving. We correctly predicted the flow inside the buckled jet and the timescales for wavelength coarsening, elucidating the underlying mechanism.}
We  {finished by comparing} our results with direct numerical simulations (Section~\ref{Comparison with Numerical Simulations}) and experiments (Section~\ref{Comparison with Experiments}), recovering good agreement,  {both in terms of the instability wavelengths and growth rates, as well as the {measured} ambient flows. Our {nonlinear} theory quantitatively captured the stretching of puller clusters, and {predicted the coarsening of three-dimensional zigzags as well as the corresponding timescale}.}

 {As shown experimentally in Ref.~\cite{shortpaper}, our analysis confirms that the active stresses {generated} by microorganisms are sufficient to produce instabilities  {in {coherently-structured} active jets}. {These instabilities} are observed for both pusher and puller microorganisms. The nature of the flows driving the instabilities reflected the small-scale dipolar flows. Pullers tend to drive extensional flow perpendicular to their swimming direction (Section~\ref{Stretching of Puller Clusters}), while pushers stretch coherent structures along their swimming direction (Section~\ref{Long-Term Dynamics of the 3D Waving Instability}). Our analysis also revealed surprising self-similarities in the breakup of a puller jet, with aligned clusters driving a flow qualitatively identical to that of individual swimmers (Section~\ref{Interactions Between Clusters}).} 

 {Theoretically, a detailed investigation of the dynamics was possible thanks to the approximations of perfect alignment between the swimmers and the external light ($\mathbf p\equiv \mathbf e_z$), and of zero diffusion. It will be interesting to generalise these results to allow for rotating swimmers or a small diffusive boundary layer outside the jet (i.e.~a finite Péclet number). While not necessary to recover the leading-order dynamics, diffusion is unavoidable in biological systems, and should therefore be accounted for in a more realistic model.
One could also allow for a position-dependent reorientation speed, in order to model shadowing in the case of phototaxis: in experiments, cells in the middle of the jet do not receive much light, and their orientations are mostly incoherent.} 
  
  Overall, this work highlights the richness and physical significance of aligned active suspensions, and we hope our study paves the way for further research on both theoretical and experimental levels.

\section{Acknowledgments}
We would like to thank Weida Liao for useful feedback. This work was partly funded by EPSRC (scholarship to MV) and supported through the Junior Research Chair Programme (ANR-10-LABX-0010/ANR-10-IDEX-0001-02 PSL; R.J.) and an ED-PIF doctoral fellowship (I.E.). TI is supported by JSPS KAKENHI (No. 21H04999 and 21H05308).



\appendix
\section{Derivation of continuum equations}\label{Matched Asymptotics}
In this first Appendix, we provide two derivations of Eqs.~\eqref{Bulk Jet Stress Balance} through \eqref{Table Boundary Continuity of Velocity} for the jet.  {In Section~\ref{Matched Asymptotic Solution}, we show how to reconcile the sharp jet boundary with the requirement of statistical homogeneity~\cite{batchelor1970stress}. In Section~\ref{Continuum equations alternative derivation}, we recover Eqs.~\eqref{Bulk Jet Stress Balance} through \eqref{Table Boundary Continuity of Velocity} from superposing the stresslet flows created by individual swimmers.}

\subsection{Solution from matched asymptotics}\label{Matched Asymptotic Solution}
 {In this subsection,} we aim to justify the assumption of a discontinuous concentration at the jet boundary, which may appear, at first,  at odds with the implicit assumption of statistical homogeneity needed to identify $S(p_ip_j-\delta_{ij}/3)$ with the particle stress (Eq.~\ref{Axisymmetric stresslet expression})~\cite{batchelor1970stress}. More precisely, this step requires a separation of {length scales} between the particle size $a_s$ and the local suspension {length scale} $L$ (defined by $\nabla_{\mathbf x}\phi\sim \phi/L$), such that there exists an intermediate {length scale} $\ell$ with
\begin{equation}
a_s\ll \ell \ll L.
\end{equation}
In other words, we should be able to find a volume around each particle containing many swimmers, and where the particle volume fraction does not change significantly. This condition is not satisfied in the presence of a step in the {concentration}, since, by definition, there is going to be an outermost particle beyond which the volume fraction drops discontinuously.

\begin{figure}[t]
 \hspace{-1cm}
\includegraphics[scale=0.25]{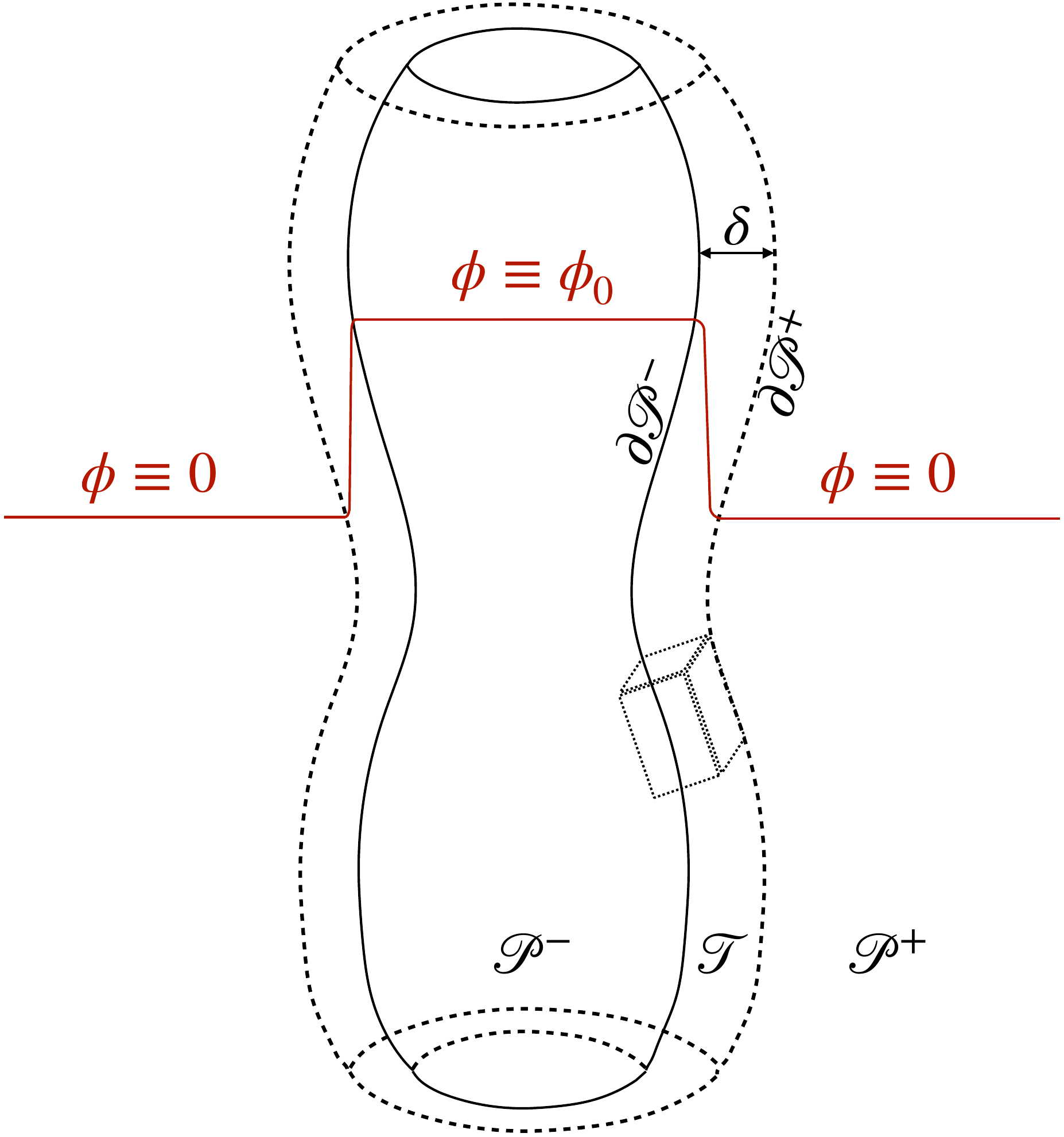}
\caption{Sketch of the smooth jet: the concentration $\phi$ varies smoothly between two constant values over a transition region $\mathcal T$ of typical width $\delta$. The inner and outer regions are denoted $\mathcal P^-$ and $\mathcal P^+$, with respective boundaries $\p\mathcal P^-$ and $\p\mathcal P^+$.}
\label{Fig: Smooth_Jet_Sketch}
\end{figure}

To address this issue, we interpret the step in the concentration as {approximating} a smoothly varying concentration field $\phi(\mathbf x,t)$, for which statistical homogeneity holds. At any point in time, the concentration is defined to be constant ($\phi\equiv\phi_0$) in the {jet} volume (which we henceforth denote $\mathcal P^-$,  {similarly to Section~\ref{Constant-Concentration Solution}}), and sharply decaying to zero on some {length scale} $\delta$ outside of $\mathcal P^-$. We let $\mathcal T$ be the transition region where the concentration changes rapidly, and $\mathcal P^+$ be the region of space where the particle concentration is zero. We denote the boundaries of these two regions by $\p \mathcal P^-$, $\p \mathcal P^+$. These are assumed to be close to each other in the sense that there exists a continuous one-to-one mapping $\chi: \p P^-\to \p P^+$ such that $\lVert \mathbf x-\chi(\mathbf x)\rVert\leq \delta$ for each $\mathbf x\in \p P^-$. A sketch of the jet is provided in Fig.~\ref{Fig: Smooth_Jet_Sketch}. Such a jet is within the continuum limit as long as the swimmers are sufficiently small, i.e.~as long as
\begin{equation}
a_s\ll \ell \ll \delta.
\end{equation}

Assuming that the swimming direction is clamped, with $\mathbf p\equiv \mathbf e_z$, the particle stress may then be coarse-grained into an active stress tensor
\begin{equation}
\Sigma_{ij}=-q\delta_{ij}+\mu\left(\frac{\p u_i}{\p x_j}+\frac{\p u_j}{\p x_i}\right) +\frac{S\phi}{V}p_ip_j.   
\end{equation}
For convenience, we denote the Newtonian part of the stress by $\sigma_{ij}$, with
\begin{equation}
\sigma_{ij}=-q\delta_{ij}+\mu\left(\frac{\p u_i}{\p x_j}+\frac{\p u_j}{\p x_i}\right).   
\end{equation}
The conditions of stress-balance, incompressibility, and swimmer conservation take the form
\begin{subequations}
\begin{align}
\p_j\Sigma_{ij}&=0,\label{Smooth Momentum Balance}\\
\p_iu_i&=0,\label{Smooth Incompressibility}\\
\frac{\p\phi}{\p t}+(\mathbf u+U_s\mathbf e_z)\cdot\nabla \phi&=0\label{Smooth swimmer conservation}
\end{align}
\end{subequations}
everywhere in space. We note that the momentum Eq.~\eqref{Smooth Momentum Balance} can be equivalently expressed as 
\begin{equation}
 -\nabla q+\mu\nabla^2\mathbf u+\frac{S}{V}\frac{\p \phi}{\p z}\mathbf e_z =\mathbf 0, \label{Smooth momentum equation expanded}
\end{equation}
where $V$ is the particle volume. If $L$ is a typical {length scale} of the system (such as the jet's wavelength), then the corresponding velocity scale is $U=SL\phi/\mu V$, obtained from matching the active and viscous term in Eq.~\eqref{Smooth momentum equation expanded}.

The idea is now that the flow should, at leading order, be independent of the exact decay of $\phi$ in the transition region $\mathcal T$. More explicitly, we want to be able to solve Eqs.~\eqref{Smooth Momentum Balance}, \eqref{Smooth Incompressibility} in each of $\mathcal P^-$, $\mathcal P^+$ and reconstruct the overall flow by means of suitable matching across $\mathcal T$. To derive the matching conditions, we note that, within $\mathcal T$,
\begin{subequations}
\begin{align}
|\p_i\p_ju_k|&\sim U/\delta L,\label{grad grad u scaling}\\
|\p_iu_j|&\sim U/L,\label{grad u scaling}\\
| u_i|&\sim U,\label{u scaling}
\end{align}
\end{subequations} 
for each $i$, $j$, $k$. Here, Eq.~\eqref{grad grad u scaling} comes from balancing the viscous and (large) active term of \eqref{Smooth momentum equation expanded}, while Eq.~\eqref{grad u scaling} and \eqref{u scaling} are then obtained by successively integrating Eq.~\eqref{grad grad u scaling} across $\mathcal T$. Intuitively, Eq.~\eqref{grad grad u scaling} through \eqref{u scaling} imply that the transition region $\mathcal T$ is characterised by rapidly-changing, although bounded, velocity gradients.

We then see, from Eq.~\eqref{grad u scaling}, that for each $\mathbf x\in \p \mathcal P^-$, 
\begin{equation}
\lVert\mathbf u|_{\mathbf x}-\mathbf u|_{\chi(\mathbf x)}\rVert=\mathcal O(U\delta/L). \label{Velocity Variation Estimate}
\end{equation}
In other words, the velocity is approximately continuous across $\mathcal T$.

Secondly, we may define a \textit{pillbox} spanning $\mathcal T$, consisting of the volume enclosed between the area patches $\mathcal A\in \p\mathcal P^-$ and $\chi(\mathcal A)\in \p\mathcal P^+$. We assume that $\mathcal A$ and, therefore, $\chi(\mathcal A)$ have typical size $(\delta L)^{1/2}$. Integrating the momentum equation $\p_j\Sigma_{ij}=0$ over such a pillbox gives
\begin{align}
0&=\int \p_j\Sigma_{ij} \mathrm dV\nonumber\\
&=\int_{\mathcal A}\Sigma_{ij}n_j\mathrm dA+\int_{\chi(\mathcal A)}\Sigma_{ij}n_j\mathrm dA +\mathcal O(\mu \delta^{3/2} U/L^{1/2})\nonumber\\
&=\left.\Sigma_{ij}n_j\right|_{\mathbf x} \cdot\lVert\mathcal A\rVert+\left.\Sigma_{ij}n_j\right|_{\chi(\mathbf x)}\cdot\lVert\chi(\mathcal A)\rVert+\mathcal O(\mu \delta^{3/2} U/L^{1/2})\nonumber\\
&=\left.\Sigma_{ij}n_j\right|_{\mathbf x} \cdot\lVert\mathcal A\rVert+\left.\Sigma_{ij}n_j\right|_{\chi(\mathbf x)}\cdot\lVert\mathcal A\rVert+\mathcal O(\mu \delta^{3/2} U/L^{1/2}). 
\end{align}
This shows that 
\begin{align}
\lVert\left.\boldsymbol{\Sigma}\cdot \mathbf n\right\rvert_{\mathbf x}-\left.\boldsymbol{\Sigma}\cdot \mathbf n\right\rvert_{\chi(\mathbf x)}\rVert=\mathcal O(\mu \delta^{1/2} U/L^{3/2}) \label{Stress Variation Estimate}.
\end{align}
In other words, stresses are approximately continuous across $\mathcal T$. 

The leading-order flow with respect to the small parameter $(\delta/L)^{1/2}$ therefore obeys
\vspace{-1.5cm}
\begin{center}
\begin{minipage}[t][4cm][t]{0.25\textwidth}
\begin{subequations}
\begin{align}
&\mathcal P^-\nonumber\\\hline
&\p_j\Sigma^-_{ij}=0\\
&\p_i u_i^-=0
\end{align}
\end{subequations}    
\end{minipage}
\begin{minipage}[t][4cm][t]{0.25\textwidth}
\begin{subequations}
\begin{align}
&\mathcal P^+\nonumber\\\hline
&\p_j\Sigma^+_{ij}=0\\
&\p_i u_i^+=0
\end{align}
\end{subequations}    
\end{minipage}
\begin{minipage}[t][4cm][t]{0.4\textwidth}
\begin{subequations}
\begin{align}
&\p \mathcal P^-\nonumber\\\hline
&\Sigma^-_{ij}n_j|_{\mathbf x}=\Sigma^+_{ij}n_j|_{\chi(\mathbf x)}\label{Stress Balance chi}\\
&u_i^-|_{\mathbf x}=u_i^+|_{\chi(\mathbf x)}\label{Velocity Continuity chi}
\end{align}
\end{subequations}    
\end{minipage}
\end{center}
 The key step is now that, if the flows in $\mathcal P^+$ and $\mathcal P^-$ do indeed vary on the large {length scale} $L$, we may evaluate the boundary {conditions}~\eqref{Stress Balance chi} and \eqref{Velocity Continuity chi} at $\mathbf x$, rather than $\chi(\mathbf x)$. Doing so yields the correct velocity fields in $\mathcal P^-$, $\mathcal P^+$ up to $\mathcal O(\delta U/L)$, which can be neglected consistently with the previous approximation. Furthermore, because the velocity varies on a {length scale} $L$, the flow in the transition region $\mathcal T$ is also determined with the same error. Imposing the boundary conditions at $\mathbf x$ is allowed, provided we check \textit{a posteriori} that the flows vary on the correct {length scale}. The simplified equations now read
 \vspace{-1.5cm}
 \begin{center}
\begin{minipage}[t][4cm][t]{0.25\textwidth}
\begin{subequations}
\begin{align}
&\mathcal P^-\nonumber\\\hline
&\p_j\Sigma^-_{ij}=0\\
&\p_i u_i^-=0
\end{align}
\end{subequations}    
\end{minipage}
\begin{minipage}[t][4cm][t]{0.25\textwidth}
\begin{subequations}
\begin{align}
&\mathcal P^+\nonumber\\\hline
&\p_j\Sigma^+_{ij}=0\\
&\p_i u_i^+=0
\end{align}
\end{subequations}    
\end{minipage}
\begin{minipage}[t][4cm][t]{0.4\textwidth}
\begin{subequations}
\begin{align}
&\p \mathcal P^-\nonumber\\\hline
&\Sigma^-_{ij}n_j|_{\mathbf x}=\Sigma^+_{ij}n_j|_{\mathbf x}\\
&u_i^-|_{\mathbf x}=u_i^+|_{\mathbf x}
\end{align}
\end{subequations}    
\end{minipage}
\end{center}
 which are the same as Eqs.~\eqref{Bulk Jet Stress Balance} through \eqref{Table Boundary Continuity of Velocity} in the main text; these  equations uniquely determine the leading-order flow. 

We must now verify that the flow indeed varies on a {length scale} $L$. We note that, because the concentration is constant in $\mathcal P^-$, $\mathcal P^+$, the simplified equation reduce to the Stokes equations with a stress-jump boundary condition given by
\vspace{-1.5cm}
 \begin{center}
\begin{minipage}[t][4cm][t]{0.23\textwidth}
\begin{subequations}
\begin{align}
&\mathcal P^-\nonumber\\
\hline \nonumber& \\[-4ex]
&\p_j\sigma^-_{ij}=0\\
&\p_i u_i^-=0
\end{align}
\end{subequations}    
\end{minipage}
\begin{minipage}[t][4cm][t]{0.22\textwidth}
\begin{subequations}
\begin{align}
&\mathcal P^+\nonumber\\
\hline \nonumber& \\[-4ex]
&\p_j\sigma^+_{ij}=0\\
&\p_i u_i^+=0
\end{align}
\end{subequations}    
\end{minipage}
\begin{minipage}[t][4cm][t]{0.45\textwidth}
\begin{subequations}
\begin{align}
&\p \mathcal P^-\nonumber\\
\hline \nonumber& \\[-4ex]
&\sigma^+_{ij}n_j|_{\mathbf x}=\sigma^-_{ij}n_j|_{\mathbf x}+\frac{S\phi_0}{V} p_ip_jn_j\\
&u_i^-|_{\mathbf x}=u_i^+|_{\mathbf x}
\end{align}
\end{subequations}    
\end{minipage}
\end{center}
Such equations have a unique solution varying on the {length scale} $L$, which completes our check \textit{a posteriori}. We remark that, although Eqs.~\eqref{Bulk Jet Stress Balance} through \eqref{Table Boundary Continuity of Velocity} yield the correct leading-order flow everywhere, they do not give the correct velocity gradients or stresses in $\mathcal T$, as in our approximation all quantities vary on a {length scale} $L$, but gradient in $\mathcal T$ vary on a {length scale} $\delta$. This {is, however, inconsequential for} the interfacial dynamics, Eq.~\eqref{Smooth swimmer conservation}, which  is completely determined by the velocity field.

\subsection{Stresslet flow superposition}\label{Continuum equations alternative derivation}
 
An alternative, flow-focused derivation of the continuum equations may be found directly from an application of the  Saintillan 
\& Shelley formalism~\cite{saintillan2008instabilitiesA, saintillan2008instabilitiesB, saintillan2013active}. As a first step, we note that the flow created by a single dipole located at $\mathbf x_0$ and oriented along $\mathbf e_z$ satisfies
\begin{equation}
\nabla q^d(\mathbf x;\mathbf x_0)-\mu\nabla^2\mathbf u^d(\mathbf x;\mathbf x_0)=-S\mathbf e_z\mathbf e_z\cdot\nabla_0\delta^{(3)}(\mathbf x-\mathbf x_0).    
\end{equation}
The velocity field $\mathbf u$ in a dilute suspension of such dipoles with local volume fraction $\phi(\mathbf x)$ can then be obtained by superposing the $\mathbf u^d$, $q^d$ for each value of $\mathbf x_0$. Therefore, the bulk flow satisfies
\begin{equation}
\nabla q(\mathbf x)-\mu\nabla^2\mathbf u(\mathbf x)=-\frac{S}{V}\mathbf e_z\mathbf e_z\cdot\int_{\mathbb R^3} \phi(\mathbf x_0)\nabla_0\delta^{(3)}(\mathbf x-\mathbf x_0)\mathrm d^3\mathbf x_0.    
\end{equation}

We want to analyse the case of a piecewise smooth concentration, i.e.~a field $\phi(\mathbf x)$ which is smooth in two disjoint open regions $\mathcal P^-$ and $\mathcal P^+$ (with $\mathbb R^3=\mathcal P^-\cup\mathcal P^+$) but discontinuous at the interface $\p\mathcal P^-=\p\mathcal P^+\eqqcolon \p \mathcal P$.
Splitting up the integral over each region and applying the divergence theorem, we find (for $\mathbf x\not\in \p\mathcal P$)
\begin{align}
\nabla q(\mathbf x)-\mu\nabla^2\mathbf u(\mathbf x)&=-\frac{S}{V}\int_{\mathcal P^-} \phi(\mathbf x_0)\mathbf e_z\mathbf e_z\cdot\nabla_0\delta^{(3)}(\mathbf x-\mathbf x_0)\mathrm d^3\mathbf x_0\nonumber\\&-\frac{S}{V}\int_{\mathcal P^+} \phi(\mathbf x_0)\mathbf e_z\mathbf e_z\cdot\nabla_0\delta^{(3)}(\mathbf x-\mathbf x_0)\mathrm d^3\mathbf x_0\nonumber\\
&=\frac{S}{V}\int_{\mathbb R^3}\delta^{(3)}(\mathbf x-\mathbf x_0)\mathbf e_z\mathbf e_z\cdot\nabla_0 \phi(\mathbf x_0)\mathrm d^3\mathbf x_0\nonumber\\&+\frac{S}{V}\oint_{\p \mathcal P}\delta^{(3)}(\mathbf x-\mathbf x_0)[\phi]_-^+\mathbf e_z(\mathbf e_z\cdot\mathbf n)\mathrm dA\nonumber\\
&=\frac{S}{V}\mathbf e_z\mathbf e_z\cdot\nabla\phi(\mathbf x)+\frac{S}{V}\oint_{\p \mathcal P}\delta^{(3)}(\mathbf x-\mathbf x_0)[\phi]_-^+\mathbf e_z(\mathbf e_z\cdot\mathbf n)\mathrm dA.\label{RHS Saintillan}
\end{align}
The final term in Eq.~\eqref{RHS Saintillan} cannot be evaluated directly, since it corresponds to the integral of the three-dimensional $\delta$ function over a surface, for which the sampling property does not apply. Eq.~\eqref{RHS Saintillan} can, however, be interpreted as a forced Stokes equation, driven by a body force $S\mathbf e_z\mathbf e_z\cdot\nabla \phi(\mathbf x)/V$ as well as a ``Stokeslet membrane'' at the interface $\p\mathcal P$. The body force stems from imbalances in active stresses caused by an uneven concentration, while the boundary term captures potential force imbalances at the interface. The strength density of the Stokeslets can be read off to be $S[\phi]_-^+\mathbf e_z(\mathbf e_z\cdot\mathbf n)/V$. We may now solve for the total flow $\mathbf u$ by superposing the flow induced by the Stokeslet membrane with the flow driven by the body force, found
from the integral formulation of the Stokes equations~\cite{happel1983low, kim2013microhydrodynamics}:
\begin{equation}
\mathbf u(\mathbf x)=\frac{S}{V}\int_{\mathbb R^3}[\mathbf J(\mathbf x-\mathbf y)\cdot\mathbf e_z][\mathbf e_z\cdot\nabla\phi(\mathbf y)]\mathrm d^3\mathbf y+\frac{S}{V}\oint_{\p \mathcal P}\mathbf J(\mathbf x-\mathbf y)\cdot\mathbf e_z(\mathbf e_z\cdot\mathbf n)[\phi]_-^+\mathrm dA.   
\end{equation}
This is the same as the solution to the forced Stokes equation with stress jump conditions~\cite{happel1983low, kim2013microhydrodynamics}
\begin{subeqnarray}
\nabla q-\mu\nabla^2 \mathbf u&=&\frac{S}{V}\mathbf e_z\mathbf e_z\cdot\nabla\phi(\mathbf x), \qquad\mathbf x\not\in \p\mathcal P\\
\left[-q\mathbf I+\mu(\nabla\mathbf u+\nabla\mathbf u^{\text T})\right]^+_-\cdot \mathbf n&=&-\frac{S}{V}[\phi]_-^+\mathbf e_z(\mathbf e_z\cdot\mathbf n), \qquad \mathbf x\in\p\mathcal P,
\end{subeqnarray}
which are precisely the forced Stokes equations that we solve in the main text. When the bulk concentration is piecewise constant, the body force in Eq.~\eqref{RHS Saintillan} disappears, as {all stresslets} cancel in the bulk. The only remaining term is the Stokeslet membrane, which arises because swimmers on one side of the interface push or pull harder than on the other.

  \section{Growth rate of   {a} three-dimensional jet}\label{Appendix A}

In this section, we delve into the calculation of the $\mathcal O(\varepsilon)$ flow determining the stability of a three-dimensional jet (Eq.~\ref{Order Epsilon Stokes Equation} through \ref{Order Epsilon Boundary Evolution}). We denote the flows outside and inside the jet as $\mathbf u_1^{\pm}=u_1^{\pm}\mathbf e_r+v_1^{\pm}\mathbf e_{\theta}+w_1^{\pm}\mathbf e_z$ (respectively), and the corresponding pressures as $q_1^{\pm}$. In order to solve for the flow, we take the Fourier transform  {of Eqs.~\eqref{Order Epsilon Stokes Equation}--\eqref{Order Epsilon Boundary Evolution}} in the $z$ and $\theta$ directions, thereby assuming
\begin{subeqnarray}
u^{\pm}_1&=&U^{\pm}_1(r){e}^{\mathrm ikz+\mathrm in\theta+st},\\
 v^{\pm}_1&=&-\mathrm iV^{\pm}_1(r)e^{\mathrm ikz+\mathrm in\theta+st},\\ w^{\pm}_1&=&W^{\pm}_1(r)e^{\mathrm ikz+\mathrm in\theta+st},\\
 q^{\pm}_1&=&Q^{\pm}_1(r)e^{\mathrm ikz+\mathrm in\theta+st}.
\end{subeqnarray}

 The calculation then proceeds as follows: we first determine the pressures $q_1^{\pm}$, and use these to derive a system of coupled ODEs  for the velocity fields, which are determined up to multiplicative constants. Imposing continuity of stress, 
 Eq.~\eqref{Order Epsilon Boundary Stress}, and  velocity,  Eq.~\eqref{Order Epsilon Continuity of Velocity}, at the jet boundary, as well as the kinematic boundary condition,  Eq.~\eqref{Order Epsilon Boundary Evolution}, results in a homogeneous linear system for the flow parameters. Such a system only admits non-zero solutions for the flow when the growth rate $s$ takes a particular value. This allows us to fully determine the growth rate as a function of the axial and azimuthal wavenumbers $k$ and $n$. 
 
In all that follows, we make use of the classical Bessel function identities \cite{arfken2011mathematical, watson1922treatise} 
\vspace{-0.5cm}
\begin{center}
\begin{minipage}{0.45\textwidth}
\begin{subequations}
\begin{align}
&K_n'-\frac{n}{z}K_n=-K_{n+1},\label{Bessel K Property 1}\\
&K_n'+\frac{n}{z}K_n=-K_{n-1},\label{Bessel K Property 2}\\
&K_{n-2}+\frac{2}{z}(n-1)K_{n-1}=K_n,\label{Bessel K Property 3}
\end{align} 
\end{subequations}
\end{minipage}
\begin{minipage}{0.45\textwidth}
\begin{subequations}
\begin{align}
&I_n'-\frac{n}{z}I_n=I_{n+1},\label{Bessel I Property 1}\\
&I_n'+\frac{n}{z}I_n=I_{n-1},\label{Bessel I Property 2}\\
&I_{n-2}-\frac{2}{z}(n-1)I_{n-1}=I_n.\label{Bessel I Property 3}
\end{align}
\end{subequations}
\end{minipage}
\end{center}
In order to include the axisymmetric case $n=0$ in the analysis, we define $I_{-n}(z)\coloneqq I_n(z)$, $K_{-n}(z) \coloneqq K_n(z)$ for $n>0$, which still obey the above properties.

Because the pressures must satisfy Laplace's equation $\nabla^2 q^{\pm}_1=0$, we  immediately conclude that
\vspace{-1cm}
\begin{center}
\begin{minipage}{0.4\textwidth}
\begin{align*}
 r^2\frac{\mathrm d^2Q_1^{\pm}}{\mathrm dr^2}+r\frac{\mathrm dQ_1^{\pm}}{\mathrm dr}-(n^2+k^2r^2)Q^{\pm}_1=0  
\end{align*}
\end{minipage}
\begin{minipage}{0.1\textwidth}
 \begin{align*}
  \hspace{1cm}\Rightarrow   
 \end{align*}   
\end{minipage}
\begin{minipage}{0.4\textwidth}
\begin{subequations}
    \begin{align}
      &Q_1^+=A\mu K_n(kr) \label{Outer Pressure Field},\\
      &Q_1^-=D\mu I_n(kr)\label{Inner Pressure Field},
    \end{align}
\end{subequations}
\end{minipage}
\end{center}
where we imposed the outer pressure field,   Eq.~\eqref{Outer Pressure Field}, to decay at $r\to\infty$ and the inner pressure field,  
Eq.~\eqref{Inner Pressure Field}, to be regular at $r=0$. 

\subsection{Outer velocity field}
We may now use  Eqs.~\eqref{Outer Pressure Field} and \eqref{Inner Pressure Field} to solve for the velocities. We start by determining the outer velocity fields, which are found by taking the $r$ and $\theta$ component of the Stokes equation, Eq.~\eqref{Order Epsilon Stokes Equation Outer}, with the pressure field from   Eq.~\eqref{Outer Pressure Field}:
\begin{subequations}
\begin{align}
& \frac{\mathrm d^2U_1^{+}}{\mathrm dr^2}+\frac{1}{r}\frac{\mathrm dU_1^{+}}{\mathrm dr}-\left(k^2+\frac{n^2+1}{r^2}\right)U^{+}_1-\frac{2n}{r^2}V_1^+=Ak K_n'(kr),\label{Outer Radial n}\\
&\frac{\mathrm d^2V_1^{+}}{\mathrm dr^2}+\frac{1}{r}\frac{\mathrm dV_1^{+}}{\mathrm dr}-\left(k^2+\frac{n^2+1}{r^2}\right)V_1^+-\frac{2n}{r^2}U_1^+=-\frac{An}{r}K_n(kr).\label{Outer Azimuthal n}
\end{align}
\end{subequations}
Adding and subtracting  
Eqs.~(\ref{Outer Radial n}), (\ref{Outer Azimuthal n}), and using properties \eqref{Bessel K Property 1} through \eqref{Bessel K Property 3}, we obtain the following uncoupled equations for $U_1^++V^+_1$ and $U_1^+-V^+_1$:
\begin{subequations}
\begin{align}
&\left[\frac{\mathrm d^2}{\mathrm dr^2}+\frac{1}{r}\frac{\mathrm d}{\mathrm dr}-\frac{(n+1)^2}{r^2}-k^2\right](U_1^++V_1^+)=AkK_n'(kr)-\frac{An}{r}K_n(kr)=-Ak K_{n+1}(kr),\label{Equation for U_1+V_1 plus}\\
&\left[\frac{\mathrm d^2}{\mathrm dr^2}+\frac{1}{r}\frac{\mathrm d}{\mathrm dr}-\frac{(n-1)^2}{r^2}-k^2\right](U_1^+-V_1^+)=AkK_n'(kr)+\frac{An}{r}K_n(kr)=-AkK_{n-1}(kr).\label{Equation for U_1-V_1 plus}
\end{align}
\end{subequations}
We may solve  Eqs.~\eqref{Equation for U_1+V_1 plus}, \eqref{Equation for U_1-V_1 plus}  exactly by exploiting properties \eqref{Bessel K Property 1} through \eqref{Bessel K Property 3}, giving
\vspace{-1cm}
\begin{center}
\begin{minipage}{0.1\textwidth}
\begin{subequations}
    \begin{align*}
   &U_1^++V_1^+=\frac{1}{2}ArK_n+BK_{n+1},\\
   &U_1^+-V_1^+=\frac{1}{2}ArK_n+CK_{n-1},
    \end{align*}
\end{subequations}
\end{minipage}
\begin{minipage}{0.1\textwidth}
 \begin{align*}
   \Rightarrow  
 \end{align*}   
\end{minipage}
\begin{minipage}{0.5\textwidth}
\begin{subequations}
    \begin{align}
   &U_1^+=\frac{1}{2}\left[ArK_n+BK_{n+1}+CK_{n-1}\right],\label{U_1^+ Expression}\\
   &V_1^+=\frac{1}{2}[BK_{n+1}-CK_{n-1}].\label{V_1^+ Expression}
    \end{align}
\end{subequations}
\end{minipage}
\end{center}
The axial flow is found from the incompressibility condition,  Eq.~\eqref{Order Epsilon Incompressibility Outer}, to be 
\begin{align}
W_1^+&=\frac{\mathrm i}{k}\left(\frac{\p U_1^+}{\p r}+\frac{U_1^+}{r}+n\frac{V_1^+}{r}\right)=\frac{\mathrm i}{2k}\left\{\left[(2-n)A-Bk-Ck\right]K_n-Akr K_{n-1}\right\}. \label{W_1^+ Expression}
\end{align}

\subsection{Inner velocity field}
An analogous procedure allows us to determine the inner velocity fields. The $r$ and $\theta$ components of the Stokes equation,  Eq.~\eqref{Order Epsilon Stokes Equation}, with pressure field from 
Eq.~\eqref{Inner Pressure Field}, are
\begin{subequations}
\begin{align}
& \frac{\mathrm d^2U_1^{-}}{\mathrm dr^2}+\frac{1}{r}\frac{\mathrm dU_1^{-}}{\mathrm dr}-\left(k^2+\frac{n^2+1}{r^2}\right)U^{-}_1-\frac{2n}{r^2}V_1^+=Dk I_n'(kr),\label{Inner Radial n}\\
&\frac{\mathrm d^2V_1^{-}}{\mathrm dr^2}+\frac{1}{r}\frac{\mathrm dV_1^{-}}{\mathrm dr}-\left(k^2+\frac{n^2+1}{r^2}\right)V_1^--\frac{2n}{r^2}U_1^-=-\frac{Dn}{r}I_n(kr).\label{Inner Azimuthal n}
\end{align}
\end{subequations}
We may decouple  Eqs.~\eqref{Inner Radial n}, \eqref{Inner Azimuthal n} by adding and subtracting them, and using properties \eqref{Bessel I Property 1} through \eqref{Bessel I Property 3}:
\begin{subequations}
\begin{align}
&\left[\frac{\mathrm d^2}{\mathrm dr^2}+\frac{1}{r}\frac{\mathrm d}{\mathrm dr}-\frac{(n+1)^2}{r^2}-k^2\right](U_1^-+V_1^-)=DkI_n'(kr)-\frac{Dn}{r}I_n(kr)=Dk I_{n+1}(kr),\label{Equation for U_1+V_1 minus}\\
&\left[\frac{\mathrm d^2}{\mathrm dr^2}+\frac{1}{r}\frac{\mathrm d}{\mathrm dr}-\frac{(n-1)^2}{r^2}-k^2\right](U_1^--V_1^-)=DkI_n'(kr)+\frac{Dn}{r}I_n(kr)=DkI_{n-1}(kr).\label{Equation for U_1-V_1 minus}
\end{align}
\end{subequations}
Once again,  Eqs.~\eqref{Equation for U_1+V_1 minus}, \eqref{Equation for U_1-V_1 minus} may be solved exactly tby exploiting properties \eqref{Bessel I Property 1} through \eqref{Bessel I Property 3}, giving
\vspace{-0.5cm}
\begin{center}
\begin{minipage}{0.1\textwidth}
\begin{subequations}
    \begin{align*}
   &U_1^-+V_1^-=\frac{1}{2}DrI_n+FI_{n+1},\\
   &U_1^--V_1^-=\frac{1}{2}DrI_n+GI_{n-1},
    \end{align*}
\end{subequations}
\end{minipage}
\begin{minipage}{0.1\textwidth}
 \begin{align*}
   \Rightarrow  
 \end{align*}   
\end{minipage}
\begin{minipage}{0.5\textwidth}
\begin{subequations}
    \begin{align}
   &U_1^-=\frac{1}{2}\left[DrI_n+FI_{n+1}+GI_{n-1}\right],\label{U_1^- Expression}\\
   &V_1^-=\frac{1}{2}\left[FI_{n+1}-GI_{n-1}\right].\label{V_1^- Expression}
    \end{align}
\end{subequations}
\end{minipage}
\end{center}
As before, the inner axial velocity follows from the incompressibility condition,  Eq.~\eqref{Order Epsilon Incompressibility Inner}, as
\begin{equation}
W_1^-=\frac{\mathrm i}{k}\left(\frac{\p U_1^-}{\p r}+\frac{U_1^-}{r}+n\frac{V_1^-}{r}\right)=\frac{\mathrm i}{2k}\left\{[(2-n)D+Fk+Gk]I_n(kr)+DkrI_{n-1}(kr)\right\}.\label{W_1^- Expression}
\end{equation}

\subsection{Determining the growth rate}
Having determined the velocity fields,   Eqs.~\eqref{U_1^+ Expression}, \eqref{V_1^+ Expression}, \eqref{W_1^+ Expression}, \eqref{U_1^- Expression}, \eqref{V_1^- Expression}, \eqref{W_1^- Expression}, and the pressure fields, Eqs~\eqref{Outer Pressure Field}, \eqref{Inner Pressure Field}, we next impose continuity of stress at the jet boundary,   Eq.~\eqref{Order Epsilon Boundary Stress}, and of velocity,  Eq.~\eqref{Order Epsilon Continuity of Velocity}, as well as the kinematic boundary condition,  Eq.~\eqref{Order Epsilon Boundary Evolution}. Doing so yields a homogeneous linear system (too cumbersome to write down explicitly) for the previously identified flow parameters $A, B, C, D,F ,G$, which must have vanishing determinant in order to admit a nonzero solution for the flow. Imposing the determinant of the linear system to vanish sets the  the growth rate to be exactly
\begin{align}
  \Re(s)=\frac{S\phi_0}{2\mu V}\frac{I_n(\xi) \left[K_n(\xi) \left(2n^2 +\xi^2 \right)-n\xi K_{n+1} (\xi)\right]-I_{n+1} (\xi)\left[\xi^2 K_{n+1} (\xi)-n\xi K_n (\xi)\right]}{\xi I_n (\xi)K_{n+1} (\xi)+\xi K_n (\xi)I_{n+1} (\xi)}, \label{Most General Growth Rate Appendix} 
\end{align}
which is Eq.~\eqref{Most General Growth Rate} in the main text.

\section{Asymptotic behaviour of the growth rate of a three-dimensional jet}\label{Growth Rate Asymptotics}

In this Appendix, we aim to thoroughly investigate the stability of a three-dimensional jet given the explicit expression in Eq.~\eqref{Most General Growth Rate Appendix} for the growth rate of a given mode $(\xi,n)$. Specifically, we aim to determine:
\begin{enumerate}
    \item A classification of the modes $(\xi,n)$ that lead to a growing or a decaying boundary perturbation for pushers and pullers. This is of interest, as it sheds light on the physical mechanism behind the observed instability.
    \item The fastest-growing mode depending on whether the jet is made of pushers or pullers. Mathematically, this corresponds to the pair $(\xi,n)$ with the largest $\Re(s)$ {(Eq.~\ref{Most General Growth Rate Appendix})}  for $S<0$ and $S>0$. This is significant, as the fastest-growing mode is the one observed in experiments and numerical simulations.
\end{enumerate}

 \subsection{Growing and decaying modes for pushers and pullers}
Based on direct numerical evaluation {(Fig.~\ref{Fig: Dispersion_Relation_6_Modes})} of the growth rate from Eq.~\eqref{Most General Growth Rate Appendix}, it appears that, for $S>0$ (pullers), a given mode $(\xi,n)$ results in a decaying perturbation for $\xi \lesssim qn$, and a growing perturbation for $\xi \gtrsim qn$, for some constant $q$. Conversely, for $S<0$ (pushers), a given mode $(\xi,n)$ results in a decaying perturbation for $\xi \gtrsim qn$, and a growing perturbation for $\xi \lesssim qn$. Inspired by these preliminary results, we may carry out a more precise analysis to identify the exact value of the constant $q$. To this end, we expand  {for $n\to\infty$} 
\begin{subequations}
\begin{align}
&I_n(qn)\sim \frac{\sqrt{2}q^n e^{n(q^2+1)^{1/2} }}{ \sqrt{\pi }{{\left(q^2 +1\right)}}^{13/4} {{\left[1+(q^2+1)^{1/2}\right]}}^n}\times\nonumber\\
&\left[\frac{\left(q^2 +1\right)^3}{2 n^{1/2}}+\frac{3\left(q^2 +1\right)^{5/2}-5(q^2 +1)^{3/2}}{48n^{3/2}}+\frac{4-300q^2 +81q^4}{2304 n^{5/2}}+\mathcal O(n^{-7/2})\right],\label{In(qn) expansion}\\
&K_n(qn)\sim\frac{\sqrt{2\pi}{\mathrm{e}}^{-n(1+q^2)^{1/2} }{{\left[1+(1+q^2)^{1/2}\right]}}^n}{q^n {{\left(q^2 +1\right)}}^{13/4}}\times\nonumber\\
&\left[\frac{\left(q^2 +1\right)^3}{2n^{1/2}}+\frac{5{{\left(q^2 +1\right)}}^{3/2} -3{{\left(q^2 +1\right)}}^{5/2}}{48n^{3/2}} +\frac{4-300q^2+81q^4}{2304n^{5/2}}+\mathcal O(n^{-7/2})\right].\label{Kn(qn) expansion}
\end{align}
\end{subequations}

Similar, although more cumbersome {(and hence omitted)} asymptotic expressions for $I_{n+1}(nq)$, $K_{n+1}(nq)$ may be obtained by substituting $q\to nq/(n+1)$ in the expansions of $I_{n+1}[(n+1)q]$, $K_{n+1}[(n+1)q]$ (evaluated via Eqs.~\ref{In(qn) expansion}, \ref{Kn(qn) expansion}), and re-expanding. Such asymptotic expressions allow us to evaluate the numerator of Eq.~\eqref{Most General Growth Rate Appendix}, with $\xi=qn$, as
\begin{align}
&\frac{2\mu V}{S\phi_0}(I_nK_{n+1}+ K_nI_{n+1})qn{\Re(s)}=I_n \left[K_n \left(2n^2 +q^2n^2 \right)-qn^2 K_{n+1} \right]-I_{n+1}\left[q^2n^2 K_{n+1}-qn^2 K_n\right]\sim\nonumber\\
& \frac{q^2(q^2-2)}{4n(q^2+1)^{5/2}} +\mathcal O(n^{-2}),
\end{align}
(all Bessel functions implicitly have argument $qn$) showing that, at leading order, $q=2^{1/2}$.

\subsection{Fastest-growing modes for pushers and pullers}
Turning now our attention to the fastest-growing modes, direct numerical evaluation {(Fig.~\ref{Fig: Dispersion_Relation_6_Modes})} of the growth rate, Eq.~\eqref{Most General Growth Rate Appendix}, {suggests} that the fastest-growing mode $(\xi,n)$ for $S>0$ (pullers) has $n=0$, while the fastest-growing mode for $S<0$ (pushers) has $n=1$. This observation is in agreement with experiments, as $n=0$ corresponds to a  {pearling} mode, while $n=1$ corresponds to a helical {mode}. 

In order to {validate} these numerical findings, we evaluate the largest possible value of ${\Re(s)}$ as a function of $n$ (depending on whether $S>0$ or $S<0$) and show that, at least asymptotically, these values decay monotonically for large $n$. Mathematically, this means showing that the function
\begin{align}
\sigma(n)\coloneqq \max_{\xi} {\Re(s)}(\xi,n)    
\end{align}
decays monotonically as $n\to\infty$. Together with {the} numerical results in {Fig.~\ref{Fig: Dispersion_Relation_6_Modes}}, such a behaviour indicates that the absolute maximum value of ${\Re(s)}$ is attained for small values of $n$, specifically $n=0$ for pullers and $n=1$ for pushers.

{For each $n\geq 1$}, we numerically find two values $\xi_1^{(n)}$, $\xi_2^{(n)}$ such that $\mathrm d\Re(s)/\mathrm d\xi=0$ at $\xi=\xi_{1,2}^{(n)}$, with {$\xi_1^{(n)}\lesssim qn\lesssim\xi_2^{(n)}$}. Such values appear to grow linearly with $n$, {i.e.~$\xi_{1,2}^{(n)}\sim q_{1,2}n$}. Expanding the Bessel functions up to and including $\mathcal O(n^{-7/2})$ in Eqs.~\eqref{In(qn) expansion}, \eqref{Kn(qn) expansion}, the numerator of $\mathrm d\Re(s)/\mathrm d\xi$ is asymptotic to
{
\begin{align}
&\frac{2\mu V}{S\phi_0}(I_nK_{n+1}+ K_nI_{n+1})q_i^2n^2\Re'[s]\nonumber\\&=2n^3(q_i+q_i^3)(K_nI_{n+1}-I_nK_{n+1})+2q_i^2n^2(2q_in+2n+1)I_nK_n\sim \nonumber\\
& -\frac{q_i^2(q_i^4-10q_i^2+4)}{4n(q_i^2+1)^{7/2}}+\mathcal O(n^{-2}),
\end{align}
}
where the Bessel functions have argument $q_in$. The leading-order term vanishes for $q_1=(5-21^{1/2})^{1/2}\sim 0.65$ and $q_2=(5+21^{1/2})^{1/2}\sim 3.10$. This means that $\xi_1^{(n)}\sim 0.65 n$, $\xi_2^{(n)}\sim 3.10 n$. The corresponding values of $\sigma(n)$, evaluated from \eqref{In(qn) expansion}, \eqref{Kn(qn) expansion}, and the expansions for $I_{n+1}[(n+1)q_i]$, $K_{n+1}[(n+1)q_i]$, are 
\begin{align}
\sigma(n)\sim
\begin{dcases}
 0.02\times\frac{S\phi_0}{\mu V n} & S>0,\\
 -0.03\times\frac{S\phi_0}{\mu V n} & S<0 . 
\end{dcases}
\end{align}
Therefore, $\sigma(n)$ decays monotonically for $n$ large. Direct numerical evaluation of the first few values of $\sigma(n)$ in Fig.~\ref{Fig: Dispersion_Relation_6_Modes} then indicates that the most unstable modes are $n=0$ for pullers and $n=1$ for pushers.

\section{Growth rate of   {a} two-dimensional sheet} \label{Appendix C: Growth Rate of the Quasi two-dimensional Sheet}

In this Appendix, we delve into the linear stability analysis of a variant of the three-dimensional setup considered in Appendix~\ref{Appendix A}. We consider the stability of a 2D sheet of swimmers (Section~\ref{Sheet of Swimmers}), initially defined by $-a\leq x\leq a$, $-\infty<y,z<\infty$ in Cartesian coordinates. We perturb the sheet boundary so that, for $t>0$, the swimmers are located in $X^-(z,t)\leq x\leq X^+(z,t)$. We consider two forms of the perturbation:
\begin{enumerate}
    \item Sinuous/in phase: $X^+=a\left(1+\varepsilon e^{st+\mathrm ikz}\right)$, $X^-=a\left(-1+\varepsilon e^{st+\mathrm ikz}\right)$.
    \item Varicose/antiphase: $X^+=a\left(1+\varepsilon e^{st+\mathrm ikz}\right)$, $X^-=a\left(-1-\varepsilon e^{st+\mathrm ikz}\right)$.
\end{enumerate}

In each case, we seek to determine the dispersion relation $s=s(k)$, and hence establish whether it is pushers or pullers that destabilize the sheet. For simplicity, we only show the calculation for the sinuous perturbation, which we expect to be unstable for pushers by analogy with the three-dimensional case. 

By the symmetry of the setup, the $x$ component of the velocity must be even in $x$, while the $z$ component of the velocity must be odd in $x$. Therefore, we only need solve for the flows in the regions $X^-<x<X^+$ (denoted $``-''$) and $x>X^+$ (denoted $``+''$). Like in the three-dimensional case, we expand the velocities, pressures, and stresses in the small parameter $\varepsilon$:
\begin{subeqnarray}
\mathbf u^{\pm}&=&\mathbf u_0^{\pm}+\varepsilon\mathbf u_1^{\pm}+\mathcal O(\varepsilon^2),\\
q^{\pm}&=&q_0^{\pm}+\varepsilon q_1^{\pm}+\mathcal O(\varepsilon^2),\\
\mathbf \Sigma^{\pm}&=&\mathbf \Sigma^{\pm}_0+\varepsilon\mathbf \Sigma^{\pm}_1+\mathcal O(\varepsilon^2),
\end{subeqnarray}
and solve order by order. The base state, like in the three-dimensional case (Eq.~\ref{3d Case Base State}), corresponds to zero net flow, or
\begin{equation}
\mathbf u_0^{\pm}=\mathbf 0, \quad q_0^{\pm}=0,  \quad \mathbf{\Sigma}^+_0=\mathbf 0, \quad  \mathbf{\Sigma}^-_0=\frac{S\phi_0}{V}\mathbf e_z\mathbf e_z.   
\end{equation}
We now expand the flow equations \eqref{Plume Bulk Stokes} through \eqref{Boundary Advection Simplified} to first order, obtaining
\vspace{-1.5cm}
\begin{center}
\begin{minipage}[t]{0.25\textwidth}
\begin{subequations}
\begin{align}
&0\leq x\leq a\nonumber\\\hline
&\mu\nabla^2\mathbf u^-_1=\nabla q^-_1\label{2d Order Epsilon Stokes Equation} \\
& \nabla\cdot\mathbf u_1^-=0 \label{2d Order Epsilon Incompressibility Inner}
\end{align}
\end{subequations}    
\end{minipage}
\begin{minipage}[t]{0.25\textwidth}
\begin{subequations}
\begin{align}
&a< x<\infty\nonumber\\\hline
&\mu\nabla^2\mathbf u^+_1=\nabla q^+_1 \label{2d Order Epsilon Stokes Equation Outer}\\
&\nabla\cdot\mathbf u_1^+=0 \label{2d Order Epsilon Incompressibility Outer}\\
&\lim_{x\to\infty}\mathbf u^+_1=\mathbf 0 \label{2d Order Epsilon Decay at Infinity}
\end{align}
\end{subequations}    
\end{minipage}
\begin{minipage}[t]{0.45\textwidth}
\begin{subequations}
\begin{align}
&x=a\nonumber\\\hline
&(\mathbf \Sigma_1^+-\mathbf \Sigma_1^-)\cdot\mathbf e_x=-\frac{S\phi_0}{V}\eta_z\mathbf e_z\label{2d Order Epsilon Boundary Stress}\\
&\mathbf u_1^+=\mathbf u_1^-\label{2d Order Epsilon Continuity of Velocity}\\
& \frac{\p \eta}{\p t}+U_s\frac{\p\eta}{\p z}=\mathbf u^{\pm}_1\cdot \mathbf e_x \label{2d Order Epsilon Boundary Evolution}
\end{align}
\end{subequations}    
\end{minipage}
\end{center}

At this order, we take the Fourier transform of the velocity and pressure fields in the $z$ direction, thereby assuming
\begin{subequations}
    \begin{align}
   u_1^{\pm}=U_1^{\pm}(x)e^{\mathrm ikz+st},\\
   v_1^{\pm}=V_1^{\pm}(x)e^{\mathrm ikz+st},\\
   q_1^{\pm}=Q_1^{\pm}(x)e^{\mathrm ikz+st}. 
    \end{align}
\end{subequations}
We now solve for the flow in a way analogous to Appendix~\ref{Appendix A}. Because the pressure is harmonic, we have
\vspace{-0.5cm}
\begin{center}
\begin{minipage}{0.2\textwidth}
\begin{align*}
 \frac{\mathrm d^2Q_1^{\pm}}{\mathrm dx^2}-k^2Q_1^{\pm}=0  
\end{align*}
\end{minipage}
\begin{minipage}{0.1\textwidth}
 \begin{align*}
\Rightarrow   
 \end{align*}   
\end{minipage}
\begin{minipage}{0.4\textwidth}
\begin{subequations}
    \begin{align}
      &Q_1^+=2\mathrm i\mu Bk e^{-kx} \label{2d Outer Pressure Field},\\
      &Q_1^-=2\mathrm i\mu kD(e^{-kx}-e^{kx})\label{2d Inner Pressure Field}.
    \end{align}
\end{subequations}
\end{minipage}
\end{center}
 {We may now determine the velocity fields.}
\subsection{Outer velocity field}
We start by determining the outer velocity fields. The $x$ and $z$ components of the Stokes equation, with pressure field \eqref{2d Outer Pressure Field}, can be readily solved to give
\vspace{-0.5cm}
\begin{center}
\begin{minipage}{0.4\textwidth}
\begin{align*}
&\frac{\mathrm d^2 U_1^+}{\mathrm dx^2}-k^2U_1^+=-2\mathrm iBk^2e^{-kx},\\
&\frac{\mathrm d^2 V_1^+}{\mathrm dx^2}-k^2U_1^+=-2Bk^2e^{-kx},    
\end{align*}
\end{minipage}
\begin{minipage}{0.1\textwidth}
\begin{align*}
\Rightarrow
\end{align*}
\end{minipage} 
\begin{minipage}{0.4\textwidth}
\begin{subequations}
\begin{align}
&U_1^+=\mathrm i(A+Bkx)e^{-kx},\label{U_1+ 2d}\\
&V_1^+=(A-B+Bkx)e^{-kx}.\label{V_1+ 2d}  
\end{align}
\end{subequations}
\end{minipage}
\end{center}
\subsection{Inner velocity field}
For the inner velocity field, the $x$ and $z$ components of the Stokes equation, with pressure field \eqref{2d Inner Pressure Field}, can likewise be solved to obtain
\vspace{-0.5cm}
\begin{center}
\begin{minipage}{0.15\textwidth}
\begin{align*}
&\frac{\mathrm d^2 U_1^-}{\mathrm dx^2}-k^2U_1^-=-2\mathrm ik^2(e^{-kx}+e^{kx}),\\
&\frac{\mathrm d^2 V_1^-}{\mathrm dx^2}-k^2V_1^+=-2k^2(e^{-kx}-e^{kx}),
\end{align*}
\end{minipage}
\begin{minipage}{0.02\textwidth}
\begin{align*}
\Rightarrow
\end{align*}
\end{minipage} 
\begin{minipage}{0.65\textwidth}
\begin{subequations}
\begin{align}
&U_1^-=\mathrm i[(C+Dkx)e^{-kx}+(C-Dkx)e^{kx}],\label{U_1- 2d}\\
&V_1^-=(C-D+Dkx)e^{-kx}-(C-D-Dkx)e^{kx}. \label{V_1- 2d}
\end{align}
\end{subequations}
\end{minipage}
\end{center}
\subsection{Determining the growth rate}
As in Appendix~\ref{Appendix A}, having determined the full flow, Eqs.~\eqref{U_1+ 2d}, \eqref{V_1+ 2d}, \eqref{U_1- 2d}, \eqref{V_1- 2d}, and pressure fields, Eqs.~\eqref{2d Outer Pressure Field}, \eqref{2d Inner Pressure Field}, we impose continuity of stress,  Eq.~\eqref{2d Order Epsilon Boundary Stress}, and velocity, Eq.~\eqref{2d Order Epsilon Continuity of Velocity}, at the jet boundary, as well as the kinematic boundary condition, Eq.~\eqref{2d Order Epsilon Boundary Evolution}. Doing so yields a homogeneous linear system for the previously identified flow parameters $A, B, C, D$, which must have vanishing determinant in order to admit a nonzero solution for the flow. Imposing the determinant of the linear system to vanish leads to the following expression for the growth rate:
\begin{equation}
{\Re(s)}=-\frac{S\phi_0}{2\mu V}\xi e^{-2\xi},\label{2d pusher jet dispersion relation}
\end{equation}
where $\xi=ak$. The sheet is therefore unstable for pushers, for which $S<0$.

The varicose perturbation case $X^+=a\left(1+\varepsilon e^{st+\mathrm ikz}\right)$, $X^-=a\left(-1+\varepsilon e^{st+\mathrm ikz}\right)$ can be handled similarly, this time making the $x$ component of velocity odd in $x$ and the $z$ component of velocity even in $x$. A similar calculation as above then yields
\begin{equation}
{\Re(s)}=\frac{S\phi_0}{2\mu V}\xi e^{-2\xi}\label{2d puller jet dispersion relation}
\end{equation}
for the varicose perturbation. The sheet is therefore unstable for pullers, for which $S>0$.

\section{{Evolution of puller clusters}}\label{Exact Solutions for Evolution of Puller Cluster}  
In this Appendix, we derive exact solutions for the long-term evolution of 3D and 2D  clusters of pullers. We show that {such solutions exist} under the assumption that the cluster is initially a sphere (in 3D) or a  {circle} (in 2D). We then {specialise} our solutions to the limit of a thin cluster. This is the most experimentally significant regime, as the initial shape of the cluster is forgotten and the dynamics is self-similar, with the rate of spreading set purely by the active stresses and the initial cluster volume (area).

\subsection{Analytical solution for the spreading of a spherical cluster}\label{Analytical Solution for the Spreading of a Spherical Cluster}
Assuming that the initial cluster is spherical with constant volume fraction $\phi_0$ and that the swimming direction is fixed to be $\mathbf p\equiv\mathbf e_z$, the subsequent evolution in the limit of a sharp concentration drop outside the cluster is governed by Eq.~\eqref{Plume Bulk Stokes} through \eqref{Boundary Advection Simplified}. These equations may be solved exactly by noting that, at any point during the shape evolution, the active stress $S\phi_0(\mathbf e_z\cdot\mathbf n)\mathbf e_z/V$ on the  boundary  of the cluster is of the same form (up to an isotropic part) as the stress exerted by an axisymmetric straining flow, $\mathbf u=\mathbf E\cdot\mathbf x=E(r\mathbf e_r-2z\mathbf e_z)$, inside the cluster. Assuming that this is indeed the internal flow at each point in time, we expect the initially spherical cluster to progressively deform into an oblate ellipsoid with semi-major axis $R(t)$. 

In order to compute the shape evolution, we need to determine the time-dependent straining rate $E=E(t)$ and internal pressure $p_0=p_0(t)$ by imposing that the internal viscous and active stresses balance the external viscous stresses at the cluster boundary. To this end, we note that the external flow $\overline{\mathbf u}$ is driven by the boundary condition $\mathbf u=\overline{\mathbf u}$ on the boundary of the cluster. {The external flow therefore} corresponds to the opposite of the perturbation flow induced by a {(fictitious)} rigid ellipsoid in strain flow. Therefore, the internal active stresses $-p_0\mathbf n+S\phi_0(\mathbf e_z\cdot\mathbf n)\mathbf e_z$ must be equal to the force exerted by a rigid ellipsoid in strain flow, given by
\begin{align}
\overline{\boldsymbol{\sigma}}\cdot\mathbf n=
-\mu E\left[\frac{1}{3F(\chi)}+\frac{G(\chi)}{3F(\chi)}\right]\mathbf n+\frac{\mu E}{F(\chi)}(\mathbf e_z\cdot\mathbf n)\mathbf e_z, \label{external stresses ellipsoid}
\end{align}
for known functions $F$, $G$ of the ellipsoid's aspect ratio $\chi$ ($0<\chi<1$)~\cite{lauga2016stresslets}. Stress continuity allows to determine the constant internal pressure $p_0$ and the stain rate, given by
\begin{equation}
 E=\frac{S\phi_0}{\mu V}F(\chi).\label{E of xi 3d}
\end{equation}
We thus obtain that the semi-major axis satisfies the evolution equation
\begin{subequations}
\begin{align}
\frac{\mathrm dR}{\mathrm dt}&=\frac{S\phi_0}{\mu V}R F(\chi),\label{R dot general 3d case}\\
F(\chi)&=\frac{1}{4(1-\chi^2)^2}\left[-3\chi^2+\frac{\chi(1+2\chi^2)}{(1-\chi^2)^{1/2}}\cos^{-1}\chi\right],\label{F definition}\\
R(0)&=R_0.
\end{align}
\label{Evolution equation ellipsoid appendix}
\end{subequations}
{Since the cluster is deformed by incompressible flow, the volume has to be conserved. Expressing therefore $\chi=R_0^3/R^3$ ($R_0$ being the initial radius), and choosing units in which $R_0=1$, Eq.~\eqref{Evolution equation ellipsoid appendix} may be recast into Eqs.~\eqref{R ode 3d main}, \eqref{ICS 3d main} from the main text.}

\subsubsection*{Thin limit of a stretching 3D cluster}\label{Thin Limit of a Stretching 3D Cluster}

As the cluster becomes more and more elongated, the initial shape is forgotten and a self-similar solution is reached, dependent only on the initial cluster volume. The stresses on the cluster boundary may be found in this limit by noting that the unit normal and radially tangent vectors $\mathbf n$ and $\mathbf t$ on the top surface $z=\chi(R^2-r^2)^{1/2}$ take the form
\begin{align}
\mathbf n&=\frac{\chi r}{(R^2-r^2)^{1/2}}\mathbf e_r+\mathbf e_z+\mathcal O(\chi^2),\\
\mathbf t&=\mathbf e_r-\frac{\chi r}{(R^2-r^2)^{1/2}}\mathbf e_z+\mathcal O(\chi^2).
\end{align}
Considering the normal and tangential components of Eq.~\eqref{external stresses ellipsoid}, by virtue of the {asymptotic behaviours} $F\sim \pi\chi/8+\mathcal O(\chi^2)$, $G\sim 2+\mathcal O(\chi)$ for $\chi\to 0$~\cite{lauga2016stresslets}, we find
\begin{align}
\mathbf n\cdot\overline{\boldsymbol{\sigma}}\cdot\mathbf n&\sim \text{const},\label{Normal Traction on Disc}\\
\mathbf t \cdot\overline{\boldsymbol{\sigma}}\cdot\mathbf n&\sim- \frac{8}{\pi}\frac{\mu E r}{(R^2-r^2)^{1/2}} \label{Tangential Traction on Disc}
.\end{align}
These expressions correspond to the normal and tangential tractions exerted by a thin rigid disk in axisymmetric straining flow. To compute the long-term evolution of the cluster, we take $\chi\to 0$ in Eqs.~\eqref{R dot general 3d case}, \eqref{F definition}, obtaining
\begin{align}
\frac{\mathrm dR}{\mathrm dt}= \frac{S\phi_0}{\mu V}\frac{3V_0}{ 32R^2}. \label{R dot equation puller drop} 
\end{align}
Integrating Eq.~\eqref{R dot equation puller drop}, and substituting the result into Eq.~\eqref{E of xi 3d}, we find
\begin{subequations}
\begin{align}
 R&\sim \left(\frac{9}{32}\frac{S\phi_0 V_0}{\mu V}\right)^{1/3}t^{1/3},\label{R large time 3d}\\
 E&\sim \frac{1}{3}t^{-1}. \label{E large time 3d}
\end{align}
\end{subequations}
Note that the results in Eqs.~\eqref{R large time 3d}, \eqref{E large time 3d} are valid for $R\gg V_0^{1/3}$, corresponding to the limit
\begin{align}
t\gg \frac{\mu V}{S\phi_0},
\end{align}
i.e.~long after the initial instability.

\subsection{Analytical solution for  spreading of a  {$2$D circular} cluster}\label{Analytical Solution for the Spreading of a Cylindrical Cluster}

A similar calculation may be carried out for a 2D   cluster. Focusing on the deformation of a single 2D slice, similar arguments as in the 3D case show that the flow inside the cluster is a pure strain, of the form $\mathbf u=\mathbf E\cdot\mathbf x=E(x,-z)$. We therefore expect the slice to deform into an ellipse of semi-major axis $R(t)$. In order to determine the time-dependent rate of strain $E(t)$ and internal pressure $p_0(t)$, we impose stress continuity at the  boundary of the ellipse. Much like in 3D, the external flow is the negative of the perturbation flow experienced by a {(fictitious)} rigid ellipse in strain flow,. The corresponding external stress exerted by the ellipse is \cite{mulchrone2006motion}
\begin{align}
\overline{\boldsymbol{\sigma}}\cdot\mathbf n=-\frac{\mu(1+\chi)^2}{\chi}\mathbf E\cdot\mathbf n \label{Traction of Ellipse}
,\end{align}
where $0<\chi<1$ is the aspect ratio. Imposing that this stress balances the internal stresses $-p_0\mathbf n+S\phi_0(\mathbf e_z\cdot\mathbf n)\mathbf e_z/V$ determines the straining rate to be
\begin{equation}
E=\frac{S\phi_0}{2\mu V}\frac{\chi}{(1+\chi)^2}. \label{E of xi 2D}
\end{equation}
Expressing $\chi=R_0^2/R^2$ from area conservation (since the cluster is deformed by incompressible flow), we obtain the evolution equation in the 2D case:
\begin{align}
\frac{\mathrm d R}{\mathrm dt}&= \frac{S\phi_0}{\mu V}\frac{\chi R}{2(1+\chi)^2}\label{R evolution 2d Appendix},\\
R(0)&=R_0. \label{ICS 2d Appendix}
\end{align}
After choosing units in which $R_0=1$, this may be recast into Eqs.~\eqref{R ode 2d main}, \eqref{ICS 2d main} in the main text.

\subsubsection*{Thin limit of a stretching 2D cluster}\label{Thin Limit of a Stretching 2D Cluster}
As the cluster becomes more and more elongated, the initial shape is forgotten and a self-similar solution is reached, dependent only on the initial cluster volume. The stresses on the cluster boundary may be found in this limit by noting that the unit normal and tangent vectors $\mathbf n$ and $\mathbf t$ on the top surface $z=\chi(R^2-x^2)^{1/2}$ take the form
\begin{align}
\mathbf n&=\frac{\chi x}{(R^2-x^2)^{1/2}}\mathbf e_x+\mathbf e_z+\mathcal O(\chi^2),\\
\mathbf t&=\mathbf e_x-\frac{\chi x}{(R^2-x^2)^{1/2}}\mathbf e_z+\mathcal O(\chi^2).
\end{align}
Considering the normal and tangential components of Eq.~\eqref{Traction of Ellipse}, we obtain
\begin{align}
 \mathbf n \cdot \overline{\boldsymbol{\sigma}}\cdot\mathbf n&\sim \text{const},\\
 \mathbf t\cdot \overline{\boldsymbol{\sigma}}\cdot\mathbf n&\sim \frac{2\mu Ex}{(R^2-x^2)^{1/2}}.
\end{align}
These expressions correspond to the normal and tangential tractions exerted by a rigid rod in 2D strain flow. To compute the long-term evolution of the cluster, we take {$\chi\to 0$} in Eq.~\eqref{R evolution 2d Appendix}, obtaining 
\begin{align}
\frac{\mathrm dR}{\mathrm dt}=\frac{S\phi_0}{\mu V}\frac{V_0}{2\pi R}\label{R dot 2d}.
\end{align}
Integrating Eq.~\eqref{R dot 2d}, and substituting the result into Eq.~\eqref{E of xi 2D}, we find
\begin{align}
R&\sim \left(\frac{S\phi_0 V_0}{\pi \mu V}\right)^{1/2} t^{1/2},\label{R(t) 2d appendix}\\
E&\sim \frac{1}{2} t^{-1}.\label{E 2d asymptotics appendix}
\end{align}
Eqs.~\eqref{R(t) 2d appendix}, \eqref{E 2d asymptotics appendix} are  again valid in the limit 
\begin{align}
  t\gg \frac{\mu V}{S\phi_0},  
\end{align}
 i.e.~long after the initial instability.


\subsection{Interactions between clusters}
In our previous analysis (Appendices \ref{Analytical Solution for the Spreading of a Spherical Cluster}, \ref{Analytical Solution for the Spreading of a Cylindrical Cluster}), we neglected interactions between clusters. We may now use the previously determined external flows to recover leading-order cluster-cluster interactions, in the limit where clusters only interact in the far field (i.e.~the typical cluster size is much smaller than the typical separation length). In this case, because the  leading-order external flow is the same as the (negative) perturbation flow of a rigid oblate ellipsoid (3D) or an ellipse (2D) in a strain flow, clusters interact primarily as stresslets. We now determine   the strength of the stresslets in each of the 3D and the 2D cases.

\subsubsection{Flow outside a 3D cluster}\label{Flow Outside a 3D Cluster}
The flow outside a 3D cluster is, at leading order, the same as the negative of the perturbation flow created by an equivalent rigid oblate ellipsoid in the strain flow $\mathbf u=\mathbf  E\cdot\mathbf x$, where $\mathbf E$ is the same as for the internal strain flow. Such a perturbation flow may be approximated by its leading-order, corresponding to a stresslet flow~\cite{lauga2016stresslets}
\begin{equation}
\mathbf u^p=-\frac{3(\mathbf x\cdot \mathbf S\cdot\mathbf x)\mathbf x}{8\pi\mu r^5}+\mathcal O(r^{-3})   
\end{equation}
where, by Eq.~\eqref{S_ij} and the divergence theorem,
\begin{align}
S_{ij}&=\frac{S\phi_0}{V}\int_{\p B}\frac{1}{2}[x_j(\mathbf e_z)_i(\mathbf e_z)_kb_k+x_i(\mathbf e_z)_j(\mathbf e_z)_kb_k]-\frac{1}{3}[x_k(\mathbf e_z)_k(\mathbf e_z)_lb_l]\delta_{ij}\mathrm dA\nonumber\\
&=\frac{S\phi_0 V_0}{V}\left[(\mathbf e_z)_i(\mathbf e_z)_j-\frac{1}{3}\delta_{ij}\right].
\end{align}
Recognizing $\phi_0V_0/V$ as the total number of swimmers, $S_{ij}$ is thus simply the sum of all internal stresslets.

\subsubsection{Flow outside a 2D cluster}\label{Flow Outside a 2D Cluster}
Likewise, for the 2D cluster, the external flow is {at} leading order~\cite{OppenheimerVShape}
\begin{align}
\mathbf u=-\frac{(\mathbf x^{\perp}\cdot\mathbf S\cdot\mathbf x^{\perp})\mathbf x^{\perp}}{2\pi \mu \rho^4}+\mathcal O(\rho^{-2}),
\end{align}
where $\mathbf x^{\perp}$ is the component of $\mathbf x$ perpendicular to the axis of the cluster, $\rho=\lVert\mathbf x^{\perp}\rVert$, and the stresslet (per unit extent in the $y$ direction) is
\begin{align}
S_{ij}&=\frac{S\phi_0}{V}\int_{\p B}\frac{1}{2}[x_j(\mathbf e_z)_i(\mathbf e_z)_kb_k+x_i(\mathbf e_z)_j(\mathbf e_z)_kb_k]-\frac{1}{2}[x_k(\mathbf e_z)_k(\mathbf e_z)_lb_l]\delta^{(2)}_{ij}\mathrm ds\nonumber\\
&=\frac{S\phi_0 A_0}{V}\left[(\mathbf e_z)_i(\mathbf e_z)_j-\frac{1}{2}\delta^{(2)}_{ij}\right],
\end{align}
where $\delta^{(2)}_{ij}$ is the identity tensor in $\mathbb R^2$ and $s$ is arclength. Once again, $S\phi_0A_0/V$ is the number of swimmers per unit length, so that $S_{ij}$ is the sum of the internal stresslets.  {In light of the far-field flows in both $3$D and $2$D, we expect puller clusters to interact like a line of stresslets (see Section~\ref{Interactions Between Clusters}).}

 \subsection{Analytical solution for a tilted puller cluster}\label{Appendix: Analytical Solution for a Tilted Puller Cluster}
 \begin{figure}[t]
\centerline{\includegraphics[width=0.6\textwidth]{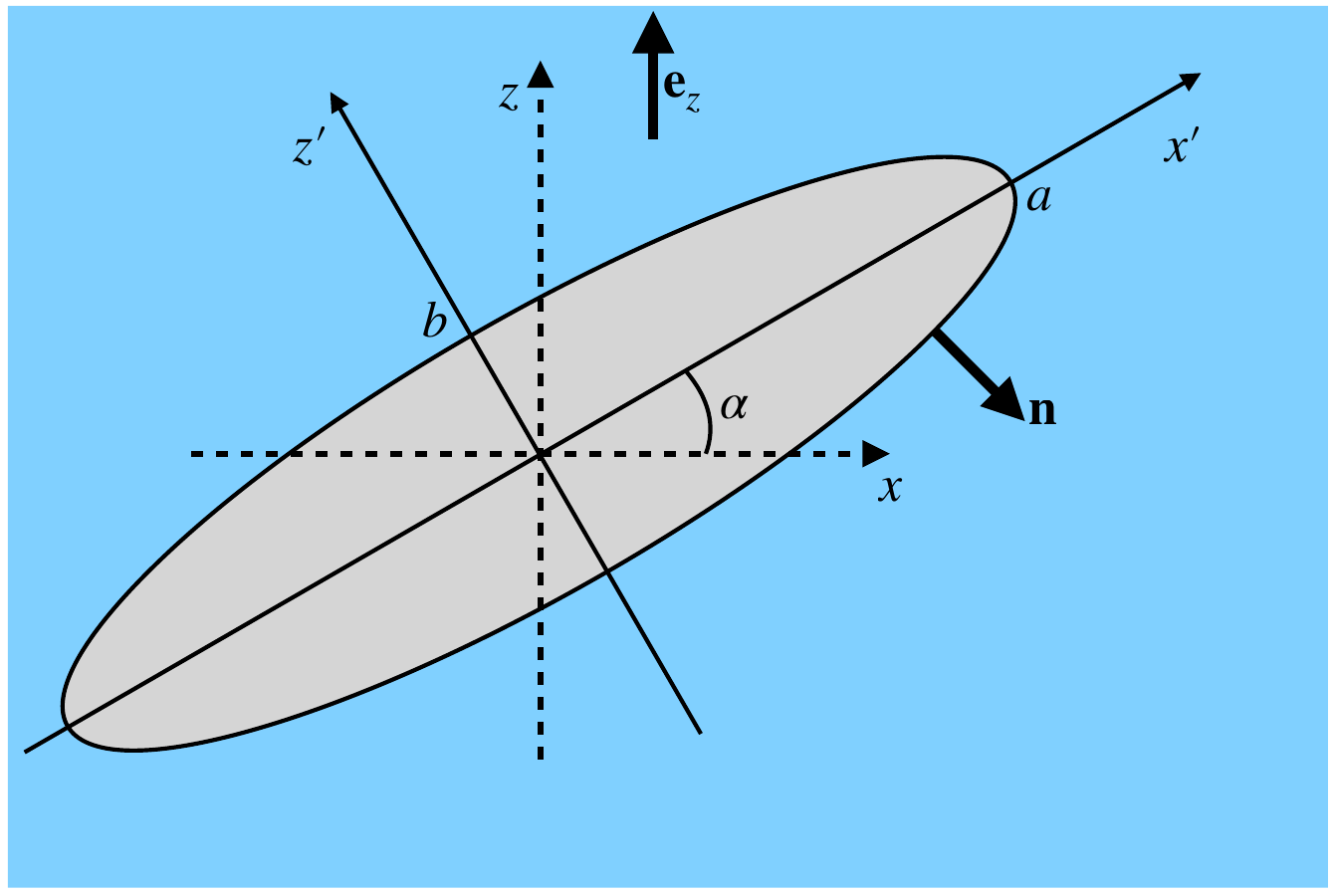}}
\caption{{Notation for dynamics of a tilted puller cluster. The deformed cluster is at every instant   an ellipse of {semi-major axis} $a(t)$ and minor semi-axis $b(t)$. The flow is described in the $x'$, $z'$ coordinate system, rotated by an angle $\alpha$ relative to the lab coordinates $x$, $z$. The outwards unit normal to the cluster is denoted  by $\mathbf n$ and the vertical direction (along which stresslets are aligned) by $\mathbf e_z$.}}
\label{Fig: Tilted_Cluster_Sketch}
\end{figure}

{In this section, we derive the analytical solution for the flow inside a tilted puller cluster, as summarised in Section~\ref{Sec: 2-dimensional Cluster Rotation and the ``V'' instability} of the main text.  We are going to assume (and later verify) that tilted clusters deform exactly as ellipses of {semi-major axis} $a(t)$ and semi-minor axis $b(t)$. We denote the angle with the $x$ axis by $\alpha(t)$, and work in the rotated coordinate systems $(x',z')$ (see Fig.~\eqref{Fig: Tilted_Cluster_Sketch} for a sketch with notation).}

{In order to match the viscous stress jump at the cluster's boundary, given explicitly by
\begin{align}
\frac{S\phi_0}{V}\mathbf e_z(\mathbf e_z\cdot\mathbf n)=\frac{S\phi_0}{V}\sin^2(\alpha) \mathbf n+\frac{S\phi_0}{V}\cos (2\alpha) \mathbf e_z'(\mathbf e_z'\cdot\mathbf n)+\frac{S\phi_0}{2V}\sin (2\alpha)(\mathbf e'_x\mathbf e_z'+\mathbf e_z'\mathbf e_x')\cdot\mathbf n,\label{Stress balance tilted cluster}
\end{align}
we assume that the flow insider the cluster takes the form 
\begin{align}
\mathbf u=E(x'\mathbf e'_x-z'\mathbf e_z')+\Omega_1\left(\frac{x'}{a^2}\mathbf e'_z-\frac{z'}{b^2}\mathbf e'_x\right)+\Omega_2(x'\mathbf e_z'-z'\mathbf e_x'),    \label{u^- ansatz tilted cluster appendix}
\end{align}
with constant pressure $p_0$. The first term (proportional to $E$) represents a straining flow aligned with the cluster. The second term (proportional to $\Omega_1$) can be viewed as a rotational (``treadmilling'') flow, corresponding to a stretched solid-body rotation matching the elliptical cluster shape. Finally, the third term (proportional to $\Omega_2$) corresponds to a rigid rotation of the whole cluster. Notice that, because the treadmilling term proportional to $\Omega_1$ satisfies the no-penetration condition (as $\mathbf n\propto x'\mathbf e'_x/a^2+z'\mathbf e_z'/b^2$), the flow $\mathbf u$ does indeed exactly deform  ellipses into ellipses.}

By continuity of velocity, the external flow can be written as
\begin{equation}
\overline{\mathbf u}=-\mathbf u'_{\text{str}}-\mathbf u'_{\text{tr}}+\mathbf u_{\text{rot}}. \label{u^+ ansatz}    
\end{equation}
In Eq.~\eqref{u^+ ansatz}, $\mathbf u'_{\text{str}}$ is the perturbation flow induced by a (fictitious) rigid ellipse $x'^2/a^2+y'^2/b^2=1$ immersed in the background straining flow $\mathbf u^{\infty}_{\text{str}}=E(x'\mathbf e'_x-z'\mathbf e_z')$, while $\mathbf u_{\text{tr}}'$ is the perturbation flow induced by the same rigid ellipse immersed in the treadmilling flow $\mathbf u^{\infty}_{\text{tr}}=\Omega_1(x'\mathbf e_z'/a^2-z'\mathbf e_x'/b^2)$. Finally, $\mathbf u_{\text{rot}}$ is the flow induced by the same rigid ellipse when rotating in a fluid at rest. By construction, the flow in Eq.~\eqref{u^+ ansatz} matches $\mathbf u$ on the ellipse's boundary and decays as $\lVert\mathbf x'\rVert\to\infty$, so it must indeed be the external flow. Denoting by $\boldsymbol{\sigma}_{\text{str}}$, $\boldsymbol{\sigma}_{\text{tr}}$, $\boldsymbol{\sigma}_{\text{rot}}$ the stresses produced by the corresponding flows outside the fictitious ellipse, continuity of stress then takes the form
\begin{equation}
-p_0\mathbf n+\boldsymbol{\sigma}_{\text{str}}\cdot\mathbf n+\boldsymbol{\sigma}_{\text{tr}}\cdot\mathbf n +\frac{S\phi_0}{V}\mathbf e_z(\mathbf e_z\cdot\mathbf n)=\boldsymbol{\sigma}_{\text{rot}}\cdot\mathbf n.
\end{equation}

Following the procedure laid out in Ref.~\cite{mulchrone2006motion}, these stresses are found to be
\begin{subeqnarray}
\boldsymbol{\sigma}_{\text{str}}\cdot\mathbf n&=&\frac{\mu E(a+b)^2}{ab}\mathbf n-\frac{2\mu E(a+b)^2}{ab}\mathbf e_z'(\mathbf e_z'\cdot\mathbf n),\\
\boldsymbol{\sigma}_{\text{tr}}\cdot\mathbf n&=&\frac{2\mu \Omega_1}{ab}(\mathbf e_z'\mathbf e_x'-\mathbf e'_x\mathbf e_z')\cdot\mathbf n-\frac{\mu(a^2-b^2)\Omega_1}{a^2b^2}(\mathbf e_x'\mathbf e_z'+\mathbf e_z'\mathbf e_x')\cdot\mathbf n,\\
\boldsymbol{\sigma}_{\text{rot}}\cdot\mathbf n&=&\frac{2\mu b\Omega_2}{a}\mathbf e'_x(\mathbf e'_z\cdot \mathbf n)-\frac{2\mu a\Omega_2}{b}\mathbf e'_z(\mathbf e'_x\cdot \mathbf n). 
\end{subeqnarray}
Stress-balance can thus be recast into
\begin{align}
 &-p_0^-\mathbf n+\frac{\mu E(a+b)^2}{ab}\mathbf n-\frac{2\mu E(a+b)^2}{ab}\mathbf e_z'(\mathbf e_z'\cdot\mathbf n) + \frac{2\mu \Omega_1}{ab}(\mathbf e_z'\mathbf e_x'-\mathbf e_x'\mathbf e_z')\cdot\mathbf n+\nonumber\\
 &-\frac{\mu(a^2-b^2)\Omega_1}{a^2b^2}(\mathbf e_x'\mathbf e_z'+\mathbf e_z'\mathbf e_x')\cdot\mathbf n +\frac{S\phi_0}{V}\sin^2(\alpha) \mathbf n+\frac{S\phi_0}{V}\cos (2\alpha) \mathbf e_z'(\mathbf e_z'\cdot\mathbf n)+\nonumber\\
 &+\frac{S\phi_0}{2V}\sin (2\alpha)(\mathbf e_x'\mathbf e_z'+\mathbf e_z'\mathbf e_x')\cdot\mathbf n=\frac{2\mu b\Omega_2}{a}\mathbf e'_x(\mathbf e_z'\cdot \mathbf n)-\frac{2\mu a\Omega_2}{b}\mathbf e_z'(\mathbf e_x'\cdot \mathbf n).
\end{align}
Matching corresponding terms, we obtain the following simultaneous linear equations for $p_0^-$, $E$, $\Omega_1$, and $\Omega_2$, in terms of the stresslet strength $S$
\begin{subeqnarray}
p_0^--\frac{\mu E(a+b)^2}{ab}&=&\frac{S\phi_0}{V}\sin^2(\alpha),\\
\frac{2\mu E(a+b)^2}{ab}&=&\frac{S\phi_0}{V}\cos(2\alpha),\\
 \frac{2\Omega_1 \mu}{ab}-\frac{\mu(a^2-b^2)\Omega_1}{a^2b^2}+\frac{2\mu a\Omega_2}{b}&=&-\frac{S\phi_0}{2V}\sin 2\alpha,\\
\frac{2\Omega_1 \mu}{ab}+\frac{\mu(a^2-b^2)\Omega_1}{a^2b^2}+\frac{2\mu b\Omega_2}{a}&=&\frac{S\phi_0}{2V}\sin 2\alpha.
\end{subeqnarray}
Solving, we recover the expressions given in Eqs.~\eqref{E Omega_1 Omega_2 tilted cluster} of the main text.

\subsection{Numerical simulations of stretching puller clusters}\label{Numerical Simulations of Stretching Puller Cluster}
\subsubsection{Evolution of a 3D cluster}\label{Evolution of a 3D Cluster}
In this section, we present the details of the numerical scheme used to solve for the evolution of a 3D cluster in Section~\ref{Stretching of Puller Clusters} of the main text. Because of the strong-alignment assumption ($\mathbf p\equiv \mathbf e_z$), the flow inside and outside the cluster obeys the Stokes equation, with the active stresses only appearing in the boundary condition $[\boldsymbol{\sigma}]^{\text{out}}_{\text{in}}\cdot \mathbf n=S\phi_0 \mathbf e_z(\mathbf e_z\cdot \mathbf n)/V$. In order to simulate the stretching of a three-dimensional  {cluster} of pullers (initially a sphere), we may therefore exploit the integral solution of the Stokes equations~\cite{happel1983low, kim2013microhydrodynamics} to write
\begin{align}
\dot{\mathbf y}=\frac{S\phi_0}{8\pi\mu V}\oint_{\p V(t)}\left[\frac{\mathbf e_z}{r}+\frac{(\mathbf x-\mathbf y)(\mathbf x-\mathbf y)}{r^3}\cdot \mathbf e_z\right](\mathrm d\mathbf S_{\mathbf x}\cdot\mathbf e_z),   \label{Interfacial evolution simulations 3d}
\end{align}
where $\mathbf y$ is a point on the boundary $\p V$, {$\mathrm d\mathbf S_{\mathbf x}$ is the inwards-pointing area vector}, and $r\coloneqq\lVert\mathbf x-\mathbf y\rVert$. In order to {integrate} Eq.~\eqref{Interfacial evolution simulations 3d} {numerically}, we notice that the solution must be axisymmetric, and thus convert Eq.~\eqref{Interfacial evolution simulations 3d} into a one-dimensional integral by performing the azimuthal integration analytically. To this end, we parametrise the axisymmetric surface at a given time point by $\{\mathbf x=R(z)\mathbf e_{\rho}+z\mathbf e_z:0\leq\theta\leq 2\pi,\ -L\leq z\leq L\}$ in cylindrical polars $(\rho,\theta,z)$. In order to determine the evolution of the half-height $L$ and the shape outline $R(z)$, we may limit our attention to the motion of points in the $xz$ plane (by rotational symmetry). The velocity at any other point on the surface may thereafter be obtained by a suitable rotation. Suppose therefore that, in Eq.~\eqref{Interfacial evolution simulations 3d}, $\mathbf y=R_0\mathbf e_x+z_0\mathbf e_z$ with $R_0\geq 0$, where $R_0\coloneqq R(z_0)$. We may evaluate the various geometrical quantities in Eq.~\eqref{Interfacial evolution simulations 3d} as
\begin{subeqnarray}
 \mathrm d\mathbf S_{\mathbf x}&=&-R\mathbf e_{\rho}+RR'\mathbf e_z,\\
r^2&=&R^2-2R_0R\cos\theta+R_0^2+(z-z_0)^2,\\
 (\mathbf x-\mathbf y)\cdot\mathbf e_z&=&z-z_0,\\
 (\mathbf x-\mathbf y)\cdot\mathbf e_x&=&R\cos\theta-R_0.
\end{subeqnarray}
By reflectional symmetry, the velocity at each point cannot have any azimuthal component. As for the remaining components, taking the components of Eq.~\eqref{Interfacial evolution simulations 3d} along $\mathbf e_x 
$ and $\mathbf e_z$ we obtain
\begin{subequations}
\begin{align}
 \mathbf e_x\cdot\dot{\mathbf y}&=\frac{S\phi_0}{8\pi \mu V}\int_{-L}^{L}\left\{\int_0^{2\pi}\frac{R\cos\theta-R_0}{r^3}\mathrm d\theta\right\} (z-z_0)RR'\mathrm dz,\label{e_x dot y'}\\
 \mathbf e_z\cdot\dot{\mathbf y}&=\frac{S\phi_0}{8\pi \mu V}\int_{-L}^{L}\left\{\int_0^{2\pi}\left[\frac{1}{r}+\frac{(z-z_0)^2}{r^3}\right]\mathrm d\theta\right\} RR'\mathrm dz. \label{e_z dot y'}
\end{align}
\end{subequations}
The integrals in curly brackets are functions of $z$ only, and may be evaluated in terms of elliptic integrals. To this end, we define
\begin{subequations}
\begin{align}
P(z)=[(R-R_0)^2+(z-z_0)^2]^{1/2},\\
Q(z)=[(R+R_0)^2+(z-z_0)^2]^{1/2},
\end{align}
\end{subequations}
and simplify Eqs.~\eqref{e_x dot y'} and \eqref{e_z dot y'} by means of the identities
\begin{subequations}
\begin{align}
&\int_0^{2\pi}\frac{\mathrm d\theta}{r}=\frac{4}{P} \mathcal K\left(\frac{P^2-Q^2}{P^2}\right),\label{Elliptic Identity 1}\\
&\int_0^{2\pi}\frac{\mathrm d\theta}{r^3}=\frac{4}{PQ^2}\mathcal E\left(\frac{P^2-Q^2}{P^2}\right),\label{Elliptic Identity 2}\\
&\int_0^{2\pi}\frac{\cos\theta}{r^3}\mathrm d\theta=\frac{8}{P^3-PQ^2}\mathcal K\left(\frac{P^2-Q^2}{P^2}\right)-\left(\frac{4}{PQ^2}+\frac{8}{P^3-PQ^2}\right) \mathcal E\left(\frac{P^2-Q^2}{P^2}\right).\label{Elliptic Identity 3}
\end{align}
\end{subequations}
In Eqs.~\eqref{Elliptic Identity 1} through \eqref{Elliptic Identity 3}, we defined
\begin{subequations}
\begin{align}
 \mathcal K(m)&=\int_0^{\pi/2}\frac{\mathrm d\phi}{(1-m\sin^2\phi)^{1/2}},\\
\mathcal E(m)&=\int_0^{\pi/2}(1-m\sin^2\phi)^{1/2} \mathrm d\phi,
\end{align}
\end{subequations}
with $m<1$, to be the complete elliptic integrals of the first and second kind, respectively. Substituting into Eqs.~\eqref{e_x dot y'}, \eqref{e_z dot y'}, we finally obtain
\begin{subequations}
\begin{align}
 \mathbf e_x\cdot\dot{\mathbf y}&=\frac{S\phi_0}{8\pi \mu V}\int_{-L}^{L}(z-z_0)\mathcal K_1RR'\mathrm dz,\label{e_x dot y' simplified}\\
 \mathbf e_z\cdot\dot{\mathbf y}&=\frac{S\phi_0}{8\pi \mu V}\int_{-L}^{L}\mathcal K_2RR'\mathrm dz,\label{e_z dot y' simplified}
\end{align}
\end{subequations}
where the integration kernels $\mathcal K_1$ and $\mathcal K_2$ take the forms
\begin{subequations}
\begin{align}
\mathcal K_1&=\frac{8R}{P^3-PQ^2}\mathcal K\left(\frac{P^2-Q^2}{P^2}\right)-\left(\frac{4R}{PQ^2}+\frac{8R}{P^3-PQ^2}\right)\mathcal E\left(\frac{P^2-Q^2}{P^2}\right)-\frac{4R_0}{PQ^2}\mathcal E\left(\frac{P^2-Q^2}{P^2}\right),\\
\mathcal K_2&=\frac{4}{P} \mathcal K\left(\frac{P^2-Q^2}{P^2}\right)+\frac{4(z-z_0)^2}{PQ^2}\mathcal E\left(\frac{P^2-Q^2}{P^2}\right).
\end{align}
\end{subequations}
The results in Eqs.~\eqref{e_x dot y' simplified}, \eqref{e_z dot y' simplified} are now amenable to numerical integration. To this end, we focus on the slice of the boundary in the $xz$ plane, with {$x\geq 0$}. Because the cluster is initially spherical, such a slice initially corresponds to a half-circle. We thereafter parametrise it via a cubic spline and evolve the location of the nodes via the explicit second-order Adams-Bashforth method. We use the numerical  {cluster} volume as a gauge for accuracy, ensuring that it never deviates by more that $1\%$ from the initial value, as expected from incompressibility.

\subsubsection{Evolution of a 2D cluster}\label{Evolution of a Quasi-2D Cluster}
In this section we present the details of the numerical scheme used to solve for the evolution of a 2D cluster in Section~\ref{Stretching of Puller Clusters} of the main text. Similarly to the 3D case, we may express the velocity of points on the boundary as
\begin{equation}
\dot{\mathbf y}=\frac{S\phi_0}{\mu V}\oint_{\Gamma(t)}[\mathbf J(\mathbf y-\mathbf x)\cdot \mathbf e_z](\mathbf n\cdot\mathbf e_z)\mathrm ds   \label{Interfacial evolution simulations},
\end{equation}
where $\Gamma$ is the corresponding horizontal slice of the cylinder, $\mathbf n$ is the unit normal pointing into the cluster, and $\mathbf J$ is the two-dimensional Green's function given by
\begin{equation}
 \mathbf J(\mathbf r)=\frac{1}{4\pi}\left[-\mathbf I \log \lVert\mathbf r\rVert+\frac{\mathbf r \mathbf r}{\lVert\mathbf r\rVert^2}\right] \label{Oseen 2d}.
\end{equation}
We numerically integrate Eq.~\eqref{Interfacial evolution simulations} by discretising the interface as a collection of $N$ points, initially uniformly spaced around the unit circle, and stepping forward with the explicit second-order Adams-Bashforth method. The unit normal and arclength element at every point are evaluated with symmetric differences, while the integrable singularity in the kernel \eqref{Oseen 2d} is regularised by approximating the Green's function with~\cite{cortez2001method}
\begin{align}
\mathbf J_{\varepsilon}(\mathbf r)&=\frac{1}{4\pi}\left[\ln\left(\sqrt{\lVert\mathbf r\rVert^2+\varepsilon^2}+\varepsilon\right)-\frac{\varepsilon(\sqrt{\lVert\mathbf r\rVert^2+\varepsilon^2}+2\varepsilon)}{(\sqrt{\lVert\mathbf r\rVert^2+\varepsilon^2}+\varepsilon)\sqrt{\lVert\mathbf r\rVert^2+\varepsilon^2}}\right]\mathbf I+\nonumber\\
&+\frac{\sqrt{\lVert\mathbf r\rVert^2+\varepsilon^2}+2\varepsilon}{4\pi (\sqrt{\lVert\mathbf r\rVert^2+\varepsilon^2}+\varepsilon)^2\sqrt{\lVert\mathbf r\rVert^2+\varepsilon^2}}\mathbf r\mathbf r.
\end{align}
We took the regularisation parameter to be $\varepsilon=0.25\times 2\pi/N$. We gauged the accuracy of the method by ensuring that the area of the deformed cluster never dropped below $95\%$ of the initial area at each step, as expected from incompressibility.

\section{Slender-body theory for 3D pusher jet}\label{Slender-Body Theory for 3D Pusher Jet}
 
In this Appendix, we provide the derivation of 
Eq.~\eqref{Straining Rate Zigzags}, \eqref{Shearing Rate Zigzags} in the main text. To this end, we must solve the {jet} equations \eqref{Plume Bulk Stokes} through \eqref{Boundary Advection Simplified}, where $\mathcal P$ is an elliptical cylinder with axis $\mathbf x(s)=\mathbf x(s_0)+s \mathbf t_1(s_0)$ ($-\infty<s<\infty)$ and constant cross-section $\mathcal C(s_0)$, given by
\begin{equation}
\mathcal C(s)=\{\mathbf x(s)+a(s)\cos(\eta) \mathbf t_1(s)+b(s)\sin(\eta)\mathbf t_2(s): 0\leq\eta<2\pi\}.    
\end{equation}
Defining local Cartesian coordinates $\mathbf x-\mathbf x(s_0)=x_1\mathbf t_1+x_2\mathbf t_2+x_3\mathbf t_3$, the local flow inside the jet associated with this configuration is assumed to be a combination of axial shear and cross-sectional strain:
\begin{subequations}
\begin{align}
\mathbf u&=\gamma x_1\mathbf t_3+E(x_1\mathbf t_1-x_2\mathbf t_2),\\
\boldsymbol{\sigma}&=-p_0\mathbf I+\gamma\mu(\mathbf t_1\mathbf t_3+\mathbf t_3\mathbf t_1)+2\mu E(\mathbf t_1\mathbf t_1-\mathbf t_2\mathbf t_2),
\end{align}
\end{subequations}
where $p_0$ is a constant. Letting $\mathbf n$ be the local outwards unit normal to this cylinder, the flow is determined by continuity of velocity and the requirement that, on the boundary, 
\begin{equation}
\boldsymbol{\sigma}\cdot\mathbf n+\frac{S\phi_0}{V}(\mathbf e_z\cdot\mathbf n)\mathbf e_z=\boldsymbol{\sigma}^{\text{ext}}\cdot\mathbf n    .
\end{equation}

\subsection{Outer velocity field}
The external flow, driven by the no-slip condition, is the negative of the perturbation flow created by a rigid elliptical cylinder in the unbounded flow $\mathbf u^{\infty}=\gamma x_1\mathbf t_3+E(x_1\mathbf t_1-x_2\mathbf t_2)$. Equating the boundary stresses, we obtain that 
\begin{equation}
\boldsymbol{\sigma}^{\infty}\cdot\mathbf n=-\frac{S\phi_0}{V}(\mathbf e_z\cdot\mathbf n)\mathbf e_z
\end{equation}
were $\boldsymbol{\sigma}^{\infty}\cdot\mathbf n$ is the traction on the rigid elliptical cylinder of cross-section $\mathcal C(s_0)$ immersed in the flow $\mathbf u^{\infty}$. Expressing $\mathbf e_z=\mathbf t_3\sin{\zeta}-\mathbf t_1\cos{\zeta}$ in terms of the local material frame, and noting that $\mathbf n\cdot\mathbf t_3=0$, we can equivalently write
\begin{equation}
\boldsymbol{\sigma}^{\infty}\cdot\mathbf n=\frac{S\phi_0}{V}(\mathbf t_1\cdot\mathbf n)\left(\frac{1}{2}\mathbf t_3\sin 2{\zeta}-\mathbf t_1\cos^2{\zeta}\right)\label{Sigma infinity dot n}.
\end{equation}
In Section~\ref{Determining the stress}, we show that
\begin{equation}
\boldsymbol{\sigma}^{\infty}\cdot\mathbf n=-p_1\mathbf n+\frac{\mu E(1+\chi)^2}{\chi}(\mathbf n\cdot\mathbf t_1)\mathbf t_1-\frac{\mu E(1+\chi)^2}{\chi}(\mathbf n\cdot\mathbf t_2)\mathbf t_2+\frac{\mu\gamma(\chi+1)}{\chi}(\mathbf n\cdot\mathbf t_1)\mathbf t_3, \label{Stress on cylinder}
\end{equation}
where $p_1$ is a constant. Assuming that Eq.~\eqref{Stress on cylinder} is true, and writing $-p_1\mathbf n=-p_1(\mathbf n\cdot\mathbf t_1)\mathbf t_1-p_1(\mathbf n\cdot\mathbf t_2)\mathbf t_2$, Eq.~\eqref{Sigma infinity dot n} yields the simultaneous equations 
\begin{align}
&p_1-\frac{\mu E(1+\chi)^2}{\chi}=\frac{S\phi_0\cos^2{\zeta}}{V},\\
&p_1+\frac{\mu E(1+\chi)^2}{\chi}=0,\\
&\frac{\mu\gamma (1+\chi)}{\chi}=\frac{S\phi_0 \sin 2{\zeta}}{2V},
\end{align}
which can be solved to yield Eqs.~\eqref{Straining Rate Zigzags} and \eqref{Shearing Rate Zigzags} in the main text.

\subsection{Determining the stress}\label{Determining the stress}
To conclude, we need to show that the stress on a rigid cylinder in the flow $\mathbf u^{\infty}$ indeed takes the form \eqref{Stress on cylinder}. The first term in Eq.~\eqref{Stress on cylinder} corresponds to an isotropic pressure contribution, while the second and third term are obtained from \eqref{Traction of Ellipse}. The final term, as we will now show, is the traction on the elliptical cylinder exerted by an external shearing flow $\gamma x_1\mathbf t_3$. To compute this traction, notice that the flow in the presence of the rigid cylinder must remain purely axial, with $\mathbf u=w(x_1,x_2)\mathbf t_3$. {In the absence of any axial pressure gradients}, we deduce that $w$ satisfies
\begin{subequations}
\begin{align}
\nabla^2 w&=0 & &\text{outside  } \mathcal C,\label{w harmonic}\\
w&=0 & &\mathbf x\in \mathcal C, \label{w no-slip}\\
w&\sim \gamma x_1 & &x_1^2+x_2^2\to\infty. \label{w asymptotics}
\end{align}
\end{subequations}
Eqs.~\eqref{w harmonic} through \eqref{w asymptotics} may be solved in elliptic coordinates $(\tau,\theta)$, defined as
\begin{subeqnarray}
   x_1&=&c\cosh \tau\cos\theta,\\
   x_2&=&c\sinh \tau\sin \theta,
\end{subeqnarray}
with $0\leq\theta<2\pi$, $0\leq\tau<\infty$. The elliptical boundary $\mathcal C$ corresponds to the level curve $\tau\equiv \tau_0$, where $\tanh(\tau_0)=\chi$ (the aspect ratio) and $c=a(1-\chi^2)^{1/2}$. Furthermore, in elliptic coordinates, Laplace's equation takes the simple form
\begin{align}
\frac{\p^2 w}{\p \tau^2}+\frac{\p^2 w}{\p \theta^2}=0.
\end{align}
The solution to Eqs.~\eqref{w harmonic} through \eqref{w asymptotics} may now be determined from separation of variables, after noting that $\displaystyle w\sim \frac{1}{2}\gamma ce^{\tau}\cos\theta$ for $\tau\to\infty$. We find
\begin{equation}
w=\frac{1}{2}\gamma c(e^{\tau}-e^{2\tau_0-\tau})\cos\theta,    
\end{equation}
with associated boundary stress
\begin{align}
\left.\boldsymbol{\sigma}\cdot\mathbf e_{\tau}\right\vert_{\tau=\tau_0}&=\frac{\mu}{c(\cosh^2\tau_0-\cos^2\theta)^{1/2}}\left.\frac{\p w}{\p \tau}\right\vert_{\tau=\tau_0}\mathbf t_3\nonumber\\
&=\frac{\mu \gamma e^{\tau_0}\cos\theta}{(\cosh^2\tau_0-\cos^2\theta)^{1/2}}\mathbf t_3\nonumber\\
&=\frac{\mu\gamma e^{\tau_0}}{\sinh \tau_0}(\mathbf n\cdot\mathbf t_1)\mathbf t_3\nonumber\\
&=\frac{\mu \gamma(\chi+1)}{\chi}(\mathbf n\cdot\mathbf t_1)\mathbf t_3.
\end{align}
This is precisely the final term in Eq.~\eqref{Stress on cylinder}, completing our analysis.

\section{Caption of supplementary movies}

\begin{itemize}[label={$\bullet$}]

\item Movie S1: Droplet instability - darkfield movie with $2 {\rm \mu m}$ polystytrene beads as tracer particles, recorded in the laboratory frame (sped up $\times 4$). {Scale bar is $200\ {\rm \mu m}$}.

\item Movie S2: Zigzag instability - darkfield movie with $2 {\rm \mu m}$ polystytrene beads as tracer particles, recorded in the laboratory frame (sped up $\times 5$). {Scale bar is $200\ {\rm \mu m}$}.

\end{itemize}

\end{document}